\documentclass[aps,rmp,showkeys]{revtex4}

\newcommand{\nn}{\nonumber}
\newcommand{\be}{\begin{equation}}
\newcommand{\ee}{\end{equation}}
\newcommand{\ba}{\begin{eqnarray}}
\newcommand{\ea}{\end{eqnarray}}
\newcommand{\eq}[1]{(\ref{#1})}
\newcommand{\1}{\'{\i}}
\newcommand{\gsim}{\raise.3ex\hbox{$>$\kern-.75em\lower1ex\hbox{$\sim$}}}
\newcommand{\lsim}{\raise.3ex\hbox{$<$\kern-.75em\lower1ex\hbox{$\sim$}}}

\def\slash#1{\setbox0=\hbox{$#1$}#1\hskip-\wd0\hbox
to\wd0{\hss\sl/\/\hss}}

\begin{document}

\preprint{hep-th/0304245}
\preprint{UGVA-DPT-2003-04-1106}

\title{The Stueckelberg Field}

\author{Henri Ruegg}%
\email{henri.ruegg@physics.unige.ch}

\author{Mart\1 Ruiz-Altaba}%
\email{ruiz@kalymnos.unige.ch}

\affiliation{%
D\'epartement de Physique Th\'eorique, Universit\'e de Gen\`eve \\
24 Quai Ernest Ansermet,
1211 Gen\`eve 4, Switzerland}


\begin{abstract}
In 1938, Stueckelberg introduced a scalar field which makes an
Abelian gauge theory massive but preserves gauge invariance. The
Stueckelberg mechanism is the introduction of new fields to reveal
a symmetry of a gauge--fixed theory. We first review the
Stueckelberg mechanism in the massive Abelian gauge theory.
  We then extend  this idea to the standard model, stueckelberging the
hypercharge
 $U(1)$ and thus giving a mass to the physical photon.
 This introduces an infrared regulator for the photon in the
standard electroweak theory, along with a modification of the
 weak mixing angle
 accompanied by a plethora of new effects. Notably, neutrinos
 couple to the photon and charged leptons have also a pseudo-vector
 coupling.  Finally, we  review  the historical influence  of  Stueckelberg's 1938 idea,
 which led to applications in many areas not anticipated by the author, such as
strings. We describe the numerous proposals to generalize the
Stueckelberg trick to the non-Abelian case with the aim to find
alternatives to the standard model. Nevertheless, the Higgs
mechanism in spontaneous symmetry breaking remains the only
presently known
 way to give masses to non-Abelian vector fields in a renormalizable and unitary theory.

\end{abstract}



\maketitle

\tableofcontents

\vskip2cm

\section{\label{sec:intro}Introduction}



Research on the theory of massive vector fields started with   \cite{Pro36},
and reached a major milestone with the standard electroweak  theory  which  is
unitary and renormalizable, and successful.

In this paper, we would like to review a contribution of Ernst C.G.
Stueckelberg in 1938, namely his introduction of the scalar
``$B$--field'' of
positive--definite metric, accompanying a massive Abelian vector field
\cite{Stu38I,Stu38II,Stu38III}. We shall see that this idea had many different
applications which went far beyond the original motivations of its author.
Stueckelberg also invented in that year the general formulation of baryon
number conservation \cite[p. 317]{Stu38III}.

We will not review other important contributions by Stueckelberg. It is,
nevertheless, noteworthy that among his discoveries one can include the
picture of  antiparticles  as particles moving backwards in time, implying pair
creation and annihilation \cite{Stu41,Stu42}, the causal
propagator \cite{Stu46}, \cite{Riv48,Riv49,Stu50} and the renormalization group
\cite{Stu51,Stu53}.

We shall mainly discuss three important topics, (1) hidden symmetry,  (2)
renormalizability, and (3) infrared divergences.

It was believed by many that only massless vector theories were
gauge--invariant. Then  \cite{Pau41} showed
 that Stueckelberg's formalism for a massive vector field satisfied a
restricted $U(1)$ gauge invariance, similar to the one encountered
in quantum electrodynamics, but with the gauge function $\Lambda(x)$ restricted
by the massive
Klein--Gordon equation.

Much later, Delbourgo,  Twisk and  Thompson found that Stueckelberg's
lagrangian for real vector fields, complemented with ghost terms, is actually
BRST invariant \cite{Del88}. The BRST symmetry   \cite{Bec74,Bec75,Tyu75}
allows for a systematic and
convenient exploration of gauge symmetries, and in fact   the
$S$--matrix elements of a
BRST--invariant theory are independent of the
gauge--fixing terms \cite{Lee76}. The BRST
symmetry facilitates considerably the effort to prove the perturbative
renormalizability of a theory, as well as its unitarity
\cite{Alv83,Bec81}, see also \cite{Zin75}, \cite{Kra98} and the textbook
\cite{Col84}.

Charged vector theories (for example
non-Abelian gauge theories) are trickier.
The electroweak theory, with spontaneously broken $SU(2)_L\times U(1)_Y$
symmetry, is renormalizable \cite{Hoo71b}. As is well known, this theory
comprises two charged and
two neutral vector bosons.  A technical problem remains:  the infrared
divergences of massless (vector) field theories.

One important aim of the present paper, therefore, is to construct a BRST
invariant $SU(2)_L\times U(1)_Y$ theory with a massive photon, which calls for
a Stueckelberg field, along with the appropriate ghost terms. This development
was suggested in \cite{Sto00}. A complication is
that the
BRST--invariant infrared regulator not only gives a mass to the
photon, but also changes its couplings to the fermions, the weak mixing angle,
and
more. But  the new terms in the lagrangian (or their modifications) are
proportional at least to the photon's mass squared, and thus very small.
Indeed, from a Cavendish experiment and the known value of the galactic
magnetic field one finds the stringent upper
limit $m_\gamma<10^{-16}$~eV \cite[p. 249]{PDBook}.

 This ``stueckelberged'' standard model has the important advantage of providing an
infrared cut-off for
 photon interactions, while preserving BRST invariance. Thus the infrared
 catastrophe should be avoided without spoiling ultraviolet renormalizability
and  unitarity. An explicit proof of this fact has not been fully worked out
yet.
 The brute force alternative, to give an explicit mass to the photon which
becomes zero at the end of the calculation and not modifying any
of the other couplings in the electroweak theory, works well for
the computation of $S$--matrix elements \cite{Pas79,Itz80}.
Dimensional regularization can be used also to provide an infrared
regulator \cite{Gas76}, but then the spacetime dimension must be
extended to a higher value, instead of a lower one as required to
regulate ultraviolet divergences, whereby the computation is
tediously long but straightforward \cite{Col84}.

 This paper is organized as follows.

 In section \ref{sec:quamasvecfi}, we review the original formalism and
 motivations of Proca and Stueckelberg.

 In section \ref{sec:brs}, we review the BRST  invariance properties of the
 massive neutral vector field together with the Stueckelberg
 $B$--field.

In section \ref{sec:uone} we consider in some detail the case of a
$U(1)$ gauge field coupled to matter, and study the Stueckelberg
mass mechanism in the absence and in the presence of spontaneous
symmetry breaking and the usual Higgs mechanism. This section is a
warm-up for the core of the paper, in section \ref{sec:sm}, where
we write out the $SU(2)_L\times U(1)_Y$ lagrangian with a
Stueckelberg field and a mass term for the hypercharge vector
boson, as well as with the usual spontaneous symmetry breaking. We
write out the full ghost sector and check the BRST invariance of
the full lagrangian. We then turn to some of the phenomenological
consequences of the model. In particular, we study mass matrices,
mixing angles and currents, and we scratch the surface of some
conundrums related to anomalies and the electric charge.
Curiously, one should now distinguish the quantum field $A_\mu$
responsible for photon scattering from the external $A_\mu^{\rm
e.m.}$ which enters in the calculation of, \sl e.g. \rm the
$(g-2)$ value of the electron: for a massless photon the two
fields coincide.

Section \ref{sec:influ} is devoted to a historical review of the influence of
Stueckelberg's three 1938 papers. In the forties and fifties, the
renormalizability of the massive Abelian Stueckelberg theory was painfully
established. There was a long debate in the sixties and seventies about the
non-Abelian case, which waned when massive
Yang--Mills with spontaneous symmetry breakdown and the Higgs mechanism was
shown to be unitary and renormalizable \cite{Hoo71b}. The problem is nowadays
essentially solved: no renormalizable and unitary
non-Abelian Stueckelberg model has been found, and \cite{Hur97} has claimed
that it is impossible to do so in perturbation theory.   Renormalizable models
of massive
Yang--Mills without Higgs mechanism nor Stueckelberg field have
been exhibited, but they are not unitary.

 On the other hand, Stueckelberg fields were introduced very early in string
 theory by Pierre Ramond and collaborators, both  for the formulation of the
antisymmetric partner to the graviton \cite{Kal74} and in
covariant string field theory \cite{Mar75,Ram86}. Stueckelberg
fields   turned out  to be crucial in the covariant quantization
of the spacetime supersymmetric  string \cite{Ber90b} and in the
destruction of unwanted $U(1)$s in string phenomenology
\cite{Ald00}. They have also proven useful in the study of
dualities in field and string theory, see  section
\ref{sec:pubel}.

\section{\label{sec:quamasvecfi}Quantization of the massive vector field}

The electromagnetic potential is described by a neutral vector field $A_\mu$
obeying Maxwell's equations. Its quantization gives rise to a massless
particle, the photon, which has only two physical degrees of freedom, its two
transverse polarizations or, equivalently, its two helicities (+1 and -1). The
vector field $A_\mu$ has, however, four components.
This is an example of how physicists   introduce
apparently unphysical entities in order to simplify the theory. Indeed, with
the four--vector $A_\mu$ one can construct a manifestly
Lorentz--invariant theory of the potential
$A_\mu$ interacting with a current $j_\mu$.

How can one reduce the four components of $A_\mu$ to the two physical degrees
of freedom of the photon? First, four becomes three in a
Lorentz--invariant way
by imposing the Lorentz subsidiary condition $\partial^\mu A_\mu=0$ (for
quantum subtleties, see below). With gauge invariance and the mass--shell
condition, only two components survive.

Contrariwise, if one adds to the wave equation of $A_\mu$ a mass term, the
gauge
invariance is lost, because the field $A_\mu$ transforms inhomogeneously  and
thus the mass term in the lagrangian is not invariant. The three components of
$A_\mu$ left by the Lorentz condition are then interpreted as belonging to a
massive vector field, that is a massive particle of spin one. This
spin--one
object has now a longitudinal polarization, in addition to the two transverse
ones.

Stueckelberg's wonderful trick consists in introducing an extra physical scalar
field $B$, in addition to the four components $A_\mu$, for a total of five
fields, to describe covariantly the three polarizations of a massive vector
field. With the Stueckelberg mechanism, which we shall exhibit in more detail
below, not only is Lorentz covariance manifest, but also, and most
interestingly, gauge invariance is also manifest. The
Stueckelberg field restores the gauge symmetry which had been broken by the
mass term.


\subsection{Proca} The original  aim of \cite{Pro36} was to describe the four
states of electrons and positrons by a Lorentz
four--vector. The motivation was to
imitate the procedure of \cite{Pau34}, who had quantized
the scalar field obeying the
Klein--Gordon equation and had
interpreted the conserved current as carrying electric charge rather than
probability. This, they thought, eliminated negative probabilities. Of course,
Proca's choice is inadequate for describing spin--$\frac 1 2$ particles, and it
didn't make much sense either a decade after Dirac's equation.
 Nevertheless, Proca's  mathematical formalism describes well
a  massive real or complex  vector field. The original papers are
devoted to the complex case, and we shall present it now, although
later on we shall consider the real case exclusively.

 Proca's equation of motion for a free complex vector field $V_\mu$ reads as
follows \cite{Pro36}:
 \be \partial^\mu F^V_{\mu\nu}(x) + m^2 V_\nu(x)  =0 \label{22} \ee
 where the field strength is
 \be F^V_{\mu\nu}=\partial_\mu
V_\nu -\partial_\nu V_\mu \label{21} \ee
with $\partial_\mu =\partial/\partial  x^\mu$ and the sign in \eq{22}
depends on the metric, which is $(+ - - -)$
throughout this paper. Differentiating the equations of motion \eq{22} with
respect to $x^\nu$ yields immediately the Lorentz condition \be \partial^\mu
V_\mu=0 \label{23} \ee Notice that a
non-zero mass is crucial for \eq{23} to
follow from the equations of motion.

Hence, $V_\mu(x)$ describes a
spin--one particle of
non-zero mass $m$.
Following the lucid presentation in \cite{Wen43,Wen48}, equation \eq{22} can be
derived from a lagrangian density for the complex $V_\mu$: \be {\cal L}= -
\frac12 F_{\mu\nu}^{V\dagger} F^{V\mu\nu} +m^2 V_\mu^\dagger V^\mu \label{24}
\ee where $\,^\dagger$ means hermitian conjugation. The hamiltonian density
 following from the above lagrangian density has three positive terms
involving the spatial components $V_i$ ($i=1,2,3$), and one negative term
depending on $V_0$. This last term can be eliminated using the Lorentz
condition \eq{23} and the resulting hamiltonian is thus positive definite.

It is then possible to use the canonical formalism to find, first, the
commutation relations for the spatial components of the vector field, and then,
using again the Lorentz condition \eq{23}, the commutations relations for its
temporal component. The result is \cite{Wen43} \be \left[ V_\mu(x) , V_\nu(y)
\right] =\left[ V_\mu^\dagger (x) , V^\dagger _\nu(y) \right] = 0 \label{25}
\ee
\be \left[ V_\mu(x) , V^\dagger _\nu(y) \right] = - i \left( g_{\mu\nu}
+\frac{1}{m^2} \partial_\mu\partial_\nu \right) \Delta_m (x-y) \label{26} \ee
Here, the massive generalization $\Delta_m(x)$ of the
Jordan--Pauli function
obeys \be \left(\partial^2 +m^2\right) \Delta_m(x)=0 \label{27}\ee where
$\partial^2=\partial_\mu\partial^\mu$.

The commutator \eq{26} differs from the corresponding expression in QED by the
second term in the
right--hand side, proportional to $1/m^2$, which is either
absent
(Stueckelberg--Feynman gauge) or with coefficient $-1/\partial^2$ instead of
$1/m^2$
(Landau gauge). After 1945 it became clear that the term $m^{-2}
\partial_\mu\partial_\nu$ gives rise to (quadratic) divergences at high
energies which cannot be eliminated by the renormalization procedure

\subsection{Stueckelberg}

Stueckelberg's formalism for the vector field differs from
Proca's. His motivations made sense at the time, so let us sketch
briefly the historical framework. Recall that   Yukawa, in order
to explain the nuclear forces, postulated the existence of a
massive particle which would mediate them, just as the photon
mediates the Coulomb force between charged particles\footnote{Late
in his life, in 1979, Stueckelberg wrote a letter to V.~Telegdi
stating  that ``I had the same idea, probably before Yukawa.'' But
it was never published.}. The first attempt \cite{Yuk35} called
for  the exchanged particle to be a component of a Lorentz
four--vector (all computations were carried in the static
approximation). Then, \cite{Yuk37}   proposed a scalar particle
instead.
  \cite{Stu38I} showed  that choosing a scalar would lead to a
repulsive instead of attractive nuclear interaction\footnote{To
find phenomenological agreement, one needs to consider an isospin
triplet of pseudoscalar intermediate particles \cite{Kem38a}.
\cite{Stu37} and \cite{Bha38} first noticed that the Yukawa
particle could decay into electrons (this is wrong, of course,
since a muon is not a pion; it was a major discovery later on
\cite{Lat47a,Lat47b,Lat47c} that
the ``$\mu$-meson'' was a different particle from the
``$\pi$-meson''. See also the criticisms of Yukawa's scalar
proposal in \cite{Kem38b,Kem38c,Ser38} and the comments in
\cite[p. 434]{Pai86}.} and then  turned  to reconsider the
exchange of a massive charged vector particle. The guiding
principle of this research  was to develop  a formalism as close
as possible to QED.

Instead of Proca's equation of motion \eq{22}, \cite{Stu38II}  wrote simply
\be \left(\partial^2 +m^2\right) A_\mu (x) =0 \label{28} \ee
which follows from the covariant lagrangian density \be {\cal L}= -\partial_\mu
A_\nu^\dagger \partial^\mu A^\nu +m^2 A_\mu^\dagger A^\mu \label{29} \ee The
difference between this lagrangian and Proca's \eq{24} (other than the change
in notation between $V_\mu$ and $A_\mu$) is a term $ \partial_\mu A_\nu^\dagger
\partial^\nu A^\mu $ which, up to total derivatives, is $( \partial^\nu
A_\nu^\dagger )(\partial^\mu A_\mu )$. This term, present in Proca's lagrangian
but not in Stueckelberg's, is responsible for being able to derive the Lorentz
condition \eq{23} from Proca's lagrangian.

So following the QED track leads, not surprisingly, to the fact that the gauge
condition must be imposed as a supplementary ingredient besides the covariant
lagrangian: a disadvantage of Stueckelberg's procedure is thus that the Lorentz
subsidiary condition does not follow from the equation of motion, as was the
case with Proca. This feature has terrible consequences on the positivity of
the hamiltonian, which is now
\be {\cal H} = - \sum_{\lambda=0,1,2,3} \left (
\partial_\lambda A_\mu^\dagger \right)\left(\partial_\lambda A^\mu \right) -m^2
A_\mu^\dagger A^\mu \label{210}\ee
The explicit sum over $\lambda$ gives no trouble, but
the implicit sum over $\mu$ does, since $A_\mu=g_{\mu\nu}A^\nu$ and thus the
contribution from $A_0$ to the energy density is negative, whereby it is not
possible to conclude that \eq{210} is positive definite. In Proca's formalism,
the negative term with $A_0$ can be eliminated with the subsidiary condition
which follows from the field equations,
but now we do not have it automatically. Where does the subsidiary condition
come from, then?

In QED, the same problem arises: $\partial^\mu A_\mu=0$ does not follow from
the
equations of motion.  \cite{Fer32} proposed   to impose instead $\partial^\mu
A_\mu \left|{\bf
phys}\right>=0$, with $\left|{\bf phys}\right>$ an admissible physical state of
the system; see also the discussion in
\cite[pp. 354--355]{Pai86}. Even this condition is too strong, however, because
it restricts
the
space of physical states to nothing. But its spirit is correct. In fact, we
only
need $\left<{\bf phys}'\right|\partial^\mu A_\mu \left|{\bf phys}\right>=0$. We
impose then \cite{Gup50,Ble50} the weaker, and sufficient, condition
\be\partial^\mu A_\mu ^{(-)}\left|{\bf phys}\right>=0 \label{212}\ee where
$A_\mu ^{(-)}$ involves only
free--field annihilation operators ($A_\mu = A_\mu
^{(-)} +A_\mu ^{(+)}$, with $A_\mu ^{(+)}$ involving only creation operators); the only
requirement on the choice of polarization is that the hamiltonian be bounded below.
The Hilbert space still has indefinite metric, but the space of physical states
is of positive definite norm.

Could the same trick be used in the massive vector field theory?
Surprisingly, because of the non-zero mass, one cannot impose the
operator condition \eq{212}, since it comes into conflict with the
canonical commutation relations. The commutation relations of
Stueckelberg's vector field are the same as those of the QED's
photon in what later was to be called the Feynman gauge: \ba
&&\left[ A_\mu(x) , A_\nu(y) \right] = 0 \nn \\ &&\left[ A_\mu(x)
, A^\dagger_\nu(y) \right] = -i g_{\mu\nu} \Delta_m(x-y)
\label{211} \ea except that $\Delta_m$ obeys the Klein--Gordon
equation with $m\not=0$. It is easy to derive from \eq{211} the
commutator \be \left[ \partial^\mu A_\mu(x) ,
\partial^\nu A^\dagger_\nu(y) \right] = i\,\partial^2 \Delta_m(x-y) \label{213}
\ee In QED, $\partial^2 \Delta_0=0$, so \eq{212} is consistent indeed. But in
the
massive vector case, since $\partial^2 \Delta_m = -m^2 \Delta_m$, the
subsidiary condition \eq{212} is inconsistent with the commutation
relations.

Stueckelberg brilliantly solved this puzzle by introducing  a new
additional scalar field $B(x)$ which is now known as the
Stueckelberg field \cite[p. 243]{Stu38I}, \cite[p. 302]{Stu38II}.
Note that the Hilbert space norm of the Stueckelberg field is
positive, a simple fact which has caused much confusion when
overlooked.

In the original formulation, Stueckelberg's $B$ field obeys the same equation
\eq{28}
as the vector field $A_\mu$; both fields are complex and with the same mass
$m$: \be \left( \partial^2
  +m^2 \right) B(x) =0 \label{214}\ee
  In close analogy with \eq{211},
  \ba &&\left[ B(x), B(y)\right] =0 \nn \\ &&
\left[ B(x), B^\dagger(y)\right] =i \Delta_m(x-y) \label{215}\ea The subsidiary
condition \eq{212} is replaced (in the
Gupta--Bleuler version), however, by \be
S(x)\left|{\bf phys}\right>\equiv \left( \partial^\mu A_\mu(x) +m\;
B(x)\right)^{(-)} \left|{\bf phys}\right> =0\label{216}\ee and one verifies
easily that $S(x)$ commutes both with $S(y)$ and with $S^\dagger(y)$. A short
explicit calculation shows that, after imposing the subsidiary condition
\eq{216}, the hamiltonian is positive definite.

 Instead of \eq{29}, the Stueckelberg lagrangian density is now
 \cite[p. 313]{Stu38III}
 \be {\cal
 L}_{\rm Stueck} = - \partial _\mu A_\nu^\dagger \partial^\mu A^\nu + m^2
 A^\dagger_\mu A^\mu + \partial _\mu B^\dagger \partial^\mu B -m^2 B^\dagger B
 \label{5Lnew} \ee which describes consistently and covariantly a free massive
 charged vector field, accompanied by the Stueckelberg scalar. An enormous
 advantage of Stueckelberg's formalism is the absence of derivatives in the
 commutation relations. These derivatives would make the theory more singular
at  higher energies.

On the other hand, the number of degrees of freedom has now been increased to
five, instead of the required three for a massive vector field. The situation
is
somewhat similar to that encountered in QED. The subsidiary condition \eq{216}
can be used to decrease the number of components to four. In addition,
Stueckelberg's theory satisfies a new gauge invariance \cite{Pau41}, a feature
that explains the lasting success of this formalism in the literature up to our
days. The fact that a supplementary condition has to be imposed on physical
states, like in QED, just means that the theory enjoys a gauge invariance which
must be fixed, like in QED. Pauli's gauge transformations are the following
(see
also the discussion of this gauge invariance in \cite{Gla53}): \ba &A_\mu
(x)\to
A_\mu'(x) = A_\mu(x) +\partial_\mu \Lambda(x) \label{2177}\\ &B(x)\to B'(x) =
B(x) +m \Lambda(x) \label{217}\ea with the complex gauge function $\Lambda$
subject to the same field equation as $B$ and $A_\mu$:
\be
\left(\partial^2 +m^2\right) \Lambda(x) =0 \label{218}\ee

The gauge invariance is manifest if we rewrite the lagrangian
\eq{5Lnew} as \ba & {\cal L}_{\rm Stueck} =& - \frac12
F_{\mu\nu}^\dagger F^{\mu\nu} +m^2 \left( A_\mu ^\dagger - \frac1m
\partial_\mu B^\dagger \right) \left( A^\mu - \frac1m \partial^\mu B \right)
\nn\\ && - \left( \partial^\mu A_\mu ^\dagger + m B^\dagger \right) \left(
\partial_\nu A^\nu + m B \right) \label{1fenetre} \ea
where \be F_{\mu\nu}(x)= \partial_\mu A_\nu(x) -\partial_\nu A_\mu(x) \ee
and we have dropped a total derivative.
The first two terms are
invariant for arbitrary $\Lambda(x)$ while the invariance of the last term
requires \eq{218}. This gauge invariance is responsible for lowering the number
of
local degrees of freedom to three, the required number for a massive vector
field. The important difference between the above gauge transformation and the
usual Abelian gauge transformation in QED is that here the gauge parameter
$\Lambda(x)$ is restricted by \eq{218}. This dynamical restriction does not
change the number of degrees of freedom, however. Note, in passing, that the
vector field in QED is real, whereas here we are following the original papers
 which dealt with complex fields. The discussion on the number of
fields and physical degrees of freedom does not change if these fields and
degrees of freedom are complex or real.

The second term in equation \eq{1fenetre}, which gives rise to the
mass term for the vector and to the kinetic term for the
Stueckelberg scalar, clearly displays another way of thinking of
the Stueckelberg mechanism, in terms of representations of the
Lorentz group. Spin one representations can be built from a vector
or, with the help of the momentum operator, from a scalar. The
Stueckelberg trick is to couple the spin one $A_\mu$ with the spin
one $\partial_\mu B$ to have enough degrees of freedom for a
gauge-invariant massive vector field. Alternatively, it
compensates the scalar piece $\partial_\mu A^\mu$ of the vector
field with the Stueckelberg scalar $B$.

To conclude the comparison between Proca's and Stueckelberg's
formalism, let us note that the latter can be brought quite close
to the former. Indeed, if one defines \cite{Pau41} \be V_\mu
\equiv A_\mu -\frac1m \partial_\mu B \label{220}\ee one sees that
Stueckelberg's lagrangian \eq{1fenetre} is the sum of Proca's
\eq{24} plus an extra term: \be {\cal L}_{\rm Stueck}= {\cal
L}_{\rm Proca} - ( \partial^\mu A_\mu^\dagger + m \, B^\dagger) (
\partial^\nu A_\nu + m \, B) \label{ssss} \ee Note that $V_\mu(x)$
is gauge--invariant under \eq{2177} ($V'_\mu=V_\mu$) and so is
$F_{\mu\nu} = F_{\mu\nu}^V$. The supplementary condition \eq{216}
is the same as Proca's \eq{23} on the Stueckelberg field's mass
shell.

 On physical states, for which \be\left< {\bf phys} \right| \partial^\nu A_\nu
 + m \, B | {\bf phys'}>=0\label{klop} \ee Stueckelberg's lagrangian
(\ref{ssss}) coincides
 with Proca's.
 As a general rule, keep in mind that for
renormalization purposes it will turn out to be  advantageous to
keep $A_\mu$ and $B$ independent as long as possible: it is not a
good policy to eliminate $B$ and recover the cumbersome Proca
lagrangian!

We shall discuss the physical relevance of Stueckelberg's $B$
field in the next section, in the context of real vector (and
Stueckelberg) fields. We have presented the complex formulation
here because that is what was originally studied by Proca and
Stueckelberg.

\section{\label{sec:brs}BRST Invariance}

We confine ourselves to real vector fields from now on.
Stueckelberg's theory  of a massive vector field $A_\mu$
accompanied by the real Stueckelberg scalar field $B$
\cite{Stu38I,Stu38II} satisfies a gauge invariance despite the
presence of the mass term for $A_\mu$ \cite{Pau41}: \ba \delta
A_\mu(x) = \partial_\mu \Lambda(x) \label{a31}\\ \delta B(x) = m\;
\Lambda(x) \label{31}\ea where the real gauge parameter
$\Lambda(x)$ is restricted by \eq{218}.

To see this, we start from the lagrangian \eq{5Lnew} specialized
to real fields \be {\cal L}_{\rm Stueck}=-\frac12 \partial^\mu
A^\nu \; \partial_\mu A_\nu +\frac12 m^2 A^\mu A_\mu +\frac12
\partial^\mu B \;\partial_\mu B -\frac12 m^2 \; B^2 \label{32}\ee
and rewrite it (up to total derivatives) as \be {\cal L}_{\rm
Stueck}=-\frac14 F_{\mu\nu} ^2 +\frac12 m^2 \left( A^\mu - \frac1m
\partial^\mu B\right) ^2 -\frac12\left( \partial_\mu A^\mu +m
\,B\right)^2 \label{33}\ee

Let us consider now the supplementary condition \eq{216} on physical states,
which
involve both the Stueckelberg and the vector fields. The Stueckelberg field
actually
participates in the definition of
asymptotic states, so there one certainly cannot ignore it \cite{Pic02}!
\cite{Gla53}
 pointed out  that the condition \eq{216} can be viewed
as a practical definition of the Stueckelberg $B$ field. The massive photon's
three physical components fix the value of $B$ through the physical state
condition, consistently. (Note that in the original literature, these massive
photons were called mesons.) See section \ref{sec:hs} below for a
more detailed discussion of these points.

Consider now the expression \be {\cal L}_{gf}=-\frac1{2\alpha} \left(
\partial_\mu A^\mu + \alpha \; m \, B\right)^2 \label{35}\ee which is just
like the 't~Hooft gauge--fixing term in the Abelian Higgs model,
where $B$ is replaced by the Goldstone boson\footnote{More generally,
one could use two different parameters,
$(2\alpha_1)^{-1}(\partial_\mu A^\mu + \alpha_2 m )^2$. This
generalization is useful in checking gauge independence of
observables in some calculations, but it has the disadvantage that
a mixing quadratic term $A^\mu \partial_\mu B$ survives in the
lagrangian.}.

We see that the Stueckelberg lagrangian \eq{33} is in fact the
Proca lagrangian supplemented by the 't~Hooft gauge--fixing term
in the Stueckelberg--Feynman gauge $\alpha=1$ \cite[p. 37]{Tay76}.
This is crucial for understanding Stueckelberg's contribution in
modern terms. The point, clarified by 't~Hooft and Veltman, is that we can
allow $\alpha$ to be any real parameter, and thus Stueckelberg's
lagrangian \eq{33} is better written for arbitrary $\alpha$ as \be
{\cal L}_{\rm Stueck}=-\frac14 F_{\mu\nu}  ^2 +\frac12 m^2 \left(
A^\mu -\frac1m \partial^\mu B\right) ^2 - \frac1{2\alpha}\left(
\partial_\mu A^\mu +\alpha\,m \,B\right)^2 \label{3333}\ee

The restriction \eq{218} on $\Lambda$ is now $(\partial^2 +\alpha m^2
)\Lambda=0$,
satisfied because of the equation of
motion for $B$, which is simply the same $(\partial^2 +\alpha m^2)B=0$.

The gauge fixing term \eq{35} could lead to non--acceptable
physics, negative-norm modes remaining coupled to physical modes,
but \cite{Del88} showed that it does not. Indeed, if one adds the
appropriate terms with the well--known ghosts \cite{Fad67} to
Stueckelberg's lagrangian \eq{32}, one obtains a theory invariant
under the BRST symmetry \cite{Bec74,Bec75,Tyu75}, as we shall show
shortly.

 For completeness, and to come closer to QED, consider also the lagrangian for
a
fermion minimally coupled to the massive vector field \be {\cal L} _f =
\bar\psi
\left[\gamma^\mu \left( i \partial_\mu +g\, A_\mu \right) - M \right] \psi
\label{36}\ee It is invariant under the  gauge transformation \ba
&& A_\mu'(x) = A_\mu(x)+\frac1m \partial_\mu B(x) \label{t7}\\  && \psi'(x) =
{\rm  e}^{iB(x)/m} \psi(x) \label{3333777} \\ &&{\bar\psi}'(x) = {\rm
e}^{-iB(x)/m}
\bar\psi(x)
\label{37} \ea

On the other hand, consider the Proca vector interaction
\be {\cal L}_{f}'= \bar \psi \left[ \gamma^\mu \left( i \partial_\mu + g V_\mu
\right) -M \right] \psi \label{34bis} \ee
The substitution $V_\mu = A_\mu + \frac1m \partial_\mu B$ yields
\be {\cal L}_{f}'= \bar \psi \left[ \gamma^\mu \left( i \partial_\mu + g A_\mu
+ \frac{g}m \partial_\mu B \right) -M \right] \psi \label{34bisbille} \ee
This shows that the bad high--energy behavior of the Proca lagrangian comes
from the term $\partial_\mu B$. However, the transformations  \eq{3333777} and
\eq{37} eliminate this term and leave us with \eq{36} \cite{Sto00}.

This  is an instance of a general $U(1)$ gauge transformation
\eq{31} where the function $\Lambda(x)$ is chosen proportional to
$ B(x)$. The elimination of $\partial_\mu B$ from the interaction
is an important step in the proof of renormalizability. Let us
stress also that the field $B(x)$ must be renormalized in a
gauge--invariant way  \cite{Gla53}. The Stueckelberg massive Abelian model was rigorously
proved to be renormalizable and unitary by  \cite{Low72}, see also \cite{Hee03}.

We now turn to the BRST symmetry for the Stueckelberg theory
coupled to a fermion, ${\cal L}={\cal L}_{\rm Stueck} +{\cal
L}_f$, given by eqs. \eq{3333} and \eq{36} above.

Let $\omega(x)$ and $\omega^*(x)$ be independent scalar anticommuting  fields.
First,
read off from the infinitesimal gauge transformations   the following  BRST
transformation ${\bf s}$: \ba &&{\bf s}\,A_\mu
=\partial_\mu \omega \label{38aa}\\ &&{\bf s}\,B=m\,\omega \\ && {\bf s}\,\psi=
 i\,g\;\omega
\psi \\ &&{\bf s}\,\bar\psi= -i\,g\; \omega \bar\psi \\ &&{\bf s}\,\omega =0
\label{38}
\ea Note that  $\omega(x)$ is an anticommuting scalar, $\omega(x)^2=0$, and
thus the BRST transformation $\bf s$ is nilpotent, ${\bf s}^2=0$, even
off--shell.

The crucial property of the BRST transformation $ {\bf s}$ is that it is
nilpotent, $ {\bf s} ^2=0$, even
off--shell, \sl i.e. \rm without using the field equations. The important point
is that the gauge parameter $\Lambda$ was
constrained by the
Klein--Gordon equation, whereas $\omega$ is free.

The fermionic lagrangian ${\cal L}_f$ and the first two terms of \eq{3333},
which we can denote
simply as ${\cal L}_{g}$, are invariant under $\bf s$. Letting
\be {\cal L}_{\rm Stueck} = {\cal  L}_{g}+{\cal L}_{\rm gf} \ee
we can consider instead of the particular
gauge--fixing term
\eq{35}, which is Stueckelberg's original one for $\alpha=1$,  a more general
one \cite{Del88} with an arbitrary local functional $\cal G$ \be {\cal L}_{\rm gf} =
{\bf s}\left[ \omega^* \left( {\cal G}(A_\mu,B,\psi,\bar\psi) + \frac\alpha2 \,
\,b \right)
\right] \label{39}\ee which is invariant under the nilpotent $\bf s$ with \ba
 &&{\bf  s}\,\omega^* = b\label{310xx}\\ &&{\bf s}\,b=0 \label{310}\ea The
auxiliary field
$b$ is just
a
Lagrange multiplier. The local functional ${\cal G}(A_\mu,B,\psi,\bar\psi)$
will be chosen in a specific form only for calculational convenience. The
global parameter $\alpha$ does not
transform under $\bf s$ and labels a family of different but equivalent gauge
slices.

In the above definition of the BRST operator $\bf s$, we have
implicitly set ${\bf s}\,\alpha=0$. One could introduce instead an
additional anticommuting scalar $\beta$ and define \cite{Pig95}
${\bf s}\,\alpha=\beta$, ${\bf s}\,\beta=0$. This extended BRST
symmetry allows one to find an extended Slavnov identity which
clarifies  the gauge-independence of observables.

To check the invariance, observe that $\bf s$ anticommutes with both $\omega$
and $\omega^*$. Using \eq{310}, the
gauge--fixing lagrangian \eq{39} can be
rewritten as \be {\cal L}_{\rm gf} = -\omega ^*({\bf s}\,{\cal G})+ b\,{\cal G}
+\frac\alpha2 \;b^2 \label{311}\ee Note that, crucially, \be {\bf s}\,{\cal
L}_{\rm gf} = \omega ^*({\bf s}\,^2 {\cal G}) -b\, ({\bf s}\,{\cal G}) + b\,
({\bf s}\,{\cal G}) =0 \ee Explicitly, \be {\cal L}_{\rm gf} =\frac12 \left(
\sqrt\alpha b + \frac1{\sqrt\alpha} {\cal G} \right) ^2 -\frac1{2\alpha} {\cal
G}^2 -\omega^* ( {\bf s}\, {\cal G}) \label{lgfexp}\ee
The auxiliary scalar $b$ field
\cite{Nak66,Lau67} has indefinite metric, in sharp contrast to Stueckelberg's
$B$ field, which
propagates and whose metric is positive. It can be eliminated using its own
algebraic
equations of motion, which is equivalent to the gaussian redefinition in the
functional formalism: \be \frac{\delta {\cal L}_{\rm gf} }{\delta b}=0
\Rightarrow b= -
\frac1\alpha \;{\cal G} \label{312}\ee so that \be {\cal L}_{\rm gf} = -
\omega^*({\bf s}\,{\cal G} ) -\frac1{2\alpha} {\cal G}^2 \label{313} \ee


There are infinitely many gauge choices for ${\cal G}$, all
providing the same $S$--matrix.  Popular  are the covariant gauge
${\cal G}=\partial^\mu A_\mu$ and the 't~Hooft--like gauge ${\cal
G}=(\partial^\mu A_\mu + \alpha m B) $, which gives \eq{35}. In
these cases, the high energy behavior of the vector field
propagator goes like $g_{\mu\nu} /k^2$, so these Stueckelberg
theories are power--counting renormalizable (being also unitary).
For ${\cal G}=0$, one recovers the Proca theory, which is not
power--counting renormalizable. The resolution of this apparent
contradiction lies in the fact that the fermion field, although
local, is not renormalizable  in the Jaffee sense  \cite{Zim69}:
 in the Proca gauge, the fermion field is something like
${\rm e}^{-\partial A} \psi$, and this field is not locally
renormalizable because the exponential kills all tempered test
functions. The Green's functions are OK, since the physics is
indeed gauge-invariant anyway, and the bothersome exponential
disappear from all fermion bilinears, but the theory is not
renormalizable.

We choose \be {\cal G}=\partial^\mu A_\mu + \alpha\, m\; B\label{314}\ee so
that \be {\bf s}\,{\cal G}=(\partial^\mu \partial_\mu \omega +\alpha \,m^2
\;\omega) \label{315}\ee and thus \be {\cal L}_{\rm gf} = -
\omega^*\left(\partial^2+ \alpha \, m^2 \right) \omega -\frac1{2\alpha}
(\partial^\mu A_\mu + \alpha\, m\; B) ^2 \label{316}\ee
The ghost term
 decouples as in QED.
 In the Stueckelberg or
Feynman gauge $\alpha=1$, we recover  the lagrangian ${\cal
L}_{\rm Stueck}$ in eq.~\eq{33}, whose BRST invariance is thus
established. It is now a canonical exercise to show the
renormalizability and unitarity of the Stueckelberg model
\cite{Low72,Pic02,Hee03}.

 \section{\label{sec:uone}Massive $U(1)$ gauge field}

Let us analyze the full $U(1) $ massive gauge field theory, including
spontaneous symmetry breakdown, before turning to the standard model in
the
next section. This will allow us to highlight the differences and similarities
between the Higgs and Stueckelberg mechanisms, which can be implemented
simultaneously. The starting lagrangian has three
components
\be {\cal L}_\circ = {\cal L}_g +{\cal L}_s+{\cal L}_f \label{lag0}\ee
where the gauge, scalar and fermion pieces are as follows: \be {\cal L}_g =-
\frac14 (\partial_\mu A_\nu -\partial_\nu A_\mu)^2 + \frac12 \left(
\partial_\mu
B-m \;A_\mu\right)^2 \label{lagg}\ee \be {\cal L}_s = \left| \partial _\mu \Phi
- i\, e\, A_\mu \Phi \right|^2 -\lambda \left( \Phi^\dagger \Phi -\frac{f^2}2
 \right)^2 \label{lags}\ee \be {\cal L}_f = \bar\psi \left(
 i \slash   \partial +g\slash A
 -M\right) \psi \label{1lagf}\ee In this
 lagrangian,  $e$, $g$, $\lambda$, $m$, $f$ and $M$ are parameters (the first
 three massless, the last three of mass dimension one). They are all customary
 except for the photon mass $m$, accompanied by the Stueckelberg field $B(x)$,
 which is a scalar commuting field with positive metric.
 It is useful to introduce the covariant derivatives
 \be
 D_\mu \Phi = \partial_\mu \Phi - i e A_\mu \Phi \ee
 and
 \be D_\mu \psi = \partial_\mu \psi -i g A_\mu \psi \ee
 The vacuum expectation
 value of the complex scalar field, $\left<\Phi\right> = f/\sqrt2$ is taken to
 be a real modulus. We will distinguish between the cases with $f$ zero or
non-zero, since in the latter case the $U(1)$ gauge symmetry is spontaneously
broken
 \cite{Eng64,Hig64,Gur64}. Since $m\not=0$, in both cases the photon is
massive.

Each of the three pieces of the above lagrangian is invariant under the BRST
transformation $\bf s$ \be {\bf s}\, {\cal L}_g ={\bf s}\,{\cal L}_s= {\bf
s}\,{\cal L}_f =0\label{sl}\ee defined by
(\ref{38aa}--\ref{38}) and \ba  &&{\bf  s}\,
\Phi  =  i\,e\;\omega \Phi \\ &&  {\bf s}\,
\Phi^\dagger  =  -i\,e\, \omega \Phi ^\dagger \label{sfie}\ea
Note also that $ {\bf s}\,
F_{\mu\nu}=0$, $ {\bf s}\, D_\mu\Phi = i\,e\,\omega \,D_\mu\Phi$, $ {\bf s}\,
D_\mu\psi = i\,g\,\omega\, D_\mu\psi$, and $f$, just like all other parameters,
is inert, $ {\bf s}\, f=0$.

In order to quantize ${\cal L}_\circ$, we must fix the gauge, as
discussed in section \ref{sec:brs}. To do so, we add to the
lagrangian a gauge--fixing piece ${\cal L}_{gf}$ given by \eq{39}.
After eliminating the auxiliary $b$, the result is a gauge--fixed
lagrangian \be {\cal L}= {\cal L}_\circ - \frac1{2\alpha} {\cal
G}^2 -\omega^* ( {\bf s}\, {\cal G}) \label{1laggg}\ee and we can
choose ${\cal G}$ as we wish. It is convenient to include a
covariant gauge condition in ${\cal G}$, as well as a term that
cancels, up to total derivatives, the quadratic mixing terms in
${\cal   L}_\circ$ involving one derivative.

Define the local current $j^\mu$ as
\be j_\mu (x)= \frac{\delta {\cal L}_\circ}{\delta A^\mu(x)}\ee
By explicit computation, this current is \be j_\mu= m(mA_\mu -\partial_\mu B)
+ie \left( \Phi^\dagger D_\mu \Phi - \Phi D_\mu \Phi^\dagger \right) +g
\bar\psi \gamma_\mu \psi \ee
This current is BRST invariant, ${\bf s}j_\mu =0$,
 and  conserved, $\partial^\mu j_\mu=0$ (from  the field equations for
$A_\mu$). Since physical states 1) contain no ghosts nor
antighosts and 2) are annihilated by the gauge condition $\cal G$,
the field equations for $A_\mu$ from the full gauge-fixed
lagrangian \eq{1laggg} imply that the expectation of the
divergence of the current vanishes between physical states: \be
\left< {\bf phys} \left | \partial^\mu j_\mu \right| {\bf phys}'
\right> =0 \ee So the current is indeed conserved in the quantum
theory.

Before ending  this section, let us note that the
Stueckelberg model  can be
viewed as a free Abelian Higgs model,
\be {\cal L_g} =-\frac14 F_{\mu\nu}^2 +\left| (\partial_\mu -i e A_\mu) \Phi
\right|^2 \ee
where the module of the complex scalar field is fixed, and its phase is the
Stueckelberg field,
 \be \Phi = \frac1{\sqrt2} \frac{m}{e}\,{\rm e}^{ieB(x)/m} \ee
 This cute formulation is due to \cite{Kib65}. We shall not exploit it in what
follows.

\subsection{Massive electrodynamics}

Let us first work out the simple case with $f=0$, so that the
$U(1)$ symmetry is unbroken in perturbation theory -- and still
the photon is massive. We choose \be {\cal G}= \partial_\mu A^\mu
+\alpha\, m\, B \label{gfffuu}\ee to cancel the cross-term between
$A_\mu$ and $B$ in ${\cal L}_\circ$, whereby \ba &{\cal L}=&
-\frac14 F_{\mu\nu}^2 +\frac{m^2}2 A_\mu^2 -\frac1{2\alpha}
(\partial\cdot A)^2\nn\\ & &+\frac12 (\partial_\mu B)^2
-\frac{\alpha m^2}2 B^2 \nn\\ &&-\omega^* (\partial^2 +\alpha\,
m^2) \omega\nn\\ &&+ {\cal L}_s+{\cal L}_f -m\,\partial_\mu ( B\,
A^\mu) \label{lnneew}\ea with ${\cal L}_s$ given by \eq{lags} with
$f=0$, and ${\cal L}_f $ given by  \eq{1lagf}. Since there is no
spontaneous symmetry breakdown, the gauge--fixing is conveniently
chosen independent of the matter fields, the complex scalar and
the Dirac fermion. The gauge sector contains first one massive
vector field, which decomposes into  three physical components of
mass $m$ (one longitudinal  and two transverse) and one spin--zero
piece $\partial A$ of mass $\sqrt\alpha m$. It also contains
  a commuting scalar Stueckelberg
$B$--field with mass $\sqrt\alpha\, m$, and a pair of
anticommuting ghost--antighost scalars, also with mass
$\sqrt\alpha\, m$. All these fields must be kept to prove
renormalizability to all orders in perturbation theory. For the
computation of $S$--matrix elements, however, we can  integrate
out the two conjugate Faddeev--Popov ghosts $\omega$ and
$\omega^*$, since they do not couple to other fields and they
never appear in external asymptotic states. We cannot, however,
integrate out the Stueckelberg $B$--field: it is a free field
 but, as discussed above, it plays a role in the definition of
physical states and it undergoes a non-trivial renormalization.
It is still possible to gauge--fix it to $B=0$, recovering a Proca
massive Abelian gauge field minimally coupled to a charged scalar
and a charged fermion.

\subsection{Spontaneously broken $U(1)$}

What happens if $f\not=0$, that is if the photon acquires a mass
both through the Stueckelberg trick and through the Higgs
mechanism? It is now convenient to choose a different
gauge--fixing functional ${\cal G}$, similar to 't~Hooft's, to
cancel not only the mixing between the photon and the Stueckelberg
field, but also that between the photon and the Goldstone. The
price we pay is a quadratic mixing term between the Stueckelberg
and Goldstone fields, which then have to be redefined through a
global rotation. We can parameterize $\Phi$ in cartesian or polar
forms.

\subsubsection{Cartesian parametrization}

Although awkward in the Abelian case, this parametrization is the one we will
use in the standard model. Recalling that the vacuum expectation value
$f/\sqrt2$ of $\Phi$ is real, we write \be \Phi= \frac1{\sqrt2} \left( \phi_1+i
\,\phi_2 +f \right) \label{k1}\ee
It is worth noting explicitly that \ba && {\bf s} \,\phi_1 = -e \omega \phi_2
\\
&& {\bf s}\,\phi_2 = e \omega (\phi_1 + f ) \ea

We choose the gauge--fixing function \be {\cal G}= \partial^\mu
A_\mu + \alpha ( m B + e f \phi_2) \ee
 Up to a
total
derivative (proportional to $ \partial_\mu \left[ A^\mu S\right]$), this
gauge--fixing function eliminates the mixing terms between the vector field and
the gradients $\partial_\mu B$ and $\partial_\mu \phi_2$ of the scalars.

We find the tremendous lagrangian \be {\cal L}= {\cal L} _2 + {\cal L} _3 +
{\cal L} _4 + {\cal L} _{gh} +{\cal L}_f \ee
 where  the quadratic piece
\ba &{\cal L}_2=&-\frac14 F_{\mu\nu}^2 -
\frac1{2\alpha} (\partial\cdot A)^2 +\frac12 m_\gamma^2 \; A_\mu^2+\frac12
(\partial_\mu G)^2\nn\\ &&+\frac12 (\partial_\mu S)^2 -\frac\alpha2
\,m_\gamma^2\;S^2 +\frac12 (\partial_\mu \phi_1)^2 -\frac12 m_H^2\,
\phi_1^2\ea
has been diagonalized into the Stueckelberg ($S$) and Goldstone ($G$)   mass eigenstate
scalar fields through
 \be \pmatrix{S\cr G\cr }
= \pmatrix{ \cos\beta & \sin\beta\cr -\sin\beta & \cos\beta \cr} \pmatrix{B\cr
\phi_2\cr} \label{krot}\ee with the angle $\beta$ given by
 \be \tan \beta =ef/m \label{1lrrr} \ee
and the short-hands for the Higgs and photon masses \be m_H^2 =2
\lambda f^2 \label{1massagists}\ee \be m_\gamma = \sqrt{ m^2 + e^2
f^2} \label{1masafo}\ee Note that $\sin\beta=ef/m_\gamma$ and
$\cos \beta = m/m_\gamma$. The gauge--fixing function is just \be
{\cal G}= \partial^\mu A_\mu +\alpha m_\gamma S\ee whereas \ba
&&{\bf s} \, S = \left( m_\gamma  + \frac {e^2 f}{m_\gamma} \phi_1
\right) \omega \\
&&{\bf s} \, G =  \frac {e m}{m_\gamma} \phi_1  \omega \ea
Note that the two contributions in quadrature to the photon mass are always
positive, so we cannot envisage a cancellation. The phase conventions are such
that $e$, $m$ and $f$ are always positive.

The cubic, quartic and ghost  lagrangians are
 \ba {\cal L}_3 &=& -e A^\mu (\phi_1 \partial_\mu \phi_2 -\phi_2
\partial_\mu \phi_1)
+e^2 f \phi_1 A_\mu^2 -\lambda f \phi_1 \left( \phi_1^2 +\phi_2^2 \right)
\nn\\ &=& -e\,A^\mu \left[ \cos\beta (\phi_1 \partial_\mu
G-G \partial_\mu \phi_1) +\sin\beta (\phi_1 \partial_\mu S-
 S \partial_\mu \phi_1) \right]\nn \\ &&+
e^2 f \phi_1 A_\mu^2
-\lambda f \phi_1 \left[ \phi_1^2 +(\cos\beta\, G +\sin\beta\,
 S)^2 \right]
\ea
\ba
{\cal L}_4&=& \frac{e^2}2 A_\mu^2
\left( \phi_1^2 + \phi_2^2\right) -\frac{\lambda}{4} \left(\phi_1^2
+ \phi_2^2\right) ^2 \nn\\ &=&
\frac{e^2}2
A_\mu^2 \left[ \phi_1^2 + (\cos\beta\, G +\sin\beta\,
 S)^2\right] -\frac{\lambda}{4}\left[\phi_1^2 + (\cos\beta \, G
+\sin\beta\, S)^2\right] ^2
\ea
\ba
{\cal L}_{gh}=-\omega^*\left[ \partial^2
+\alpha\,m_\gamma^2 +\alpha  e^2 f\, \phi_1 \right]\omega \label{kcartt}\ea
The couplings of the
two scalars $G$ and $S$  (identical except for a
$\sin\beta$ or $\cos\beta$ weight) are both derivative and non-derivative.
Furthermore, the Faddeev--Popov ghosts couple to the Higgs field $\phi_1$.

In the limit $m\to0$, $\beta\to\pi/2$ and the massless $G$
decouples, whereas the surviving $S$ coincides with the original
$\phi_2$. In this limit, of course, the photon mass is due only to
the Higgs mechanism. Contrariwise, when $f\to0$, $\beta\to0$, the
field $S\to B$ decouples, and only $G$ remains coupled. Curiously,
in this limit, the surviving field is again $\phi_2$. So in both
extreme limits, $f\to0$ and $m\to0$, the original Stueckelberg
field decouples. Also, the lagrangian in both limits is identical,
except, of course, that the photon mass \eq{1masafo} is either
$ef$ or $m$. In general, for $f\not=0$ and $m\not=0$, there are
altogether three propagating scalar fields with different masses,
the Higgs $\phi_1$, the Goldstone $G$, and the Stueckelberg $S$.

\subsubsection{Polar parametrization}

Letting \be \Phi(x)=\frac1{\sqrt2} {\rm e}^{i\theta(x)/f} (H(x)+f)
\label{1polio}\ee the scalar part of the lagrangian is \be {\cal
L}_s= \frac12 (\partial_\mu H)^2 +\frac12 \left( \partial_\mu
\theta - e \,f\, A_\mu \right)^2 \left(1 + \frac{H}{f} \right)^2 -
\lambda \, H^2 \left(f + \frac{H}{2}\right)^2 \label{lsss}\ee Note
that the Goldstone field $\theta$ is massless and couples only
through its derivatives. Due to spontaneous symmetry breaking, the
photon $A_\mu$ has acquired a mass $ef$, which adds in quadrature
to the Stueckelberg mass $m$.

It would be tempting  to carry out the  gauge transformation
 \ba &&\Phi \to
{\rm e} ^{ -i\theta/f} \Phi = \frac1{\sqrt2}
(f+H) \\
&&A_\mu \to A_\mu -\frac1{e\,f} \partial_\mu \theta \\
&& B\to B-\frac{m}{e f} \theta \\ && \psi \to {\rm e} ^{ -i\,
g\theta/(e\, f)} \psi \label{1llu} \ea whereby the Goldstone
$\theta$ would disappear completely from the classical lagrangian:
\ba {\cal L} & =& -\frac14 F_{\mu \nu}^2 +\frac12 (m
A_\mu-\partial_\mu B) ^2 +\frac12 (\partial_\mu H)^2 \nn\\ &&
+\frac{e^2}2 A_\mu ^2 (H+f)^2 -2 \frac\lambda4 \left( H^2 + 2f H
\right)^2 +{\cal L}_f \ea
 Note, however, that the choice $\Lambda=-\theta/(ef)$ in \eq{1llu} means
that we must require $(\partial^2 +m^2 )\theta=0$, which is not consistent with
the field equations  for $\theta$ from the original lagrangian \eq{lag0}.
Thus, this ``unitary'' gauge is not allowed due to the presence of the
Stueckelberg field. Scalars are, indeed, trickier than
fermions.

Therefore,  we choose \be {\cal G}= \partial_\mu A^\mu +\alpha\,
m\, B + \alpha\, e\, f\, \theta \label{ffff}\ee whereby the quadratic piece of
the gauge-fixed lagrangian is
\ba &{\cal L}_2=& -\frac14 F_{\mu\nu}^2
+\frac{m_\gamma^2}2 A_\mu^2 -\frac1{2\alpha} (\partial_\mu A^\mu)^2
=\bar\psi(i\slash\partial -M)\psi \nn\\
&&+\frac12 (\partial _\mu B)^2 +\frac12 (\partial _\mu \theta)^2
-\frac\alpha2 \left(m \,B +e \,f\, \theta\right)^2\nn\\ &&+\frac12
(\partial _\mu H)^2- \frac{m_H^2}2 H^2\nn\\ &&+\omega^*
\left(\partial ^2 +\alpha\; m_\gamma^2\right) \omega -\partial
_\mu \left[ A_\mu (e \,f\, \theta + m \,B)\right]\label{1lqqua}\ea
where   the photon and Higgs masses  are given by
 \eq{1masafo} and \eq{1massagists}.
To derive this expression, it is useful to keep in mind that in
the polar parametrization, $ {\bf s}\, H=0$ and $ {\bf s}\,
\theta=e \,f\, \omega$ (and, as usual, ${\bf s}\, f=0$). Note that
the last line of \eq{1lqqua} can be ignored in the computation of
$S$--matrix elements.

Observe that the photon mass \eq{1masafo} squared is the sum of
two contributions, one from the Stueckelberg mechanism and the
other from the Higgs mechanism. Note also that the $B$ and
$\theta$ fields mix, so we must rotate them into mass eigenstates.
One of these is massless, and the other has its mass squared equal
to the gauge parameter times the photon mass squared. We will call
them $G'$ and $S'$ (after Goldstone and Stueckelberg,
distinguishing them from $G$ and $S$ in the cartesian parametrization above): \be
\pmatrix{S'\cr G'\cr } = \pmatrix{ \cos\beta &
\sin\beta\cr -\sin\beta & \cos\beta \cr} \pmatrix{B\cr \theta\cr}
\label{lrot}\ee with the angle $\beta$ defined by eq. \eq{1lrrr}

Dropping the total derivative $-m_\gamma\;\partial_\mu(A^\mu\, S')$ and the
non-interacting ghost-antighost system, we end up with the following
gauge-fixed
quantum lagrangian, appropriate for perturbative evaluations of
$S$--matrix
elements: \ba &{\cal L}=& -\frac14 F_{\mu\nu}^2 +\frac{m_\gamma^2}2 A_\mu^2 -
\frac1{2\alpha} (\partial_\mu A^\mu)^2 +{\cal L}_f \nn\\ &&+\frac12 (\partial
_\mu S')^2 -
\frac\alpha2 m_\gamma^2\; S^{'2} +\frac12 (\partial _\mu G')^2 \nn\\ &&+\frac12
(\partial _\mu H)^2-\frac{m_H^2}2 H^2-\frac{\lambda}{4}  H^4  -\lambda f
\,H^3
\nn\\ &&+\left( \frac{H}f +\frac{H^2}{2f^2} \right) \left[
\frac1{m_\gamma} (e \,f\,\partial_\mu S' +m\, \partial_\mu G') -e\, f\, A_\mu
\right]^2 \label{liifi}\ea
The most salient feature of this lagrangian is that
there are two independent fields with derivative couplings to the Higgs.
The
two scalars $G'$ and $S'$ have the same masses as $G$ and $S$ in the
cartesian parametrization, but now their couplings (identical except for a
$\sin\beta$ or $\cos\beta$ weight) are only derivative.
Furthermore, as customary in the Abelian case, the Faddeev--Popov ghosts
decouple.

Note that in the limit $m\to0$, that is when the photon mass is
due solely to the spontaneous symmetry breakdown, the massless
field $G'$ decouples and the massive field $S'$ coincides with the
original Goldstone $\theta$. This reproduces, of course, the usual
Higgs mechanism. The limit $f\to0$ is rather singular in this
polar parametrization \eq{1polio}, but in this limit the massive
$S'$ coincides with $B$ and decouples whereas the massless $G'$ is
just $\theta$ and remains coupled.

\section{\label{sec:sm}Electroweak theory with a massive photon}

  We now come to one of the nain points of this paper which is to allow a
consistent regularization of the infrared divergences due to the photon by
giving it a finite mass. To achieve this, we use the Stueckelberg mechanism
described earlier of introducing an auxiliary scalar field $B$ of positive
metric while preserving the BRST invariance of the standard electroweak theory
  \cite{Sal68,Wei67,Gla61a} .
  This allows a separate treatment of infrared and ultraviolet divergences in
the perturbative expansion. It does not suffice to add an explicit
mass term for the photon, however, even with the Stueckelberg
trick, because it would spoil the $SU(2)\times U(1)$ symmetry.
Instead,  following \cite{Sto00} and \cite{Gra01}, we give a
Stueckelberg mass to the vector field $V_\mu$ corresponding to the
hypercharge Abelian factor $U(1)_Y$of the gauge group
\footnote{Raymond Stora informed us that his original idea to use
the Stueckelberg field in the standard model arose in the course
of a conversation with Tobias Hurth.}.
  After the symmetry breakdown $SU(2)_L \times U(1)_Y \to U(1)_{\rm em}$, the
photon field $A_\mu$ inherits a mass proportional to the original Stueckelberg
mass for the hypercharge vector boson.  Empirically, this mass is strictly
bound \cite{Luo03,PDBook}.
  As we shall show, it is necessary to modify many of the
parameters in the electroweak theory, albeit by very small
amounts. The spontaneously broken electroweak theory is still BRST
invariant \cite{Sto00,Gra01}, and in addition it is free of
infrared divergences. Of course, infrared divergences in QCD
remain.

It is perhaps worth stressing at this early point that the
Stueckelberg mechanism is implemented in the Abelian factor of the
standard model gauge group to give mass to the Abelian gauge boson
without any symmetry breaking.   We do not know of a mechanism for
generating a Stueckelberg mass if none is present to start with.
Thus, if there is a grand unification of the standard model into a
gauge group without Abelian factors, then there is no reason to
expect a Stueckelberg mechanism at low energies. On the other
hand, if at high energies the gauge group contains a $U(1)$
factor, its gauge boson can acquire a Stueckelberg mass without
symmetry breakdown and this mass could then tumble down. In the
framework of string theory, such an Abelian mass generation is not
forbidden. String phenomenlogists use the Stueckelberg trick
assiduously to get rid  of spurious $U(1)$'s (see section
\ref{sec:pubel}).

Our approach is rather that since gauge invariance allows for an
Abelian vector mass term, it should be considered. Accordingly,
the Stueckelberg mass $m$ for the Abelian $U(1)_Y$ factor is a
free parameter of the standard model, just like $\theta_{QCD}$. We
are aware, however, that the introduction of this new mass scale
entails unavoidably a new hierarchy problem,  notably with respect
to the electroweak breaking scale $f$ ($m << f$). Of course,
dimensional transmutation yields an intrinsic mass scale for
strong interactions, $\Lambda_{QCD}$, so it is somewhat symmetric,
perhaps, that the Stueckelberg mechanism produces a mass scale for
the Abelian factor of the standard model gauge group. These three
unrelated masses, associated with each of the three gauge group
factors of the standard model,  accompany generically  the three
distinct phases of a gauge theory (confinement, spontaneous
symmetry breaking, or Abelian). This could bring up an additional
problem, since this third phase is usually called Coulomb
precisely because the static potential is of infinite range. But a
massive photon implies that the electrostatic potential deviates
from Coulomb, so it is not of infinite range. In fact, we are not
quite clear about what properties the external ``classical''
electromagnetic fields inherit from the modified massive quantum
photon. The issue does not arise in Stueckelberg QED, but it does
in the Stueckelberg standard model: the Noether current leaving
invariant the vacuum does not couple to the asymptotic (and
massive) photon. We shall return to these tricky issues below.








Other authors have investigated in different directions.
\cite{Cve91} exploited the Stueckelberg formalism, but they constructed a
standard
model with the gauge symmetry realized non-linearly, which is equivalent to
taking the Higgs mass to infinity. In a similar vein,
\cite{Gro93a}  integrated out the
Higgs field from the standard model  and obtained an effective lagrangian
close
to Proca's \cite{Gro93b,Gro94,Gro95,Dit95a,Dit95b,Dit96}. Our approach is
orthogonal to
these viewpoints, since we keep the physical Higgs field and our photon is not
massless. Also, we ensure exact
quantum unitarity and renormalizability.

The hypercharge is normalized such that $Q=T_3 +Y/2$. We follow
the careful notation of \cite{Tay76}, and leave to the Appendix some details of
our conventions, as well as the long formulas of interest only to the conscientious
scholar.

 For starters, we concentrate in section \ref{sec:gausec} on the gauge sector,
which consists of the
 vectors, scalars and ghosts. We turn to   the
 fermion matter fields, with their lagrangian and   BRST
transformations, in
 section \ref{sec:matter}. For simplicity, we
present here the Stueckelberg modification of the minimal standard theory,
without any neutrino masses: we do not include  right--handed
gauge singlets.
 Including them, with their phenomenologically required large Majorana mass, does
not change the analysis in any substantial way: it would be tantalizing to
speculate that neutrino masses are related to the Stueckelberg mechanism.

\subsection{\label{sec:gausec}The gauge sector}

The starting gauge-invariant  lagrangian is a sum of pieces associated with the
gauge, the scalar, and the fermion fields: \be {\cal L}_\circ={\cal L}_g+{\cal
L}_s  +{\cal L}_f\label{m0}\ee

The gauge lagrangian contains the usual kinetic terms for  the
vector fields $\vec W^\mu$ and $V^\mu$, as well as the
Stueckelberg mass for $V^\mu$, along with the kinetic term for the
Stueckelberg field $B$ (of zero hypercharge and weak isospin):
\be{\cal L}_g =-\frac14 \vec{F}_{\mu\nu} ^2 -\frac14
{F}_{\mu\nu}^2 +\frac12 \left(m\; V_\mu -\partial_\mu B\right)^2
\label{1mag}\ee where the gauge potential field strengths are \ba
&{F}_{\mu\nu}= \partial_\mu V_\nu -\partial_\nu V_\mu \\
&\vec{F}_{\mu\nu}= \partial_\mu {\vec W} _\nu -
\partial_\nu {\vec W}_\mu +g {\vec W}_\mu \times {\vec W}_\nu \label{mmg}\ea

The scalar lagrangian, including Higgs and Goldstones,  is \be
{\cal L}_s=\left| D_\mu \Phi \right|^2-\lambda \left( \Phi^\dagger
\Phi - \frac{f^2}2 \right)^2 \label{ms}\ee where the scalar weak
isodoublet is \be \Phi = \frac1{\sqrt2} \left( H+f +i\,\vec\tau
\cdot \vec \phi\right) \pmatrix{ 0 \cr 1 \cr} =  \pmatrix{ i\,
\phi_- \cr \frac1{\sqrt2}\left( H-i\,\phi_3 +f\right) \cr}
\label{1mfi}\ee and its covariant derivative is \be D_\mu \Phi =
\left( \partial_\mu -i\frac{g}2 \vec\tau \cdot \vec W_\mu -i
\frac{g'}2 V_\mu \right) \Phi \label{mcvd}\ee The minima of the
potential are located at $|\Phi^\dagger\Phi|=f^2/2$, and we choose
the vacuum to be given by $<\Phi>=f/\sqrt2$, with $f$ real. The
scalar lagrangian is spelled out in section \ref{a:scal} of the
Appendix.

The gauge lagrangian \eq{1mag} is invariant under the BRST
transformation $\bf s$ defined by \ba && {\bf s}\, V_\mu =
\partial_\mu \omega \\ && {\bf s}\, B = m\; \omega \\ && {\bf s}\,
{\vec W}_\mu = \partial_\mu \vec \omega -g \,\vec\omega \times
\vec W _\mu \label{msvbw}\ea where $\omega$ and $\vec\omega$ are
anticommuting scalars. The BRST operator $\bf s$ is nilpotent,
${\bf s}^2=0$, with \ba & &{\bf s} \,\omega=0 \\ && {\bf s}
\,\vec\omega = -\frac{g}2 \vec\omega \times \vec \omega
\label{somega}\ea Note that ${\bf s}\, F_{\mu\nu}=0$ and ${\bf
s}\, {\vec F}_{\mu\nu}=-g \, \vec\omega \times {\vec F}_{\mu\nu}$.

The scalar lagrangian \eq{ms} is also BRST invariant with the
additional definition \be {\bf s}\, \Phi =\frac{i}2 \left(g\,
\vec\tau \cdot \vec \omega +g'\, \omega \right)\Phi \label{sfi}\ee
which implies that ${\bf s}\, (D_\mu\Phi) =(i/2) (g\, \vec\tau
\cdot \vec \omega +g'\, \omega )D_\mu\Phi$. The BRST transforms of
the components of the scalar doublet defined in equation \eq{1mfi}
are \ba &&{\bf s}\, \phi_\pm = \frac{g}2 (\pm i \omega_\pm \phi_3
\mp i \omega_3 \phi_\pm +\omega_\pm  H) \mp i \frac{g'}2 \omega
\phi_\pm
+\frac{f\,g}2 \omega_\pm \\
 &&{\bf s}\, \phi_3 =
\frac{g}2 ( i\omega_- \phi_+ -i\omega_+ \phi_- +\omega_3 H) -\frac{g'}2 \omega
H
+\frac{f}2 (g\omega_3 -g' w) \label{sficomp}\ea
and \be {\bf s}\, H = -\frac{g}2
\vec\omega\cdot\vec \phi +\frac{g'}2\omega \phi_3 \label{shigs}\ee

The fermion lagrangian, discussed in the next section, is also BRST--invariant.
 The nilpotency of the BRST operator $\bf s$ (without having to use the
 equations of motion)
 and the invariance of the full lagrangian under
$\bf s$ ensure the renormalizability of the theory to all orders in
perturbation theory, hopefully.
To check this explicitly, the full Slavnov--Taylor
identities would have to be proved. This is
a significant piece of work which has not been done.

We add to ${\cal L}_\circ$ a
gauge--fixing lagrangian \be {\cal L}_{gf} = {\bf
s}
\left[ \vec\omega^* \cdot \left(\vec{\cal G}+\frac\alpha2 \vec
b\right)+\omega^* \left({\cal G}+\frac{\alpha'}2 b\right) \right]
\label{slgf}\ee with more Faddeev--Popov ghosts $\omega^*$, $\vec\omega^*$ (independent
from $\omega$ and $\vec\omega$) and Nakanishi--Lautrup ghosts $b$, $\vec b$, subject to

\ba &{\bf s}\, \vec\omega^* = \vec b \\ &{\bf s}\, \vec b
=0\\ &{\bf s}\, \omega^* = b \\ &{\bf s}\, b =0\label{sghost} \ea

Notice that we keep two different gauge parameters $\alpha$ and $\alpha'$ for
each of the factors in the electroweak gauge group, namely $SU(2)_L$ and
$U(1)_Y$, respectively. Below, we will equate them to simplify some
tree--level
expressions, but in general they must be kept distinct because under
renormalization they behave differently, since there is no symmetry which
favors their equality.

Eliminating the auxiliary Nakanishi--Lautrup ghosts $b$ and $\vec b$ is
not a very good idea, since once they are gone the BRST operator
is nilpotent only on-shell. Of course, at tree level it's safe and
instructive to get rid of them  algebraically,  finding the
lagrangian \be {\cal L}_{\rm ph}= {\cal L}_\circ + {\cal L}'_{\rm
gf} + {\cal L}_{\rm gh} \label{1omonosov} \ee with \be {\cal
L}'_{\rm gf}= - \frac1{2\alpha'} {\cal G}^2 -\frac1{2\alpha}
{\vec{\cal G}}^2 \label{mlll}\ee and \be {\cal L}_{\rm gh}=
-\omega^* \;{\bf s}\,{\cal G} -\vec\omega^* \cdot {\bf
 s}\,\vec{\cal G} \label{mlll2}\ee

Following 't~Hooft, we choose the following gauge functions \cite{Gra01} \ba
&&{\cal G} = \partial_\mu V^\mu +\alpha'\,m\,B -\alpha'\frac{g'}2 f\, \phi_3 \\
&&\vec{\cal G} = \partial_\mu \vec W^\mu +\alpha\frac{g}2 f\, \vec\phi
\label{kmkm}\ea Notice that the $ SU(2)$ gauge function is just 't~Hooft's,
whereas the $U(1)$ function $\cal G$ contains also a term involving the
Stueckelberg field. These gauge functions have been chosen to give total
derivatives when combined with the terms in the lagrangian with one gauge
boson,
one scalar and one derivative. All total derivatives in the lagrangian we just
drop.

\subsubsection{Mass eigenstates}

Let us collect terms in the lagrangian ${\cal L}_{\rm ph}$ in eq.
\eq{1omonosov} into three pieces: the fermionic lagrangian ${\cal
L}_f$ that we have not even written out yet, a quadratic
lagrangian ${\cal L}_{2}$ with at most two fields, and the rest,
which we call the interaction lagrangian ${\cal L}_{\rm    int}$.
They are all spelled out in the Appendix. We concentrate now on
${\cal L}_2$ and diagonalize it.

Just like in the usual standard theory, the charged vector fields $W_\mu^\pm $
have mass
\be M_W = \frac {f\;g}2\label{masw}\ee whereas the charged scalars, $\phi_\pm
$, and the charged ghost-antighost pairs, $\omega^{(*)}_\pm $, have mass
$\sqrt\alpha\, M_W$.

Neutral boson fields mix in pairs. Explicitly, in the bases $(W_3^\mu, V^\mu)$,
$(\phi_3,B)$ and $(\omega_3,\omega)$, the respective square mass matrices for
vectors, scalars and ghosts are \ba &&M_v^2=\frac{f^2}4 \pmatrix{ g^2 & -
g\;g'\cr -g\; g'& g^{\prime2}+\mu^2\cr} \\ &&M_s^2=\frac{f^2}4 \pmatrix{
\alpha\;g^2 + \alpha' g^{\prime2} & -\alpha' \;g'\;\mu \cr -\alpha'\; g'\;\mu&
\alpha'\; \mu^{2}\cr} \\ &&M_g^2=\frac{f^2}4 \pmatrix{ \alpha\;g^2 & -\alpha'
\;g\;g' \cr -\alpha \;g\;g' & \alpha'\; (g^{\prime2}+\mu^{2})\cr} \ea where the
last matrix is understood to be sandwiched between $(\omega^*_3,\omega^*)$ and
$(\omega_3,\omega)$,  and  we have introduced the  rescaled Stueckelberg mass
of the
hypercharge vector boson \be \mu = 2\frac{m}{f} \label{mustu}\ee which behaves
like a coupling constant. These mass matrices reduce to the usual ones of the
standard model when $\mu=0$.

The mass eigenstates are obtained by rotations
\ba &\pmatrix { Z^\mu \cr A^\mu\cr} = \pmatrix
{\cos
\theta_w & -\sin\theta_w \cr \sin \theta_w & \cos\theta_w \cr} \pmatrix{W_3^\mu
\cr V^\mu \cr} \\ &\pmatrix { G\cr S\cr} = \pmatrix {\cos \beta & -\sin\beta
\cr
\sin \beta & \cos\beta \cr} \pmatrix{\phi_3 \cr B \cr} \\ &\pmatrix { \chi_Z
\cr
\chi_A\cr} = \pmatrix {\cos \tilde\theta_w & -\sin \tilde\theta_w \cr \sin
\tilde \theta_w & \cos\tilde \theta_w \cr} \pmatrix{\omega_3 \cr \omega \cr}
\\ &\pmatrix { \chi^*_Z
\cr
\chi^*_A\cr} = \pmatrix {\cos \tilde\theta_w & -\sin \tilde\theta_w \cr \sin
\tilde \theta_w & \cos\tilde \theta_w \cr} \pmatrix{\omega^*_3 \cr \omega^*
\cr}
\label{rotata} \ea
where the mixing angles are defined by
\ba & &\tan 2\theta_w =
\frac{2\,g\,g' }{ g^2 -g^{\prime2} -\mu^2 } \label{rotatapatate}\\ & &\tan
2\beta= \frac{2\,\mu\,g'
}{ \frac{\alpha}{\alpha'}\, g^2 +g^{\prime2} -\mu^2 } \\ & &\tan 2\tilde
\theta_w = \frac{2\,g\,g' }{ g^2 -\frac{\alpha'}\alpha \left( g^{\prime2}
+\mu^2 \right)} \ea The last two expressions simplify if we set
$\alpha'=\alpha$, which we are allowed to do at tree level. In this simpler
case, $\tilde\theta_w =\theta_w$.

Let us point out the main differences with the usual electroweak theory. Since
the hypercharge gauge field $V_\mu$ has a Stueckelberg mass, the weak mixing
angle
$\theta_w$ is modified. Just as in the usual theory, the mixing angle between
the associated ghost fields is also the weak mixing angle if the two gauge
parameters $\alpha$ and $\alpha'$ are equal, but not otherwise.  The
new mixing angle is $\beta$, between the longitudinal degrees of freedom of the
$SU(2)_L$ and of the $U(1)$ neutral vector bosons. This is reasonable, since
the latter is the Stueckelberg field, which does not exist in the minimal
electroweak theory. The angle $\beta$ is tiny, proportional to the ratio of the
Stueckelberg mass to the electroweak vacuum expectation value. Again, let us
stress that the charged sector does not change.

To expand in powers of the Stueckelberg mass $m$ or, better, in terms of
the rescaled Stueckelberg mass $\mu=2m/f$,  it is useful to introduce the
convenient  parameter
\be
\epsilon=\frac{\mu^2}{g^2+g^{\prime2}} = 4 \frac{m^2}{f^2(g^2+g^{\prime2})} \label{combini}
\ee
Note that $\mu$ has mass dimension zero but behaves as a coupling constant,
whereas $\epsilon$ is truly dimensionless.
 The following trigonometric functions of the  modified weak mixing angle are
handy:
\be \tan \theta_w= \frac{g'}g \left( 1+\epsilon  \right) +{\cal O}(\epsilon^2)
\ee
\be {\rm s}_w= \sin \theta_w =
\frac{g'}{\sqrt{g^2+g^{\prime2}}}\left( 1+ \epsilon
\,\frac{g^2}{g^2+g^{\prime2}}  \right) + {\cal  O}(\epsilon^2)  \ee
\be {\rm c}_w= \cos \theta_w = \frac{g}{\sqrt{g^2+g^{\prime2}}} \left(
1- \epsilon \, \frac{g^{\prime2}}{g^2+g^{\prime2}} \right)+{\cal
 O}(\epsilon^2)  \ee

Finally, the small mixing angle $\beta$ between the Goldstone $\phi_3$ and the
Stueckelberg  $B$ is approximately
\be \beta = g' \frac{\sqrt{g^2 +g^{\prime2} }}
{\frac\alpha{\alpha'} g^2 +g^{\prime2} } \sqrt\epsilon
+ {\cal O} \left(\epsilon ^{3/2}\right)
\simeq \frac{g'}{\sqrt{g^2 +g^{\prime2} }}
 \sqrt\epsilon
\simeq \sin\theta_w \sqrt\epsilon
\ee
where the last expressions hold at tree level if we choose $\alpha=\alpha'$.

As was to be expected, the photon has a non-vanishing mass $M_A$
proportional, to first order, to the original Stueckelberg mass
$m= \mu f/2$ of the hypercharge vector boson: \be M_A  = m \cos
\theta_w+{\cal O}(m^3 )  = M_W  \sqrt\epsilon +{\cal
O}(\epsilon^{3/2}) \ee The $Z^\mu$ vector boson mass $M_Z$ differs
slightly from the usual one because the weak mixing angle is slightly
different. Now it becomes \ba & M_Z &= \frac{f}2 \sqrt{g^2
+g^{\prime2}} \left(1+\frac\epsilon2 \frac{g^{\prime2} }{g^2
+g^{\prime2}} \right) +{\cal O}(\epsilon^2) \nn\\ && =
\frac{M_W}{\cos\theta_w} \left( 1-\frac\epsilon2 \sin^2 \theta_w
\right) \ea The exact mass eigenvalues are given in the Appendix,
eq. \eq{1mma}.

Very nicely, after   rotating by $\beta$  the Goldstone
$\phi_3$ and Stueckelberg $B$ fields, the mass eigenstates $G$ and $S$ have
exactly the same masses as the anticommuting ghosts $\chi_Z$ and $\chi_A$,
obtained by rotating through $\tilde \theta_w$ the ghosts $\omega_3$ and
$\omega$.

 If we set $\alpha'=\alpha$ (valid at tree level), then we
find
the simple formulas
\ba M_S=M_{\chi_A} = \sqrt\alpha \; M_A \\ M_G= M_{\chi_Z} = \sqrt\alpha \; M_Z
\ea
The full expressions are given in section \ref{a:mass} of the Appendix.

\subsection{\label{sec:matter}Matter}
 The fermion lagrangian is the sum of a lepton and a quark lagrangians.  Added
to the gauge and scalar lagrangians discussed above, it yields the full
classical lagrangian.

The lepton lagrangian is
\ba &{\cal L}_\ell=& \bar R \left( i\, \slash \partial - g' \slash
V \right) R \nn\\ &&+ \bar L \left( i\, \slash\partial -
\frac{g'}2
\slash  V + \frac{g}2 \vec\tau \cdot {\vec {\slash W}} \right) L \nn\\
&&-\left( y_e \bar R (\Phi^\dagger L) + {\rm \; h.c.} \right)
\label{1ml}\ea where $y_e$ is a Yukawa matrix, the hypercharges of
$R$ and $L$ are $-2$ and $-1$, we have suppressed family indices,
and \be R = e_R = \frac{1-\gamma_5}2 e \ee \be L = \pmatrix{\nu_L
\cr e_L \cr} = \frac{1 + \gamma_5}2 \pmatrix{\nu \cr e \cr}
\label{mlr} \ee

The BRST transformations of the lepton fields are
\ba && {\bf s}\, R = -i g' \omega R \\
&& {\bf s}\, L = \frac{i}2 \left( g  \vec \tau \cdot \vec \omega - g' \omega
\right) L \ea

The quark lagrangian is
\ba &{\cal L}_q=
& i\,\bar Q  \left(  \slash \partial -i \frac{g'}6 \slash V
-i \frac{g}2 \vec\tau \cdot {\vec  {\slash W}} \right) Q \nn\\
&&+ i\,\bar U \left(  \partial_\mu -i \frac{2g'}3  \slash V \right) U \nn\\
  &&+ i\,\bar D  \left(  \slash \partial+i\frac{g'}3  \slash V \right) D \nn\\
  &&  -\left( y_d \bar D (\Phi^\dagger  Q)
  + y_u (\bar Q i \tau^2 \Phi) U + {\rm \;
h.c.} \right) \label{1mq}\ea where \ba
Q=\pmatrix {u _L\cr d _L\cr } \\
U= u_R \\ D= d_R \label{mqud} \ea with
hypercharges $1/3$, $4/3$ and $-2/3$, and we have
suppressed both family and color indices.

The BRST transformations of the quark fields are
\ba && {\bf s}\, U =  i \frac{2 g'}3 \omega U \\
&& {\bf s}\, D =  -i \frac{g'}3 \omega D \\
&& {\bf s}\, Q = \frac{i}2 \left( g  \vec \tau \cdot \vec \omega + \frac{ g'}3
\omega \right) Q \ea

The Yukawa interactions in \eq{1ml} and \eq{1mq} give the usual
mass matrices to the fermions, \be M_e=\frac{y_e \; f }{\sqrt2}\ ,
\quad M_u=\frac{y_u \; f } {\sqrt2}\   , \quad M_d=\frac{y_d \; f
}{\sqrt2} \ee so that the free fermion lagrangian is \be {\cal
L}_{ff}=i \bar \nu_L  \slash \partial \nu_L +\sum_{\psi=e,d,u}
\left\{ i\bar \psi_L \slash \partial \psi_L + i\bar \psi_R
\slash\partial \psi_R -M_\psi \bar \psi_R \psi_L -M_\psi^\dagger
\bar \psi_L \psi_R \right\} \ee They also give rise to the
interactions between the scalars and the fermions: \ba & {\cal
L}_y=&-\frac{1}{\sqrt 2} (H+i\cos \beta G +i \sin \beta S ) \left(
y_e \bar e_R e_L +y_d \bar d_R d_L +y_u \bar u_R u_L \right) \nn\\
&& +i\phi_+ \left( y_e \bar e_R \nu_L +y_d \bar d_R u_L -y_u \bar
u_R d_L \right) \nn\\ && +{\rm h. c.} \ea where we have already
eliminated $\phi_3$  and $B$ in favor of  the mass eigenstates $G$
and $S$.

The interaction between the fermions and the gauge bosons is cute. From the
covariant kinetic terms for the fermions we find the usual charged current
lagrangian \be {\cal L}_{cc}= \frac{g}{\sqrt 2}
\left( \bar\nu_L \slash W_- e_L + \bar u_L \slash W_- d_L \right)
+ {\rm h.c.}  \label{chargedc}\ee

The neutral currents are rather funny due to the massiveness of
the photon and the modified weak mixing angle. Indeed, the neutral current
lagrangian can be
written as \be {\cal L}_{nc} =\sum_\psi \bar \psi \left(
n_\psi^A \slash A +n_\psi ^Z \slash Z \right) \psi \ee where the
sum runs over all the two-component fermionic fields with non-zero isospin,
$\psi\in\{\nu_L,e_L,e_R,d_L,d_R,u_L,u_R\}$. All the exact couplings
and their expansion to first order in $\epsilon$ are given in section
\ref{a:coup} of the Appendix.

To illustrate the novel features, it is instructive to display the
approximate lepton couplings to the photon, with the help of
\eq{combini} and the traditional
 \be e=
\frac{gg'}{\sqrt{g^2+g^{\prime2}}}  \label{cargae}\ee
 which has no particular physical meaning when the Stueckelberg mass does not
vanish:
\ba & n_\nu^A &   \simeq
{\frac\epsilon2} e
\\ & n_{e_L}^A &
\simeq - e
\left(1+\frac\epsilon2 \frac{g^2-g^{\prime2}}{g^2+g^{\prime2}}
\right) \\ & n_{e_R}^A &  \simeq -
e \left(1-\epsilon
\frac{g^{\prime2}}{g^2+g^{\prime2}} \right)      \ea

These   neutral currents can be rewritten in Dirac spinor notation as follows:
\be
{\cal L}_{nc} = \sum_\psi \left\{
\bar \psi \slash  A (v_\psi^A +a_\psi^A \gamma_5) \psi
+  \bar \psi \slash  Z (v_\psi^Z +a_\psi^Z \gamma_5) \psi
\right\}  \ee
where the sum runs over $\psi\in\{\nu,e,u,d\}$. The leptonic couplings to the
photon are
\ba
&&v_\nu^A= a_\nu^A=-a_e^A
\simeq \frac\epsilon4 e \\
&&v_e^A \simeq  -e \left(1 +
\frac\epsilon4 \frac{ g^2 -3g^{\prime2} } {g^2 +g^{\prime2} } \right)  \ea

Notice the universality of the fermionic axial coupling to the photon,
  only of order $\epsilon$.
  There is also a universality in the fermionic axial couplings to
   the $Z$, which
remain  unchanged at first order from the standard value.
   These universalities extend  to the quarks, with the couplings
   shown in section \ref{a:coup}.

A BRST-consistent mass for the photon in the standard electroweak
theory implies then that it has not only the customary vector
couplings (slightly modified) but also a small non-zero axial
coupling! Indeed, the photon couples differently to left and right
electrons. This means that the vector coupling to the photon of
the electron is not quite (minus) one or, rather, and even more
curiously, that the left electron's coupling to the photon is
different from the right electron's. Even more surprisingly, the
photon couples also to the neutrinos.

The expansion to first order in  $\epsilon$ is useful for getting
a flavor of what is going on, and will be commented upon in
section \ref{sec:phenol} below.

\subsection{\label{sec:anomal}Anomalies}

The theory we have constructed has exact   BRST symmetry, even at the quantum level, so
it is
free of anomalies.  In fact, since the only change in the BRST structure with respect to
that of the standard theory is the addition of the doublet ${\bf s}\,B=\omega$, ${\bf
s}\,\omega=0$ with trivial cohomology, the anomalies in the Stueckelberg modified standard
model are the same as those in the usual  standard model. But we can check the vanishing of
all anomalies in the modified model directly.

It is quite remarkable that anomalies cancel
directly in the basis of physical vector bosons (the propagating degrees of
freedom, namely the massive photon, $W^\pm$, $Z$, and massless gluon).
The triangle graph with three external photons, for instance, is proportional
to the sum of the cubes of the photonic couplings of the left-handed fermion
fields
minus the sum of the cubes of the  photonic couplings of the right-handed
fields.
Using the two-component form
(\ref{nnnu}-\ref{nnnv}) of the fermion couplings to the neutral vector bosons,
the charged current \eq{chargedc}, and the fact that quarks come in triplets of
$SU(3)$ whereas leptons are singlets thereof, the following exact relationships
between the fermion couplings are verified, where we note on the left the three
gauge bosons at the vertices of the fermion loop,
 including the gluons $G$ and the gravitons $h$:
\ba
& AAA \qquad & (n_\nu^A)^3+ (n_{e_L}^A)^3- (n_{e_R}^A)^3
+3  (n_{u_L}^A)^3+3  (n_{d_L}^A)^3
-3  (n_{u_R}^A)^3-3  (n_{d_R}^A)^3 =0 \\
& AAZ \qquad & (n_\nu^A)^2(n_\nu^Z)+ (n_{e_L}^A)^2 (n_{e_L}^Z)
- (n_{e_R}^A)^2(n_{e_R}^Z)
+3  (n_{u_L}^A)^2(n_{u_L}^Z) \nn\\
&& \qquad\qquad +3  (n_{d_L}^A)^2(n_{d_L}^Z)
-3  (n_{u_R}^A)^2(n_{u_R}^Z)-3  (n_{d_R}^A)^2 (n_{d_R}^Z)=0 \\
& AZZ \qquad & (n_\nu^A)(n_\nu^Z)^2+ (n_{e_L}^A) (n_{e_L}^Z)^2
- (n_{e_R}^A) (n_{e_R}^Z)^2
+3  (n_{u_L}^A) (n_{u_L}^Z)^2 \nn\\
&& \qquad \qquad +3  (n_{d_L}^A)(n_{d_L}^Z) ^2
-3  (n_{u_R}^A) (n_{u_R}^Z)^2-3  (n_{d_R}^A) (n_{d_R}^Z)^2 =0 \\
& ZZZ \qquad & (n_\nu^Z)^3+ (n_{e_L}^Z)^3- (n_{e_R}^Z)^3
+3  (n_{u_L}^Z)^3+3  (n_{d_L}^Z)^3
-3  (n_{u_R}^Z)^3-3  (n_{d_R}^Z)^3 =0\\
& AWW \qquad & (n_\nu^A)+ (n_{e_L}^A)
+3  (n_{u_L}^A)+3  (n_{d_L}^A)  =0\\
& ZWW \qquad & (n_\nu^Z)+ (n_{e_L}^Z)
+3  (n_{u_L}^Z)+3  (n_{d_L}^Z) =0 \\
& AGG \qquad &
   (n_{u_L}^A)+   (n_{d_L}^A)
-   (n_{u_R}^A)-   (n_{d_R}^A) =0\\
& ZGG \qquad &
  (n_{u_L}^Z)+   (n_{d_L}^Z)
-   (n_{u_R}^Z)-   (n_{d_R}^Z) =0\\
& Ahh \qquad &
 (n_{e_R}^A)+   3   (n_{u_R}^A)+3   (n_{d_R}^A) =0\\
& Zhh \qquad &
   (n_{e_R}^Z)+  3   (n_{u_R}^Z)+3   (n_{d_R}^Z) =0\ea

Triangle graphs other than those listed here vanish trivially.  Note that the
actual condition from  the cancellation of the  $Ahh$ and $Zhh$ anomalies
\cite{Wit85}
is
really the displayed relation minus the relation from $AWW$ (or $ZWW$).

This exact cancellation takes place independently of the value of the
Stueckelberg mass, and independently of the values of $g$ and $g'$ or, more
accurately, independently of the value of the ``massive'' weak mixing angle
$\theta_w$.

All these anomalies cancel family by family.
The cancellation of anomalies requires, of course, the color factor 3 for
quarks.

\subsection{\label{sec:courant}Currents}

Recapitulating, the classical lagrangian for the
Stueckelberg--modified standard electroweak theory is \be {\cal
L}_\circ = {\cal L}_ g+{\cal L}_ s+{\cal L}_\ell +{\cal L}_q
\label{llkk}\ee where the gauge, scalar, lepton and quark
lagrangians are given by eqs. \eq{1mag}, \eq{ms}, \eq{1ml} and
\eq{1mq}, respectively.

\subsubsection{\label{sec:ccurr}Classical currents}

 Let us define the classical currents as
 \ba
 && j_\mu  = \frac{\delta \cal L_\circ} { \delta V^\mu}\\
 && \vec J_\mu = \frac{\delta \cal L_\circ} { \delta \vec W^\mu}\ea

The equations of motion for $V_\mu$ and $ \vec W_\mu$ then read as follows:
\ba &&  \partial_\mu F^{\mu\nu} = -j^\nu \\
 &&  \partial_\mu \vec F^{\mu\nu} = -\vec J^\nu \ea
and all four currents are conserved ($ \partial_\mu j^\mu = \partial_\mu \vec J
^\mu =0$).

Explicitly,
 \ba & j_\mu =&   m(mV_\mu-\partial_\mu B) \nn\\ &&+ \frac{i g'}2
 \left( \Phi ^\dagger  D_\mu \phi - (D_\mu \Phi)^\dagger \Phi \right)
 \nn\\ &&+\sum_{U,D,R,Q,L} g' Y_f \bar \psi \gamma_\mu \psi \ea
and the customary
 \ba & \vec J_\mu =&  g \; \vec F_{\mu\nu}\times  \vec  W^{\nu}    \nn
 \\ &&+ \frac{i g}2
 \left( \Phi ^\dagger \vec \tau D_\mu \phi - (D_\mu \Phi)^\dagger \vec\tau \Phi
\right)
 \nn\\ &&+\frac{g}2  \sum_{ Q,L}   \bar \psi  \vec \tau \gamma_\mu \psi \ea
  It turns out that
  $ \partial j = \partial \vec J =0 $
  and $ {\bf s }j  =   0 $,
 whereas 
   $ {\bf s }   {\vec  J} \not =0 $.

   Of course, it is not surprising that
   the conserved  $SU(2)$ current $\vec J$   is not  BRST--invariant
   (neither is  the   covariantly
    conserved current):
   it  forms an  $SU(2)$ triplet! Under the restricted BRST transformations
     with $\omega^1 =\omega^2=0$, the third component $J_3$ of the $SU(2)$
current
     is invariant.


 After spontaneous symmetry breaking,  the current coupled to the physical
massive photon is
  \be J_{\rm A}^\mu =     {\rm c}_w j^\mu + {\rm s}_w J_3^\mu   \ee

We are interested in linear combinations of $j$ and $J_3$.
 Note that any linear combination of $j_\mu$ and $J^3 _\mu$ is conserved, in
particular the Noether current associated with global $SU(2)\times U(1)$
transformations leaving the vacuum invariant
 \be J_{\rm em}^\mu = \frac1{\sqrt{g^2 +g^{\prime2}}} \left( g j^\mu + g'
J_3^\mu \right) \ee

  Somewhat more explicitly,  the fermionic parts of these currents are
   \ba && J_{\rm em}^{(f) \mu} = e
\sum_{U,D,R,Q,L} \bar \psi Q_\psi \gamma^\mu \psi   \\ &&
   J_{\rm A}^{(f) \mu}  =    \sum_{U,D,R,Q,L} \bar \psi
\left( g'{\rm c}_w Y_\psi/2 + g {\rm s}_w  T_\psi^3 \right)  \gamma^\mu \psi
\ea
  with $e$ given by eq. \eq{cargae},
 $Q_\psi =Y_\psi/2 +T^3_\psi $,
and $T^3_Q=T^3_L=\tau^3/2$,
$T^3_R=T^3_U=T^3_D=0$. Since $\tan \theta_w = (1+\epsilon)g'/g + {\cal O}
(\epsilon ^2 ) $, the two currents differ for
 a non-zero $\epsilon = 4 m^2 /(f^2 (g^2 +g^{\prime2}
))$.

  The gauge and scalar pieces of these two currents are related just like the
above. Indeed, from the expressions
  \ba  &
   J_{\rm A}^{(g) \mu} & =    m  {\rm c}_w \left( mV^\mu -\partial^\mu B
\right)  +i g {\rm s}_w
   \left( F_{\mu\nu} ^+ W^{\nu -} -  F_{\mu\nu} ^- W^{\nu + } \right)  \\ &
   J_{\rm A}^{(s) \mu}  & =
   \frac{ {\rm c}_w g' + {\rm s}_w g }2
   \left( g' V^\mu +g W_3^\mu \right) \phi_+ \phi_-
   -\frac{ {\rm c}_w g g'   }2 \phi_3 \left( W_-^\mu \phi_+ + W_+ ^\mu \phi_-
\right) \nn\\ & & +
   \frac{ {\rm c}_w g' - {\rm s}_w g }2
   \left[ \frac12
    \left( \phi_3^2 +(H+f)^2 \right) \left( g' V^\mu -g W_3^\mu \right)
    + H \partial^\mu \phi_3 - \phi_3 \partial^\mu H + f\partial^\mu \phi_3
   \right]
   \ea
   it suffices to replace the weak mixing angle  by its traditional or
Stueckelberg--free value, ${\rm c}_w \to g/\sqrt{ g^2 +g^{\prime2}} $ and
  ${\rm s}_w \to g'/\sqrt{ g^2 +g^{\prime2}} $ to find the expressions for
$J_{\rm em}^{(g) \mu}$ and $J_{\rm em}^{(s) \mu}$. Note the
enormous simplification in the scalar current, where the second line drops off,
including in particular the term linear in $\partial_\mu \phi_3$.

  The Noether current which is associated with the global transformation
leaving invariant the vacuum expectation value of the scalar field
is just $J_{\rm em}^\mu $, and thus the {\it v.e.v.} is invariant
under the action of the electric charge $Q=\int {\rm d}^3 x J_{\rm
em}^0 =Y/2 +T_3 $, but not under that of the charge associated
with $J_A$.

  \subsubsection{\label{sec:qcurr}Quantum currents}

  Adding the gauge fixing terms to the lagrangian \eq{llkk}, so that
  \be {\cal L}= {\cal L}_\circ + {\cal L} _{\rm gh} + {\cal L} _{\rm gf} \ee
 where the last two terms are given by eqs. \eq{mlll} and \eq{mlll2},  the
field equations for $V_\mu$ and $ W^3_\mu$ are now
 \ba &&  \partial_\mu F^{\mu\nu} +\frac1{\alpha'} \partial^\nu  {\cal G} =
j^\nu \\
 &&  \partial_\mu   F_3^{\mu\nu}+\frac1{\alpha} \partial^\nu {\cal   G}_3
   -i g \partial^\nu \left( \omega^{*+}
\omega^- - \omega^{*-} \omega^+ \right) =   J_3^\nu +i g \left(
\omega^{*+} \partial^\nu \omega^- - \omega^{*-} \partial^\nu
\omega^+ \right)\ea with the usual          \ba  &&  {\cal G}   =
           \left( \partial
V + \alpha' m B - \frac{\alpha'g'f}2 \phi_3 \right) \\ &&
        {\cal \vec G} =            \left( \partial
W_3 +  \frac{\alpha g f}2 \phi_3 \right)\ea

 Since physical states satisfy the supplementary conditions
  \be \left< {\rm phys}' | {\cal G} | {\rm phys} \right> =0 \ee
 \be \left< {\rm phys}' | {\cal \vec G} | {\rm phys} \right> =0 \ee
and they do not contain any ghosts, it is clear that the divergence of the
currents $j^\mu$ and $\vec J^\mu$ are zero sandwiched between physical states:
   \be \left< {\rm phys}' | \partial j | {\rm phys} \right> =0 \ee
 \be \left< {\rm phys}' | \partial \vec J | {\rm phys} \right> =0 \ee

Of course, this result is fully expected. Since the gauge--fixing
lagrangian is ${\bf s}$ of something, the difference between the
quantum and classical currents is again the ${\bf s}$ of
something, which cannot make any difference for physical states.

  \subsubsection{\label{sec:cconc}Conclusions}

  From the above analysis, it would seem that the current associated with
external classical fields (the background fields?) is $J_{\rm em}^\mu $,
whereas the current coupled to the quantum asymptotic photon field is $J_{\rm
A}^\mu$.  The latter can be measured in scattering processes like  Bhabha,
Compton or bremsstrahlung. The former's physical relevance stems from the fact
that it is the Noether current of the global $U(1)_{\rm em}$ symmetry leaving
invariant the vacuum.

\subsection{\label{sec:phenol}Some phenomenology}

Before proceeding to an overview of some of the phenomenological
issues associated with our modification of the standard model, it
is perhaps wise to recall the stringent experimental limits on the
mass of the photon. The analysis of experimental data is based on
the Maxwell--Proca equations \ba
&& \partial_\mu F^{\mu\nu}_V = j^\nu + m_\gamma^2 V^\nu \\
&&       \partial_\mu \tilde F^{\mu\nu}_V =      0 \ea where in
addition to the Lorentz gauge following from the field equations
one has imposed the Proca gauge $B=0$. These equations are not
gauge-invariant, but they are the gauge-fixed version of
gauge-invariant field equations, as we have discussed in section
\ref{sec:quamasvecfi}. Since gauge-fixing does not change the
physics, they are perfectly valid starting points for
experimentalists. Interestingly, in an experiment of size $L$,
photon mass effects scale like $(m_\gamma L )^2$, without any
resonance effects in the photon frequency \cite{Gol71}.
Measurements of the energy density with a Cavendish torsion
experiment lead to the strong limit \cite{Luo03} \be m_\gamma <
1.2 \times 10^{-17} \;{ \rm eV} \ee Direct measurements of the
speed of light are five orders of magnitude worse \cite{Sch99}.
Other limits and methods, as well to references to the early
literature, notably \cite{Sch49}, can be found in
\cite{PDBook,Lak98}.

It is important to keep in mind that the direct limit on the photon's mass is
very
strong, so that the modifications to the standard model stemming from the
consistent application of the Stueckelberg mechanism to the hypercharge Abelian
factor, in particular the modified weak mixing angle and fermion couplings to
the photon
and the $Z$, are not expected to be competitive.

It is useful  to view the introduction of the BRST--consistent mass for the
photon in the standard model  as a  tiny modification of the latter. Charged
currents do not change, whereas to lowest order in the Stueckelberg mass
parameter, the weak neutral currents remain essentially undisturbed: the photon
acquires a mass and changes its couplings without affecting much the rest of
the theory. This is, of course, fortunate, since the experimental success of
the standard model constitutes the culmination of the quantum understanding of
nature.

There are no flavor--changing neutral currents in the theory.

A major problem arises in the computation of charges for bound states. Consider
for example the neutron, or rather the baryon with valence quarks $udd$.
Depending on the chirality of the three quarks, we find different charges or,
more precisely, couplings to the photon.
The results can be summarized in terms of the neutrino's coupling to the
photon,
\be
Q^\epsilon_\nu =n_\nu^A=\frac12 (g{\rm s}_w - g'{\rm c}_w ) \simeq
     \frac{e}2 \epsilon \ee with $e$ defined in eq. \eq{cargae}. In the
following table,  we
denote by $Q({\cal B})$ the coupling of the bound state $\cal B$ (labelled by
its valence
quarks) to the physical asymptotic massive photon.
\ba && Q(u_Ld_Ld_L)=   -Q^\epsilon_\nu
 \\
 && Q(u_Ld_Ld_R)= 0 \label{neu1}\\
 && Q(u_Ld_Rd_R)=  - Q^\epsilon_\nu \\
 && Q(u_Rd_Ld_L)= 2  Q^\epsilon_\nu  \\
 && Q(u_Rd_Ld_R)=  Q^\epsilon_\nu  \label{neu2} \\
 && Q(u_Rd_Rd_R)= 0 \ea
 That the charge of $\Delta ^0$ is not exactly zero is of no particular
experimental relevance, but one should have serious problems accepting the fact
that a neutron's charge depends on its spin. It is comforting that \eq{neu1},
one of the true neutron states, is neutral, but disquieting that the coupling
of
\eq{neu2} to the photon does not vanish.

 The $uud$ bound state (proton) has similarly three different couplings to the
photon
  depending
on the handedness of the valence quarks, with $u_Lu_Ld_L$ and $u_Lu_Rd_R$
degenerate.
 The fact that left- and right-handed electrons have different couplings to the
photon
  has the
same origin as the difference in couplings to the photon  for the
various bound states of three valence quarks. This situation  is
very problematic not only conceptually, but also for the stability
of matter. Indeed, it is very hard to escape catastrophic and
observable consequences (for examples, electric fields near
grounded metallic conductors) if matter is not neutral. The total
charge of the  hydrogen atoms can be read off from the following
table, showing the total  coupling to the photon  of the bound
states of a $uud$ baryon  and an electron.

 \ba  & e_L & e_R \nn\\
 u_Lu_Ld_L &  0 & -Q^\epsilon_\nu  \\
 u_Lu_Ld_R &  -Q^\epsilon_\nu  &  2 Q^\epsilon_\nu  \\
 u_Lu_Rd_L &   Q^\epsilon_\nu   &  0 \label{22218} \\
 u_Lu_Rd_R &  0 & -Q^\epsilon_\nu  \label{22219} \\
  u_Ru_Rd_L &  2  Q^\epsilon_\nu    &  Q^\epsilon_\nu  \\
  u_Ru_Rd_R &  Q^\epsilon_\nu   &0\ea

 Equations \eq{22218} and \eq{22219} could be interpreted as follows. If one
writes a ``left--handed proton wave--function'' as
 \be p_L = \left[ u_L(1) u_R(2) -u_R(1) u_L(2) \right] d_L(3) \ee
 its photon charge $g'{\rm c}_w$ is equal and opposite to the charge of the
right-handed electron $e_R$. Similarly, if
   \be p_R= \left[ u_L(1) u_R(2) -u_R(1) u_L(2) \right] d_R(3) \ee
 its charge $\left( g'{\rm c}_w+  g{\rm s}_w \right)/2$ is compensated by the
 $e_L$ charge. For the neutron, only the photonic charge of $n_L$ vanishes.
  The neutrality of normal matter is thus assured.
  This calculation is too naive, however, in the absence of a realistic
three--quark model.

\section{\label{sec:influ}The influence of Stueckelberg's 1938 papers}

Stueckelberg's 1938 papers, rather difficult to read then and now, have been
continuously cited from 1941 to the present. The following domains of influence
will be reviewed: renormalization of massive vector field interactions, hidden
symmetry, and  electroweak theory without spontaneous symmetry breaking.

Other topics worthy of attention which have developed from Stueckelberg's 1938
papers include baryon number, as emphasized in \cite[footnote on p.25]{Wig67},
broken chiral symmetry,  and  electromagnetic properties of vector
mesons. We shall not review them here.

We must distinguish two aspects of
Stueckelberg's formalism for massive gauge vector fields (vector mesons in
those days):

1) The decomposition of the massive Proca vector field $V_\mu$, as
in equation \eq{220} above, namely \be V_\mu = A_\mu - \frac1m \:
\partial_\mu B \label{61}\ee
The $\partial_\mu B$ term is
responsible for the singular character of Proca's theory: the
commutation relations of the massive vector $A_\mu$ and of the
Stueckelberg field $B$ are local, but the presence of the
derivative of $B$ makes the commutation relations of $V_\mu$
non-local.

2) The replacement of Proca's free lagrangian by Stueckelberg's:
\ba & {\cal L}_{Stueck} (A_\mu, B) &= {\cal L} _{Proca} (A_\mu, B)
+ {\cal L}_{gf} \nn\\ && = -\frac12 (\partial_\mu A_\nu)^2
+\frac12 m^2 A_\mu^2 +\frac12 (\partial_\mu B)^2 -\frac12 m^2 B^2
\label{62}\ea (We consider the neutral case \eq{32}, which follows
from \eq{3333} with $\alpha=1$; for charged vector fields the
Proca and Stueckelberg lagrangians are given by the historical
\eq{29} and \eq{5Lnew}, respectively). Previously, it was believed
that a massive vector theory could not be gauge invariant, but
\cite{Pau41} showed that ${\cal L}_{Stueck}$ was a counter-example
to such belief, still surprisingly common nowadays. We have seen
in section \ref{sec:brs} that the theory of real massive vector
fields is even BRST invariant, which is relevant for its
renormalizability \cite{Del88}.

We now review the historical development of these ideas in three
different but complementary directions: (A) renormalizability, (B)
hidden symmetries, and (C) massive theories without Higgs. We also
mention (D) some related applications of the Stueckelberg trick.

\subsection{The question of renormalizability}

\subsubsection{Power--counting renormalizability}

It was found in the 1930s that the quantum field theory of electrons and
photons (quantum electrodynamics, QED) was plagued by infinities, already at
low orders in perturbation theory.

In 1949, Dyson showed that, in QED, renormalization of mass and charge of the
electron and renormalization of the wave-functions (or better, the rescaling of
the field operators) removed all the divergences from the $S$--matrix to all
orders in perturbation theory \cite{Dys49a,Dys49b}. This is now known as the
power--counting procedure, because it is based on counting the powers of
four--momenta over which one integrates. Dyson's proof was later made more
rigorous
by  
 \cite{Wei60},  \cite{Bog76} and others.

After Dyson, it was natural to ask whether massive vector field interactions
were also renormalizable. Vector mesons were first considered, following
Yukawa, as mediators of strong nuclear interactions, with little
phenomenological success.

\cite{Miy48} was the first to use Stueckelberg's lagrangian
\eq{62} extensively in the theory of charged massive vector mesons interacting
with nucleons, mimicking the  treatment of QED in the
super--many--time formalism \cite{Tom46}, which has the advantage of being
manifestly Lorentz invariant. Tomonaga and collaborators applied this formalism
to the
interaction of electrons with photons \cite{Kob47a,Kob47b} and of mesons with
photons \cite{Kan48a,Kan48b}. In the latter case, an additional interaction
term was necessary to satisfy relativistic invariance. Miyamoto showed that,
for the interaction of mesons and nucleons, the additional term is provided
automatically in the Stueckelberg formalism by the scalar $B$--field.

Miyamoto then derived the generalized Schr{\"o}dinger equation, the
integrability conditions, Stueckelberg's auxiliary condition, and
the passage to the Heisenberg picture. Proca's theory is not well
adapted to the many-time formalism, and  it has problems with the
integrability conditions. Miyamoto's careful and extensive work
paved the way for further research.

In a different vein, using   the many--time formalism of \cite{Dir32},
\cite{Pod48} considered a
non-renormalizable modification of QED with higher
derivatives of the massless $A_\mu$ field. They claimed to get a finite
self-energy for a point source. In the process of trying to quantize the
theory, they introduced the Stueckelberg field $B$ in order to get a consistent
subsidiary condition, similar to Stueckelberg's \eq{klop}.

\subsubsection{1949--1954 : lessons from QED}

The first   definite answer to the question of
renormalizability of vector interactions with the nucleons was provided by
\cite{Mat49a,Mat49b}. This problem depends crucially on the high--energy
behavior of the $S$--matrix, which depends in turn on the power of the
energy--momentum factors.
Four--momenta are the Fourier transform of derivatives, and they may appear in
the commutation relations of the quantized
fields, and hence in the propagators, as well as in the interaction lagrangian.

This point can be illustrated by comparing  Proca's commutation relations for
real massive vector fields,
\be \left[ V_\mu(x) ,V_\nu(y) \right] = -i \left( g_{\mu\nu} + \frac1{m^2}
\partial_\mu \partial_\nu \right) \Delta_m(x-y) \label{63}\ee with
Stueckelberg's: \be\left[ A_\mu(x) ,A_\nu(y) \right] = -i g_{\mu\nu}
\Delta_m(x- y) \label{64}\ee Similarly,  \eq{26} can be compared with \eq{211}
for charged (\sl i.e. \rm non-hermitian) vector fields.
 Proca's theory is clearly more divergent at high energies than
Stueckelberg's. On the other hand, the interaction of Proca's massive vector
field with a charged fermion field $\psi$ is the harmless
\be {\cal L}^I_{Proca} = e\bar\psi \gamma^\mu V_\mu \psi  \label{65}\ee
whereas Stueckelberg's is \be
{\cal L}^I_{Stueck} = e \bar\psi \gamma^\mu \psi \; \left( A_\mu - \frac1m
\partial_\mu B \right) \label{66}\ee The last vertex diverges like $p_\mu$, so
it seems that Stueckelberg's interacting theory is also singular.

According to Matthews, however, the bad terms with $\partial_\mu B$ can be
eliminated from the interaction by a unitary transformation, as follows.
Working
in the Dirac or interaction picture, Matthews (quoting Miyamoto) writes,
instead
of \eq{66}, the interaction term \be {\cal L}^I_{Stueck} = j^\mu\; \left( A_\mu
- \frac1m \partial_\mu B \right) +\frac1{2m^2} (j^\mu n_\mu)^2\label{67}\ee
where
$j^\mu =e \bar\psi \gamma^\mu \psi$ and $n_\mu$ is a normal unit vector to a
general space-like surface. The point is that the last term,
quadratic in $j^\mu$, is absolutely necessary for what they called
``integrability''
 in those days; note that it does not look renormalizable. The
physical states are defined using Stueckelberg's
subsidiary condition \eq{216},
 \be (\partial^\mu A_\mu +m B)^{(-)} |{\bf phys}>  =0 \label{68}\ee
  Now Matthews performs the unitary transformation
\cite{Dys48,Cas49} \ba && |{\bf phys}> \to |{\bf phys}'> = {\rm
e}^{-i G}|{\bf phys}> \label{69}\\ && G=\frac{1}m \int d\sigma^\mu
j_\mu(x) B(x) \label{610}\ea where the integral is over the
space--like reference surface. Note the similarity to the gauge
transformation (\ref{t7}--\ref{37}). This redefinition eliminates
the last two terms in \eq{67} and thus  we end up with an
interaction lagrangian exactly like that of the massless photon
interacting with the electron current in QED, which is
renormalizable \cite{Dys48}. For charged (non hermitian) vector
fields, an additional term spoils the renormalizability
\cite{Cas49}. See also \cite{Bel49a,Bel49b} and \cite{Gup51}.

\cite{Phi54} worked in the same framework as Matthews
(Stueckelberg lagrangian, with the unitary transformation
\eq{69}), but criticized a technicality concerning the
integrability conditions. Introducing a ``quasi interaction
representation", in which the massive field $A_\mu$ obeys free
field equations but the ``free'' equations for the $B$ and $\psi$
fields include the term $j_\mu(x) \partial^\mu B(x)$, these can be
eliminated just as proposed by Matthews.

The review
\cite{Mat51}  established to which meson interactions Dyson's
proof of finiteness of QED could be applied. The result is that
the only interaction of vectors or pseudovectors with fermions
satisfying Dyson's criteria is the vector interaction of a neutral
vector \cite{Mat49a}. On the other hand, the scalar interactions
of scalars and the pseudoscalar interactions of pseudoscalars
require only a finite number of renormalizations, as in the case
of QED.\footnote{In 1951 parity conservation was still
unquestioned.} In addition to the counterterms analogous to those
occurring in QED, one needs a quartic (pseudo)scalar term and, in
the case of scalar mesons, a further cubic term. See also the
textbook presentation of \cite{Ume56} and the discussions of
\cite{Fuj59}.

\subsubsection{1960--1962 : equivalence theorems}

In the late 1950s and early 1960s, the interest in intermediate
vector theories was revived by work  on an isospin $SU(2)$
gauge--invariant theory of massless vector fields \cite{Yan54}.
For a history of gauge fields, see \cite{Ora00}. Furthermore,
 \cite{Fey58}, \cite{Sak58} and \cite{Sud58}  proposed the
universal $V-A$ theory of weak nuclear interactions, which ``can most
beautifully be formulated by assuming an intermediate vector particle,'' as
stated by \cite{Kam60}.  Indeed, \cite{Fey58,Sud58} suggested that the $V-A$
interaction could be mediated by a charged spin--one particle. \cite{Blu58}
proposed to add
a neutral field in the framework of an $SU(2)$ Yang--Mills invariance, and
\cite{Gla61a} added yet another neutral particle to achieve $SU(2)\times U(1)$
invariance. On the other hand, \cite{Fuj59} proposed a massive vector meson to
mediate strong interactions, and \cite{Sak60} identified it with the $\rho$.
\cite{Lee49} had already proposed the idea that the exchange of a boson of
non-specified spin could explain the approximate equality of the
$\beta$--decay and muon interaction couplings (the universality of weak
interactions). \cite{Sch57} had  ``freely invented'' an intermediate vector
boson in strong and weak interactions (with some hints from experiment).
 It was tempting to identify the latter with the
Yang--Mills field. But for empirical reasons, related to
the Fermi theory of weak interactions, this particle ought to be massive. And
yet, the theory of massive charged vector fields seemed not to be
renormalizable. The main reason for this singularity seemed to be the lack of
gauge invariance of such theories.

In this context, \cite{Gla59}  conjectured  that a ``partially conserved"
vector current could lead to a renormalizable theory. This proposal was
refuted independently by \cite{Sal60} and  \cite{Kam60}. Both used the
decomposition \eq{220} $V_\mu =A_\mu -
m^{-1}\partial_\mu B$, with $m$ the mass of the vector meson, and found a
general equivalence theorem for vector meson interactions, from which they
deduced ``a precise criterion for renormalizability in the conventional sense"
 \cite{Kam61}.
They were inspired by
 \cite{Dys48}, who  had already shown that the pseudovector interaction
 \be g \bar\psi \gamma^\mu \gamma_5 \partial_\mu \tilde B \psi \label{D2}\ee of
a
{\sl pseudoscalar } field $\tilde B$ with a nucleon field $\psi$ was equivalent
to an exponential pseudoscalar interaction of $\tilde B$ with $\psi$, using the
unitary transformation \be \psi\to \psi' = {\rm e}^{i g \gamma_5 \tilde B}\psi
\label{D1}\ee which eliminated the pseudovector interaction  and took the mass
term $M\bar\psi\psi$ of the nucleon into \be M \bar\psi'\left( 1-{\rm e}^{-2ig
\gamma_5 \tilde B} \right) \psi' \label{D3}\ee

\cite{Sal60}  applied the same procedure   to a real pseudovector Proca
field $\tilde V_\mu=\tilde A_\mu -\frac1m \partial_\mu \tilde B$. The
pseudovector interaction of $\tilde V_\mu$ with fermions contains the
pseudovector interaction of $\tilde B$. The elimination of the latter yields,
again, a non-renormalizable exponential interaction. The current \be j_\mu = g
\bar \psi \gamma_\mu \gamma_5 \psi \ee is ``partially conserved:" \be
\partial^\mu j_\mu = 2i g M \bar\psi \gamma_5 \psi \ee

 \cite{Kam60} treated the general case  of the interaction of the
neutral vector field $V_\mu$ with an arbitrary complex field of
spin 0, 1/2, or~1, and assumed that the interaction hamiltonian
was of the form $H_1+H_2$, with $H_1$ gauge invariant and $H_2$
not gauge invariant. Writing $V_\mu = A_\mu - \frac1m \partial_\mu
B$ and applying the analogue of \eq{D1}, the $B$ field is
successfully eliminated from $H_1$, but reappears in an
exponential in $H_2$.

Both Salam's and Kamefuchi's examples contradict Glashow. They had
only considered, nevertheless, the case of Abelian $U(1)$ gauge
invariance.

Shortly thereafter, \cite{Ume61} generalized the equivalence theorem of
\cite{Sal60} and \cite{Kam60} to the
non-Abelian isospin $SU(2)$ Yang--Mills
gauge theory, with massive vector mesons. They found that the mass terms spoil
renormalizability.

To prove the equivalence theorems, they used Stueckelberg's lagrangian, which
they introduced in a new and elegant way. Then they extended it to the
Yang--Mills case.


Their starting point is the Proca lagrangian for a real vector field $V_\mu$,
with interactions. Introduce in addition to the real scalar Stueckelberg field
$B(x)$
an extra real scalar field $C(x)$, with the wrong energy and the wrong metric
in Hilbert
space: \ba &{\cal L}_{UK} =&-\frac14 F_{\mu\nu}(V)^2 +\frac12 m^2 V_\mu^2
+\frac12 (\partial_\mu B)^2 -\frac12 m^2 B^2 \nn\\ && -\frac12 (\partial_\mu
C)^2 +\frac12 m^2 C^2 + {\cal L}_{int} (U_\mu, \phi) \label{U1}\ea The
non-vanishing commutation relations are \ba && \left[ V_\mu (x) , V_\nu (y)
\right]
= -i \left( g_{\mu\nu} +\frac1{m^2} \partial_\mu \partial_\nu \right)
\Delta_m(x-y) \nn\\ &&\left[ B (x) , B(y) \right] = i \Delta_m(x-y)
\label{U2}\\
&& \left[ C (x) , C (y) \right] = -i \Delta_m(x-y) \nn\ea and the interaction
lagrangian depends on some other ``matter'' fields $\phi(x)$ and on the vector
field \be U_\mu(x) = V_\mu (x) +\frac1m \partial_\mu(B(x) -C(x)) \label{U3}\ee

It follows from \eq{U1} and \eq{U3} that $E=B-C$ is a free massive
field: \be (\partial^2 + m^2 ) E(x) =0\label{U4}\ee To ensure
positivity of the physical Hilbert space, one can impose the
subsidiary condition \be E^{(-)}|{\bf phys}> =0 \label{U5}\ee for
physical states in the Heisenberg representation, since it is
consistent with the field equations. Define now $A_\mu$ by \be
V_\mu(x) = A_\mu(x)  -\frac1m \partial_\mu C (x) \label{U6}\ee
Then \eq{U3} implies that the interacting vector field is \be
U_\mu(x) = A_\mu(x) + \frac1m \partial_\mu B (x) \label{U7}\ee
Substituting \eq{U6} into \eq{U1} we find, after integrating by
parts and dropping total derivatives, \be {\cal L}'_{UK} =
-\frac12 (\partial_\mu A_\nu)^2 +\frac12 m^2 A_\mu^2 +\frac12
(\partial_\mu B)^2 -\frac12 m^2 B^2  -\frac12 D^2 + {\cal L}_{int}
(U_\mu, \phi) \label{U8}\ee where the auxiliary field \be D(x)=
\partial_\mu A^\mu (x)+ m C (x) \label{U9}\ee satisfies the
algebraic equation of motion $D=0$. Using this fact, the
subsidiary condition \eq{U5} reads now \be (B + \frac1m
\partial^\mu A_\mu)^{(-)} |{\bf phys}> =0 \label{U10}\ee so that
both Stueckelberg's lagrangian and Stueckelberg's subsidiary
condition are recovered.

Umezawa and Kamefuchi proceeded to extend the Stueckelberg lagrangian to
isovector fields. They decomposed the lagrangian into a free piece: \be {\cal
  L}_0 = -\frac12 (\partial_\mu \vec A_\nu)^2 +\frac12 m^2 \vec A_\mu^2
+\frac12
(\partial_\mu \vec B)^2 -\frac12 m^2 \vec B^2 \label{U12}\ee and an
interaction:
\be  {\cal L}_I = \frac12 (m^2 -m^{\prime2}) (V_3^\mu)^2 -\frac{g}2 \vec
{\cal A}_{\mu\nu} \cdot \vec V^\mu \times \vec V^\nu - \frac{g^2}4 (\vec V^\mu
\times
\vec V^\nu )^2 \label{U13}\ee where \be \vec V_\mu = \vec A_\mu -\frac1m
\partial_\mu \vec B \ee and the short-hand \be \vec {\cal A}_{\mu\nu} =
\partial_\mu \vec A_\nu - \partial _\nu \vec A_\mu \ee is {\sl not} the
non-Abelian field strength of $\vec A_\mu$.

Notice the astute first term in the interaction, with $m'$ a free parameter.
For $m=m'$, the full isospin symmetry is restored, and the isovector current
\be \vec j_\nu =\partial^\mu \vec {\cal A}_{\mu\nu} - m^2 \vec V_\nu
\label{U15}\ee is conserved, $\partial^\mu \vec j_\mu =0$.

The derivation of the equivalence theorems for the isovector
lagrangians, for $m=m'\not=0$ or for $m\not= m'$, is rather
lengthy. It turns out that, if $m\not=0$, the theory is not
renormalizable, even when the current \eq{U15} is conserved.


In the classic paper \cite{Sal62} entitled {\sl Renormalizability of Gauge
Theories}, Salam gave a simpler and more general discussion of the
renormalizability condition, which we now summarize. Salam's paper is also
remarkably modern in its notation and outlook. Consider a set of spinor fields
$\psi$ on which acts a Lie
group with generators $T_i$: \be \psi(x) \to \psi'(x) = {\rm e}^{ig T_i b^i(x)
} \psi(x) \equiv U(x) \psi(x) \label{S1}\ee with  structure constants
defined
by \be [T_i, T_j] =i f_{ij}^{\quad k} T_k \label{S2}\ee and a set of vector
fields \be V_\mu(x) = T_i V_\mu^i (x) \label{S3}\ee transforming
inhomogeneously: \be V_\mu(x) \to V_\mu'(x) = U^{-1}(x) V_\mu(x) U(x)
+\frac{i}g
U^{-1}(x) \partial_\mu U(x) \label{S4}\ee

The following lagrangian is invariant under the gauge
transformation $U(x)$: \be {\cal L}_S (\psi, V_\mu) = i \bar \psi
(\slash \partial  -i g \slash V ) \psi +M \bar\psi \psi -\frac14
{\rm tr}\, F_{\mu\nu} F^{\mu\nu} \label{S5}\ee with \be F_{\mu\nu}
= (\partial _\mu -i g V_\mu) V_\nu -(\partial _\nu -i g V_\nu)
V_\mu \label{S6} \ee the covariant field strength, which
transforms homogeneously: \be F_{\mu\nu}(x) \to F_{\mu\nu}'(x)
=U^{-1}(x)F_{\mu\nu}(x) U(x) \label{S7}\ee Add now a vector mass
term, which is not invariant under the gauge transformation
\eq{S4}, \be {\cal L}_{mass} = -\frac12 m^2 \:{\rm tr}\, V_\mu^2
\label{S8}\ee

To study the renormalizability of this theory, Salam proposed two steps. First,
introduce the Stueckelberg fields $B^i$ through $V_\mu = A_\mu -\frac1m
\partial_\mu B$, with $A_\mu =T_i A_\mu^i$ and $B =T_i B^i$. Secondly, change
$\psi$ to $\psi'$ and $V_\mu$ to $V_\mu'$, using the gauge transformations
\eq{S1} and \eq{S4}, with the
gauge parameters chosen as $b^i=B^i$. Under this transformation, ${\cal
L}_S(\psi, V_\mu)$ is invariant, but ${\cal L}_{mass}$ is not.

On the other hand, it follows from \eq{S4} that $V^\prime_\mu = A_\mu +{\cal
O}(g)$, and hence, in the weak coupling limit $g\to0$ where asymptotic states
are defined, one obtains \be V_\mu ^{\prime in} =A_\mu^{in} \label{S10}\ee The
$S$--matrix
in the new variables has contributions from two pieces: those from ${\cal
L}_S(A^{in}_\mu , \psi^{in})$ yield only renormalizable infinities (the
derivative couplings of $B$ with $\psi$ have been eliminated), whereas
those from ${\cal L}_{mass}(A_\mu^{in}, B^{in})$  produce exponential
infinities {\sl unless} either
\be m=0 \label{S11}\ee
or the following two conditions hold:
\be {\rm tr} \left[ \frac{m^2}{g^2} (\partial_\mu U) (\partial^\mu U^{-1}) -
(\partial_\mu B^{in})^2 \right] =0 \label{S12} \ee
and
\be {\rm tr} \left[ A_\mu^{in} \left( U^{-1}\partial^\mu U - i\frac{g}{m}
\partial^\mu
B^{in}\right) \right] =0 \label{S13} \ee

This is a powerful theorem. For a massive neutral vector field
interacting with fermions, for example with the nucleons, there is
only one $B$ field, and $U$ is Abelian. Then both \eq{S12} and
\eq{S13} are satisfied, and the theory is renormalizable even with
a massive vector.\footnote{This result can be understood easily in
terms of the BRST invariance of section \ref{sec:brs}.}

In general, \eq{S12} and \eq{S13} can be satisfied provided ${\rm tr}\, T_i
T_j=0$. However, for simple Lie groups ${\rm tr}\, T_i T_j=\lambda
\delta_{ij}$, with the normalization $\lambda\not=0$, and thus the last two
conditions are not satisfied.

The only term in the lagrangian considered which is not gauge invariant is the
vector mass term. Clearly, any other non-invariant term in the lagrangian, of
the generic form ${\cal L}(\psi)$, will transform into ${\cal L}(S\psi')$, with
$S$ containing
non-renormalizable exponentials of Stueckelberg's $B$ field.
This checks, for instance, with the fermion mass term $M\bar\psi \psi$ under
the transformation \eq{D1} above, which yielded the horrible \eq{D3}.

Salam concluded that renormalizability of a gauge theory requires
vanishing masses. Salam then developed the ideas of \cite{Nam61}
to get masses in a self--consistent way, and later co-birthed the
concept of broken symmetry \cite{Gol62}, which eventually gave
mass to the vectors of a broken gauge invariance
\cite{Eng64,Hig64}. The first discussion of a non-Abelian
spontaneous symmetry breakdown is in \cite{Kib67}, whereas
\cite{Hoo71a,Hoo71b} provided the proof of renormalizability.

Earlier,  \cite{Kom60} had calculated explicitly to lowest order
the self--energy and vertex correction of the isospin $SU(2)$
Yang--Mills theory, in agreement with the above result
\cite{Sal62}. However, to prove non-renormalizability one must
consider not the Green's functions in general, but the Green's
functions on-shell, {\sl i.e.}  the $S$-matrix elements
\cite{Vel68}. \cite{Zim68} used Stueckelberg's lagrangian and its
invariance under the Pauli gauge transformations to study the
renormalization of masses of real vector fields. The
renormalization  of massive chiral $U(1)$ theory was carried out
by \cite{Lee69}. For a generalization of the Stueckelberg
formalism see \cite{Fuj64}. A later discussion of these subjects
can be found in \cite{Ito76}.

\subsection{\label{sec:hs}Hidden symmetries}

As mentioned above, it was widely assumed that giving a mass to the photon
would spoil gauge invariance. But   the Stueckelberg lagrangian
with a physical scalar field $B$ in addition to the massive photon
\cite{Stu38I,Stu38II} enjoys indeed gauge invariance \cite{Pau41} and even
BRST invariance \cite{Del88}. The Proca lagrangian does not have this symmetry,
so it is absolutely necessary to include the Stueckelberg field $B$ which does
not play, however, a dynamical role. One may call this state of affairs a
``hidden symmetry." As we shall see below, the same trick has been used by
several authors in more general contexts.

Glauber, who visited Pauli at Z{\"u}rich in 1950, was the first to
emphasize the close relationship between Stueckelberg's
gauge--invariant scheme and QED \cite{Gla53}. We also owe him the
remark that the \cite{Dys48} transformation is just a gauge
transformation, which eliminates the $B$--field from the
interaction. For vanishing photon mass $m$, the Stueckelberg field
$B$ disappears as well from the supplementary condition, which is
then the same as in QED, and eliminates the longitudinal
polarization of the vector field $A_\mu$. For non-vanishing mass
$m\not=0$, this condition can be considered as a definition of the
$B$--field, while the massive photon is no longer restricted to be
transverse.

Glauber then proceeds to calculate radiative corrections to the
photon mass. These can be separated in two parts, according to
whether they are gauge--invariant or not. Both pieces are formally
divergent. As in electrodynamics, the non-gauge--invariant
integrals must be presumed to vanish (to second order, the
contribution is identical to the one of QED). To renormalize the
theory, one must remember that the $B$--field is still present in
the free--field hamiltonian. To preserve gauge invariance of the
corrections, one has to reintroduce $B$ through the supplementary
condition in the last step. The photon mass correction is
logarithmically divergent, and vanishes when $m$ goes to zero.

 Aware of Glauber's preprint, \cite{Ume52} generalized these
results to tensor representations,   studied the transition when the photon's
mass vanishes [see also \cite[pp. 113 and 204]{Ume56}], and  gave a
classification of the renormalizable and non-renormalizable interactions of
neutral and charged particles of spin 0, 1/2 and~1.  \cite{Bon63} considers the
gauge invariance of massive vector theories in a five--dimensional formalism
and shows the connection between Stueckelberg's formalism and that of
\cite{Ogi61}. The latter start from the lagrangian for the real field $A_\mu$
interacting with a conserved (Dirac) current
\be {\cal L}= -\frac12 \partial_\mu A_\nu \partial^\mu A^\nu -\frac{m^2}2
A_\mu^2 +j_\mu A^\mu \ee
This is invariant under the transformation $\delta A_\mu=\partial_\mu
\Lambda(x)$ subject to $(\partial^2-m^2)\Lambda=0$. The $A_\mu$ field can be
split into an invariant spin--1 part, and a non-invariant spin--zero part. They
show that the scalar has no interactions, so one can forget it. Furthermore,
the total energy operator  is positive definite up to an irrelevant constant.
Hence, the supplementary condition \eq{216} imposed by Stueckelberg to ensure
positivity is no longer required.

Many other papers deal with the relation of massive to massless QED, focusing
on a variety of questions, independently of Stueckelberg's $B$ field. For
example, \cite{Coe51} and \cite{Stu57} find that
after a suitable canonical transformation, the contributions of the scalar and
longitudinal components of the vector field to the
$S$--matrix compensate each other in the
limit $m\to0$. \cite{Sch62,Sch62b} exhibits
a gauge invariant massive field theory which has no continuous limit to QED
when $m\to0$. \cite{Bou62} invent a soluble field theory in which they then
carry out the limit as the bare mass vanishes, but in which the vector particle
remains massive. This toy model is gauge invariant, since   a
Stueckelberg massless scalar field is also introduced. \cite{Fel63} also show
that ``gauge
invariance does not require the bare photon mass to be zero''.
\cite{Kam64} use the original formalism of
\cite{Stu38I,Stu38II,Stu38III} to show that the representation of gauge
transformations for massive vector fields is inequivalent to that for zero
mass. As  discussed below, the limit $m\to0$ was also studied by
\cite{Dam70} and \cite{Sla71}.

\cite{Ram86}  applied Stueckelberg's scheme to a completely new
domain, in order  ``to obtain the fully covariant and gauge
invariant
field theory for free open bosonic strings in [the critical] 26 dimensions.
[This]
approach [$\ldots$] is based on very simple analogies with local field theory.
[$\dots$] Stueckelberg fields arise naturally and are shown to be unrestricted
for the most general gauge transformations."

Ramond remarks that ``in any theory which  is known in a specific gauge, one
can always reconstruct the original gauge invariant theory provided one knows
the form of the gauge transformations and the gauge conditions."  This was
precisely the situation for the first quantized string, where the gauge
symmetry is given by (half of) the Virasoro algebra.

In massless QED, the gauge transformation is of course $\delta A_\mu(x) =
\partial_\mu \Lambda(x)$, and the gauge condition is $\partial^\mu A_\mu=0$.
From this, and the equation of motion $\partial^2 A_\mu=0$, one can deduce the
Lorentz invariant and gauge invariant equation $\partial^\mu(\partial_\mu A_\nu
- \partial_\nu A_\mu)=\partial^\mu F_{\mu\nu}=0$.

In the Proca theory, one would start with \be (\partial^2 +m^2 )
A_\mu(x) =0 \label{R1}\ee and \be \partial^\mu A_\mu =0
\label{R2}\ee One could try the gauge transformation \be \delta
A_\mu = \partial_\mu \Lambda(x) \label{R3}\ee The variation of
\eq{R2} gives \be \partial^\mu A_\mu + \partial^\mu \partial_\mu
\Lambda =0 \label{R4}\ee However, the variation of \eq{R1} implies
\be (\partial^2 +m^2 ) A_\mu + \partial_\mu (\partial^2 +m^2)
\Lambda=0 \label{R5}\ee which is not compatible with \eq{R4}. Now,
Ramond rewrites \eq{R4} as \be \partial^\mu A_\mu + (\partial^2
+m^2) \Lambda -m^2 \Lambda =0 \label{R6}\ee and interprets the
last term as the variation of the Stueckelberg scalar field $-mB$.
Equation \eq{R6} then becomes \be \partial^\mu A_\mu + m B +
(\partial^2 +m^2) \Lambda =0 \label{R7}\ee On the mass-shell,
$(\partial^2 + m^2 )\Lambda=0$, hence the supplementary condition
is now \be <{\bf phys'}| \partial^\mu A_\mu + m B |{\bf phys}> =0
\label{R8}\ee which is gauge invariant provided one completes
\eq{R3} with \be \delta B = m \Lambda \label{R9}\ee Substituting
\eq{R7} into \eq{R5}, one gets the covariant equation of motion
\be \partial^\mu F_{\mu\nu} +m^2 A_\nu -m \partial_\nu B =0
\label{R10}\ee ``which is the Stueckelberg equation for a massive
vector field."

Ramond proceeds ``to apply these tricks" to the string equation of
motion \be (L_0-1) \Phi =0 \label{R11}\ee with gauge conditions
\be L_n \Phi =0 \qquad (n\ge1) \label{R12}\ee and gauge
transformation \be \delta \Phi = \sum_{n\ge1} L_{-n} \Lambda^{(n)}
\label{R13}\ee The Virasoro operators $L_n$ satisfy the algebra
\be \left[ L_n, L_m \right] = (n-m) L_{n+m} +{D\over12} n(n^2-1)
\delta_{n,-m} \label{R14}\ee where $D=26$ is the number of
spacetime dimensions. The Stueckelberg fields $\Phi_n^{(p)}$ are
then introduced, with variations \be \delta \Phi_n^{(p)} = -L_n
\Lambda^{(p)} + (2n+p) \Lambda^{(n+p)} \label{R15}\ee and
equations of motion \be L_0 \Phi_n^{(p)} = - L_0 \left( L_n
\Lambda^{(p)} + (2n+p) \Lambda^{(n+p)} \right) \label{R16}\ee

We leave to the reader the pleasure of exploring the rest of the paper, which
concludes as follows: ``It should be clear that the subsidiary (Stueckelberg)
fields lead to much simpler looking expressions." For technical details, see
\cite{Pfe86} and its superpartner \cite{Kle89}.

The ten--dimensional ``superstring'' \cite{Gre84} is equivalent to the
``fermionic string'' \cite{Ram71,Nev71}. Its covariant quantization turns out
to be tricky, so people started by quantizing the
ten--dimensional superparticle
\cite{Cas76,Bri81}, which describes the dynamics of the
zero--modes of the
ten--dimensional superstring. As shown by \cite{Ber90a,Ber90b}, Stueckelberg
symmetries appear also in this context.

The lagrangian of the classical superparticle in first order formalism is
given by \be {\cal L}_{cl} = P_\mu \dot{X}^\mu - \Theta \Gamma^\mu P_\mu
\dot\Theta -\frac12 g P^2 \label{B1}\ee Here, $(X^\mu, \Theta)$ are the
classical coordinates of the superparticle, with $X^\mu$ a vector of the
ten--dimensional Lorentz group and $\Theta$ a
16--component
Majorana--Weyl spinor of
positive chirality, $P_\mu$ is the canonical momentum conjugate to $X^\mu$, $g$
is the einbein, and the dot denotes a time derivative. In order to quantize
\eq{B1}, \cite{Ber90a} introduce an infinity of ghosts, antighosts, and
Lagrange multipliers. They propose a new lagrangian which is BRST invariant.
Besides, it is also invariant under Stueckelberg transformations.

To illustrate these symmetries, we write down the transformation of the
coordinate $X^\mu$: \be \delta_{St} X^\mu = \sum_{p\ge0} \theta_{p+1,0}
\Gamma^\mu \epsilon^{p,0} \label{B2}\ee where $\theta_{p,0}$ are some of the
ghosts, and $\epsilon^{p,0}$ are local parameters with have the same
commutation properties and reality and chirality conditions as the antighosts
$\bar\theta^{p,0}$.

The Stueckelberg symmetries show that, effectively, the antighosts
$\bar\theta^{p,0}$ do not occur in the new lagrangian. Hence, they can be
eliminated by field redefinitions.

Commenting on a previous paper on the quantization of the superparticle,
\cite{Ber90a,Ber90b} showed that ``the mysterious gauge symmetry found by
\cite{Fis89} is a Stueckelberg symmetry \eq{B2}."

The final result of \cite{Ber90b} is a free quadratic lagrangian, BRST
invariant without any constraints. The Noether BRST charge $Q$ is nilpotent
($Q^2=0$)
off--shell.

\subsection{\label{subsec:nohiggs}Massive gauge theories without Higgs}

\subsubsection{Successes and problems of the standard theory}

The standard theory of electroweak interactions
\cite{Wei67,Sal68,Gla61a} has many virtues. It is a gauge theory
with BRST invariance \cite{Bec74,Bec75,Tyu75} and its gauge group,
$SU(2)\times U(1)$, is spontaneously broken through the  Higgs
mechanism \cite{Hig64,Eng64,Gur64,Kib67} to the $U(1)$ invariance
of quantum electrodynamics. The theory is unitary and
renormalizable \cite{Hoo71b,Bec76,Bec81}. The massive gauge vector
bosons $W^\pm$ and $Z$ corresponding to the broken symmetries have
been discovered \cite{Arn83a,Arn83b,Arn83c,Ban83,Bag83} and have
been abundantly produced at LEP and SLAC, and of course the $U(1)$
gauge boson is the massless photon. The standard theory is well
suited for perturbative computations, allowing detailed
calculations of cross--sections and decay rates, in remarkable
agreement with experiment; a good review is  \cite{Alt00}.

In spite of its extraordinary and complete success, the standard theory has
some weaknesses, though what they are is somewhat a matter of taste. The
remarkable agreement of all known data with the standard theory has prompted
theoreticians to look for alternatives of it which preserve such valuable
virtue and overcome its shortcomings.

To begin with, one should point out that the
spin--zero Higgs particle has not
yet been discovered, although indications exist of  $M_{\rm Higgs}=115 \,{\rm
GeV}$  \cite{Ale02,Del01,Opa01,L301}. Even if this result is falsified, this is
not worrisome, and the experimental results available to date can be
reinterpreted as bounding $M_{\rm Higgs}>114\,{\rm GeV}$. The
 LEP
measurements are of such precision that they verify the radiative corrections
of the standard theory, and thus bound the  Higgs mass (through its logarithm)
to around $  M_{\rm   Higgs}<215 \,{\rm GeV})  $  at 95\% C.L. \cite{PDBook},
so not finding the Higgs at the Tevatron would not be catastrophic, in sharp
contrast to what would happen if it were not found at the LHC. At any rate, the
theory does not predict the Higgs mass, which is quite an independent parameter
 (subject to more or less educated bounds, less stringent than the
experimental ones). Let us note, however, that if the Higgs were
heavier than around 300~GeV, then it would be strongly coupled, so
we could not calculate in perturbation theory
\cite{Cas88,Cas96,Cas97} and the above bounds would have to be
reinterpreted. It is important to stress that there is a logical
difference between the Higgs mechanism and the existence of the
Higgs particle: the latter provides an elegant and simple
implementation of the former. In terms of parameters, the non-zero
vacuum expectation value is independent of the (fundamental or
effective) scalar field's mass. Nevertheless, let us emphasize
right away that all efforts to implement the Higgs mechanism of
the standard model without a physical Higgs boson have failed so
far.

There are other theoretical misgivings about the standard theory.
Paramount is the hierarchy problem. What stabilizes the energy
scale of electroweak symmetry breaking, $m\sim O(10^2)$~GeV, with
that of gravity, $M\sim O(10^{18})$~GeV? Equivalently, the likely
unification of the electroweak and strong couplings, should  take
place at a comparably remote energy \cite{Geo74}. Why should it be
so different from the electroweak scale? Typically, the radiative
corrections to the $Z$ mass would be of the order of log$M/m$, so
in order to reach phenomenological agreement and keep $m<<M$, one
needs a careful and unnatural fine tuning of the mass parameters
in the theory, order by order in perturbation theory. (The more
dramatic quadratic divergence problem is an artifact of
regularizations breaking gauge invariance.) The hierarchy problem
is, thus, the wide difference between the Higgs vacuum expectation
value and the superstring (or quantum gravity, or unification)
scale. Supersymmetry stabilizes the hierarchy problem but does not
solve it. It could well be that the hierarchy problem is related
to the cosmological constant problem, the solution to which has
not yet been found, even in the framework of string theory.

Another problem, called the infrared catastrophe, arises in
quantum chromodynamics, the non-Abelian gauge theory of strong
interactions. Its gauge group, $SU(3)$, is unbroken, and its
massless vector bosons, the gluons, interact with colored
particles, namely quarks and themselves. When the energy of the
scattering process becomes small, the number of gluons emitted by
a colored particle diverges. The same problem occurs in quantum
electrodynamics, where the number of photons of low energy emitted
by a charged particle diverges. In QED, a remedy to this situation
has been found, since the photon is neutral. But in QCD, the
gluons carry color and therefore interact with themselves. This
makes the infrared problem untractable, and the divergences are
laboriously removed only at the very end of the computation of
suitable observables \cite{Wei65}, see however \cite{Sla81}. On
the other hand, a theory of massive vector bosons would not be
plagued by these infrared divergences; the Higgs mechanism does
not help since it would break asymptotic freedom
\cite{Hoo78,Nie95,Oji82}.

\subsubsection{Alternative models}

For the above reasons, among others, models have been proposed where the vector
boson
mass is put in by hand, in a variety of more or less astute ways. One could
call them ``genuine" massive
non-Abelian gauge theories. All such models
proposed so far are either not unitary or not renormalizable. This could well
be the end of the story. Nevertheless, the history of physics is full of
no--go
theorems which turned out to be wrong. Hence the continued theoretical interest
in these theories.  On the other hand,  massive vector
boson models which are renormalizable but not unitary could still provide
valuable hints towards a solution of the infrared problem.

On the more theoretical side, one might ask whether a
``non-renormalizable''
theory is really useless. For instance, it is conceivable that the divergences
in the Green's functions would cancel in the
on--shell
$S$--matrix elements
\cite{Nie95}: this happy situation does not seem to be realized in any of the
models constructed so far. A quite unconventional proposal is  to
 work in the Euclidean region of
space--like external momenta, to consider a massive scalar theory with an
exponential self--interaction which seems
non-renormalizable, and then to suggest a method for constructing an
$S$--matrix finite to all orders in perturbation theory \cite{Efi65}. See also
\cite{Fra63,Del69,Gho72, Gin75}. This method was developed in the
Yang--Mills
case by \cite{Sal71,Tay71,Leh71} and later by \cite{Fuk81,Fuk82,Fuk83}.

 In a different vein,      \cite{Geo93} discussed  composite Higgs fields,
 chiral symmetry, and  technicolor, whereas \cite{Nie96} discussed how massive
 Yang--Mills could arise from massless
 Yang--Mills coupled to a  topological
 field theory.

As emphasized, among others, by \cite{Vel68,Rei69,Vel70} and
\cite{Sla72b}, the problem of unitarity and renormalizability is
very delicate without having at our disposal a consistent and
parameter--free method of regularizing the perturbative expansion.
Also, as epitomized by the neutral vector boson theory, hidden
symmetries may be responsible for  the cancellation of divergences
in individual graphs. Let us just emphasize in passing that the
standard theory is indeed unitary and renormalizable in the
dimensionally regularized perturbative expansion
\cite{Bol72,Hoo72}, see also
\cite{Aky73a,Aky73b,Aky74a,Aky74b,Bre77}.

Our aim now is not to provide an exhaustive history of ``genuine" massive
vector models, but only to focus on the influence of Stueckelberg's seminal
papers \cite{Stu38I,Stu38II,Stu38III} in this general line of research. His
crucial contribution was the introduction of a \sl bona fide \rm auxiliary
scalar field with positive metric and positive energy, modifying Proca's
original model.

We shall limit ourselves to sketching the context in which
Stueckelberg's idea has been generalized or applied to various
domains. In the following, we rely mostly on the lucid reviews by
\cite{Del88} and by \cite{Nie95}. Let us recall that \cite{Del88}
proved that the original Stueckelberg theory for neutral massive
vector fields (with the addition of Faddeev--Popov ghosts and the
Nakanishi--Lautrup Lagrange multiplier) was invariant under
nilpotent BRST transformations. This ensures unitarity and
renormalizability. The upshot of both reviews for existing massive
Yang--Mills theories without a Higgs mechanism is the following:
those which use a suitable generalization of the Stueckelberg
mechanism are unitary order by order, but not perturbatively
renormalizable, due to the non-polynomial interaction, whereas
those which do not use the Stueckelberg mechanism are
renormalizable but not unitary, because of physical ghosts
\cite{Cur76,Boe96}.

\subsubsection{1962--1986: generalized Stueckelberg trick}

What is the generalization of the Stueckelberg trick to
non-Abelian massive
gauge theories? Remember that \cite{Sal62} and \cite{Ume61} kept the
 substitution $A_\mu\to A_\mu - m^{-1} \partial_\mu B$ of the Abelian case,
letting  $A_\mu=A_\mu^i T_i $, $B= B^i T_i$ with $T_i$ the generators of a Lie
algebra. The new lagrangian was then
gauge--invariant. The divergences due to the Proca vector propagator were then
explicit in the $\partial_\mu B^i$ terms. The latter was transformed away by a
unitary redefinition of the fields which yielded a
non-polynomial (actually, exponential) term in the Stueckelberg field $B$.

\cite{Kun67} proceeded in a slightly different way, starting from
\be {\cal L}= -
\frac14 \left( \partial_\mu A_\nu^i -\partial_\nu A_\mu^i + g f_{ijk} A_\mu^j
A_\nu^k \right)^2 +\frac{m^2}2 {\rm Tr}\, \left[ A_\mu^i T^i -\frac{i}g U^{-1}
\partial_\mu U \right]^2 \label{Kuku}\ee with \be U= {\rm exp} \left( i
\frac{g}m B^i T_i\right) \ee The lagrangian \eq{Kuku} is invariant under the
gauge transformation \ba &&\delta A_\mu^i = \left(D_\mu \Lambda\right)^i =
\partial_\mu \Lambda^i + g f^{ijk} A_\mu^j \Lambda^k \nn\\ && \delta B^i = m
\Lambda^i \ea The end result is very similar to that of \cite{Sal62} and
\cite{Ume61}.
A common weakness of these schemes is the absence
of ghosts  in the lagrangian \eq{Kuku}, as emphasized by  \cite{Sla72b}: in the
language of path integrals,  the  ghosts compensate the propagation of
unphysical states of the gauge  fields \cite{Fad67}. It turns out that the same
 $S$--matrix is obtained with or without the
 Stueckelberg fields. Indeed, using
 Pauli--Villars regularization and the
 preprint \cite{Hoo71b}, Slavnov showed that the
 $S$--matrix is independent of
 the longitudinal part of the Green's function.  Crucially,
  the arguments for the renormalizability of the massive neutral vector
 theory do not apply to the massive
 Yang--Mills case. The symmetry of the
 theory, however, ensures a partial cancellation of divergences. Indeed, as
 shown by \cite{Vel68, Sla71}, the
 one--loop diagrams of the massive
 Yang-- Mills theory do not generate any divergences other than the usual ones
 associated with mass, charge, and
 wave--function renormalization. It is interesting to note that
\cite{Vel68b} introduced a triplet of scalar fields to give mass
to the $W$ vector bosons which look like  non-Abelian Stueckelberg
fields.

What about higher orders? Since the massless Yang--Mills field
theory is renormalizable, one could expect that the massive theory
is also renormalizable, if the $m\to0$ limit exists. Alas, this
limit is sick \cite{Sla71,Bou70,Dam70}: the matrix elements of
massive Yang--Mills theories are discontinuous in the limit
$m\to0$. The reason for this singularity is easy to understand
from counting physical fields: a massive vector particle has three
physical degrees of freedom whereas a massless one has only two.
If the vector field happens to be neutral, as in QED, all the
matrix elements which are not diagonal in the number of
longitudinal photons vanish as $m\to0$, and thus the massless
limit is well--defined. In the non-Abelian case, however, where
the gauge fields interact with themselves, this is not so, and the
limit contains a charged massless scalar field in addition to the
transverse vector modes.

One would be tempted to conclude that the massive Yang--Mills
theory is not renormalizable in the usual sense, and indeed
\cite{Rei69} found new divergences at two loops. To settle this
issue satisfactorily, an invariant regularization is required.
Attempts have been made \cite{Del69} to apply the method of
\cite{Efi65,Fra63}, but they have been criticized by Slavnov:
ambiguity in the summation procedure, unreliable transition to the
pseudo--Euclidean region, and the fact that the solutions obtained
do not, in general, have the symmetry built into the original
lagrangian.

The extremely clear paper \cite{Sal70} precedes \cite{Sla72b} and covers
roughly the same ground as it. After recalling the Stueckelberg formalism for
neutral vector
fields, they  recast it in the language of path integrals. The advantage is
that field redefinitions can be tracked more carefully, including
non-trivial
Jacobians in the measure \cite{Fad67}. The Stueckelberg substitution \be V_\mu
\to A_\mu
=V_\mu +\frac1m \partial_\mu B \ee of the neutral Proca field $V_\mu$ by a
vector field $A_\mu$ and a scalar $B$ yields the generating functional \ba& Z
\left[ I^\mu,\eta ,\bar\eta\right] =&\ \int {\cal D} A^\mu{\cal D} B {\cal D}
\bar\psi{\cal D} \psi \; {\rm exp}  i \int \Biggl\{\frac12 A^\mu \left(
\partial^2 +m^2 \right) A_\mu -\frac12 B \left( \partial^2 +m^2 \right) B \nn
\\ &&+ {\cal L}_f + I^\mu \left( A_\mu -\frac1m \partial_\mu B \right)
+ \bar\psi\eta+\bar\eta\psi \Biggr\} \ea where ${\cal L}_f$ contains all the
terms involving fermions. One can now introduce a Lagrange multiplier $C$ to
get an equivalent expression, using the functional identity \be \int {\cal D}C
\;\delta \left( \partial_\mu V^\mu +\frac1m \partial^2 B \right) =1 \ee
Dropping, for the sake of notational convenience, the fermionic fields and
their  sources, the above generating functional is equivalent to \be Z \left[
I^\mu\right] = \int {\cal D} A^\mu{\cal D} B {\cal D} C \; {\rm exp}  i \int
\left\{{\cal L}[A^\mu -\frac1m \partial^\mu B ] +C \partial_\mu A^\mu  + I^\mu
\left( A_\mu -\frac1m \partial_\mu B \right) \right\} \ee The propagators are
now \ba &&\left< T A_\mu(x) A_\nu (y) \right> = -\left( g_{\mu\nu} +
\frac{\partial_\mu \partial_\nu}{\partial^2} \right) \Delta_m (x-y) \nn\\
&&\left< T A_\mu(x) B (y) \right> = 0 \\ &&\left< T B(x) B (y) \right> =
\Delta_0 (x-y) \nn\ea and thus \be \left< T (A_\mu(x)-\frac1m \partial_\mu B(x)
)
(A_\nu(y)-\frac1m \partial_\nu B(y) )\right> = -\left( g_{\mu\nu} +
\frac{\partial_\mu \partial_\nu}{\partial^2} \right) \Delta_m (x-y) \ee

This language was then generalized by \cite{Sal70} to charged
vector fields. The first example is provided by an isotriplet
$V_\mu^i$ of vector fields. Let $A_\mu$ and $\Omega$ denote the
transverse and longitudinal parts of $V_\mu$, defined by \be
V_\mu^i \tau^i =A_\mu^i \Omega \tau^i \Omega^{-1} + \frac{i}{g}
\Omega \partial_\mu \Omega^{-1} \label{slauni}\ee  with $\tau^i$
the Pauli matrices and \be \partial^\mu A_\mu^i=0\ee Equation
\eq{slauni} can be viewed as a non-Abelian gauge transformation,
where $\Omega $ is a $2\times2$ unitary matrix, conveniently
expanded  as \be \Omega(x) = \frac{g}m \left[ \sigma(x) -i \tau^i
B^i(x) \right] \ee in terms of the constrained field variables
subject to \be \sigma(x)^2 + \vec B(x)^2 =m^2 /g^2 \ee The change
of field variables is done according to \be {\cal D}V=\int {\cal
D}A {\cal D} \Omega J(A^\Omega) \exp \left\{-\frac{i}2 \int
(\partial A + m B )^2 \right\} \ee where the Jacobian is given by
\cite{Fad67} \be (J(A^\Omega))^{-1} = \int {\cal D} \Omega \exp
\left\{-\frac{i}2 \int (\partial A^\Omega + m B )^2 \right\}\ee
and $A^\Omega$ is defined by the right--hand side of \eq{slauni}.
The generating functional is then \ba Z[I] &&= \int {\cal D} V
\exp i\int \left( {\cal L}[V] +I V \right) \nn\\ &&= \int {\cal D}
A {\cal D} \Omega J(A^\Omega) \exp i \int \left( L[A^\Omega]
-\frac12 (\partial A +m B)^2 + I A^\Omega \right)\ea and the
chronological pairings read as follows:   \ba &&\left< T
V^i_\mu(x) V^j_\nu (y) \right> = -\left( g_{\mu\nu} +
\frac{\partial_\mu \partial_\nu}{\partial^2} \right) \Delta_m
(x-y) \delta^{ij} \nn\\ &&\left< T A^i_\mu(x) A^j_\nu (y) \right>
= - g_{\mu\nu}  \Delta_m (x-y) \delta^{ij} \nn\\ &&\left< T
A^i_\mu(x) B^j (y) \right> = 0 \\ &&\left< T B^i(x) B ^j(y)
\right> =  \Delta _0(x-y) \delta^{ij}\nn\ea From this,
\cite{Sal70} developed a perturbation theory.

Note that the above procedure can be applied to a subset of the charged vector
fields. For example, if $V_\mu^3$ is not present, then merely replace ${\cal
D}V \to {\cal D} V \delta (V^3)$.

\cite{Fuk81} describe a quantum theory of massive
Yang--Mills fields. They
start from the lagrangian of \cite{Kun67}, to which they add
Faddeev--Popov
ghosts and the
Nakanishi--Lautrup Lagrange multiplier. The latter is quantized
in a Hilbert space of indefinite metric, following \cite{Kug78,Kug79}. Their
scalar field $\xi$ is in fact the generalized Stueckelberg field $B$ defined by
\cite{Kun67} and used by \cite{Sla72b}. \cite{Fuk81} introduce, however, a
different subsidiary condition, again following \cite{Kug78}. The resulting
theory is claimed to be invariant under a nilpotent BRST transformation and
unitary; this statement is questioned by \cite[p. 442]{Del88}.  Its
lagrangian contains an exponential in the scalar field, just like \cite{Kun67}
and hence, it is not renormalizable in the conventional sense, although
\cite{Fuk81} claim, quoting \cite{Sal71}, that it is renormalizable in the
 sense of \cite{Efi65}. An inconclusive  extension to massive
two forms in this direction has been considered in \cite{Lah92}.

Based on \cite{Kun67} and \cite{Fuk81,Fuk82,Fuk83}, \cite{Son84}
introduce a Stueckelberg scalar for the $U(1)_Y$ vector boson in
the usual way, plus an isovector Stueckelberg field for the
$SU(2)_L$ sector. They find, interestingly, that the ratio of the
masses of these scalars had to be proportional to the weak mixing
angle, that is to $g'/g$. The theory is, however, not
renormalizable.


\cite{Bur86a} formulates the Abelian theory of massive vector
bosons in a gauge invariant way, without Higgs fields, but using
the theory of constrained systems \cite{Dir64}. The Proca and
Stueckelberg formulations appear as particular gauges, unitarity
being obvious in the former, and renormalizability in the latter
\cite{Mat49a,Mat49b}.  BRST invariance is also discussed, after
introducing Faddeev--Popov ghosts.

 \cite{Bur86b} extends this method to the
 non-Abelian case, constructing a
 gauge invariant lagrangian, following \cite{Kun67,Sla71,Fuk81}. There exists a
 gauge with only physical particles, which can be  called the unitary gauge.
However,
 it is not power--counting renormalizable. In a different gauge, there are
 Faddeev--Popov ghosts and a Stueckelberg scalar field. Due to the nilpotent
 BRST and
 anti--BRST invariance, the fundamental
 Ward--Takahashi--Slavnov--Taylor  identities are satisfied
\cite{War50,Tak57,Sla72a,Tay71b}, and thus
 unitarity is ensured, but not renormalizability. In yet another gauge, the
 theory is renormalizable but the BRST transformation is not nilpotent, and
 unitarity is not satisfied \cite{Cur76a,Cur76,Cur79}; see also \cite{Car88,Cab92}.
This emphasizes the importance  of the nilpotency of the BRST
invariance. \cite{Cur79} apply this model (renormalizable but not
unitary) to the infrared problem.

  \cite{Bur86b} then abandons the  lagrangian formalism and extends the
  Stueckelberg Abelian  field equations
  \ba & & \partial^\mu F_{\mu\nu} + m^2 A_\nu +
  \partial_\nu B =0 \\ && \partial^\mu A_\mu = \alpha B \ea to the
  non-Abelian
  case as follows: \ba & & D_{ij}^\mu F^j_{\mu\nu} + m^2 A^i_\nu + D^{ij}_\nu
  B_j =0 \\ && \partial^\mu A_\mu^i = \alpha B ^i \ea With canonical
commutation
  relations, the gauge field propagator is now \be \Delta_{\mu\nu}^{ij}= -
  i\delta^{ij} \left\{ \frac{g_{\mu\nu} -k_\mu k_\nu /m^2 }{k^2 -m^2
  +i\epsilon} + \frac{k_\mu k_\nu /m^2 }{k^2 -\alpha m^2 +i\epsilon} \right\}
  \ee Although it is possible to maintain gauge invariance, unitarity cannot be
  satisfied. The conclusion, therefore, is rather pessimistic. \cite{Bur86b},
  however, ends the paper with the following remark: ``We have also
  emphasized that the nonrenormalizable couplings always involve unphysical
fields. Since, in general, ghosts of the type used here do not
contribute at all to physical amplitudes, it is quite plausible
that the physical sector of massive Yang--Mills theory be
renormalizable although it is not by power counting and although
it is not unitary at each order of the perturbation expansion.
This is particular to massive Yang--Mills theory and the existence
of such a renormalizability is still an open question which merits
further attention especially if Higgs bosons remain experimentally
undetected."

\subsubsection{\label{subsec:delb}1988: unitarity versus renormalizability
reassessed}

We now present the important review \cite{Del88}. We have already mentioned
several times their treatment of the Stueckelberg theories of the Abelian
massive vector field. In their discussion of a variety of massive
gauge--invariant
non-Abelian models without Higgs mechanism, the themes are unitarity,
renormalizability and BRST invariance. They also provide a rich bibliography up
to 1988.

Their first example is the non-Abelian generalization of the
Stueckelberg formalism by \cite{Kun67}, supplemented by
\cite{Sla71, Sla72b}. Following the latter, they establish
unitarity for the gauge propagator to one loop, working in the
Landau gauge. They show explicitly that one half of the
contribution of the Faddeev--Popov ghosts to the imaginary part of
this propagator is compensated by the spin--zero part of the gauge
vector field, whereas the other half is compensated by the
Stueckelberg scalar field. This means that in the zero--mass limit
one does not recover the massless Yang--Mills theory, as already
emphasized by \cite{Sla71}. Repeating their analysis of unitarity
for fermion--antifermion scattering to order $g^4$, \cite{Del88}
find again that the Stueckelberg scalar contributes the essential
factor of $1/2$ with the correct sign. They then turn to the
high--energy behavior of longitudinally polarized vector bosons,
computing their elastic scattering. In a theory with Higgs bosons,
the amplitude is bound, in agreement with unitarity. In the
Stueckelberg case, even if $S^\dagger S=1$ is satisfied order by
order in $g^2$, it turns out that the amplitude in increasing
orders of $g^2$ scales with an increasing power of $E^2/m^2$ (they
show this explicitly up to $E^4/m^4$). They conclude that
renormalizability is not satisfied perturbatively in the
generalized Stueckelberg scheme. Nevertheless, \cite{Del88} point
out that \cite{Shi75,Shi75a,Shi77} established in a
non-conventional manner the renormalizability of two--dimensional
massive Yang--Mills, elaborated upon by   \cite{Bar78}.

A more complete discussion of unitarity bounds can be found in
\cite{Cor73,Cor74,Lle73}. They introduce the concept of ``tree
unitarity'', holding when the $N$--particle $S$--matrix elements
in the tree approximation diverge no more rapidly than $E^{4-N}$
in the high--energy limit, and discuss its relation to gauge
invariance and  renormalizability.  They remark  that ``a big
advantage of the Stueckelberg formalism is that all bad behavior
is now isolated in the vertices." Curiously, they quote
\cite{Stu57} instead of \cite{Stu38I,Stu38II}\footnote{A recurrent
problem in the historical record is that many citations to
Stueckelberg's work are incorrect, a sad reflection of the fact
that his papers have not been read.}. They conclude that tree
unitarity is only satisfied in models with spontaneously broken
symmetry.

The second model discussed by \cite{Del88} was proposed by \cite{Fra69} and
\cite{Cur76} as a possible candidate for a theory of massive
Yang--Mills
fields; see also \cite{Oji80} and the particularly clear \cite{Oji82}. The
lagrangian involving no Stueckelberg fields but including
Faddeev--Popov ghosts
is  \be {\cal L}= -\frac14 F^2 + \frac{m^2}2 A^2 -\frac1{2\alpha} (\partial
\cdot A)^2 +  \omega^* \partial \cdot D \omega + \alpha m^2  \omega^* \omega
+ \frac\alpha8 (\bar\omega \times \omega)^2 \label{uuyy}\ee where all fields
carry $SU(2)$ indices. In \cite{Fra69}, the Landau gauge    $\alpha=0$ is
chosen. This lagrangian is
invariant under the extended ``BRST transformation'' \ba && \delta A_\mu =
D_\mu
\omega \nn\\ && \delta \omega = \frac12 \omega \times \omega \\ && \delta
 \omega^* = -\frac1\alpha \partial\cdot A +  \omega^* \times \omega \nn \ea
which is not nilpotent: \be \delta^2 \not= 0 \ee The theory is
gauge invariant and has a good high--energy behavior, albeit only
in the Landau gauge. Indeed, thanks to the gauge--fixing term, the
vector propagator has a $k_\mu k_\nu/k^2$ term, instead of $k_\mu
k_\nu/m^2$. Hence, the model is power--counting renormalizable. It
is not unitary, for at least three reasons:

 1) The proof of unitarity for massive gauge theories in \cite{Kug79} rests on
 the nilpotency of the BRST charge $Q$, to whose cohomology physical states
 belong. Nilpotency could be enforced using the
 Nakanishi--Lautrup Lagrange
 multiplier $b$, such that $\delta  \omega^*= b+ \omega^* \times \omega$ and
 $\delta b= b\times \omega$. This modification would spoil the invariance,
 however.

2) The ghost and
gauge--fixing terms, that is the last four terms of
\eq{uuyy},  are not the $\delta$ of something, and the physical lagrangian is
not by itself gauge invariant. Hence, the ghosts cannot be eliminated from the
physical
$S$--matrix.

3) In the Landau gauge, $\cal L$ is just the effective action in
\cite{Kun67,Sla71}, without the Stueckelberg terms, which were shown to be
crucially necessary for unitarity at one loop.

A pretty variant of the Stueckelberg model was presented by two of
the authors in \cite{Del86}. Using the field equations, the scalar
Stueckelberg field $B$ was eliminated in favor of a gauge--fixing
functional of the vector field in such a way that the gauge
invariance of the mass terms was preserved; the inherent
non-polynomiality could then be disregarded in some particular
gauge. Whereas the renormalizability of the scheme was not cast in
doubt, it turns out that unitarity is violated \cite{Kub87,Kos87}.
For example, \cite{Kub87} shows that the one--loop correction to
the imaginary part of the gauge boson self--energy is the same as
in the spontaneously broken theory in the Landau gauge, except
that the would--be Goldstone boson is missing. But its
contribution is necessary to compensate the Faddeev--Popov ghosts.

\cite{Del88} reexamine this model in an equivalent formulation. They find that
it is invariant under a nilpotent BRST transformation. It is also
 power--counting renormalizable and gauge covariant (that is to say, the
$S$--matrix
does not depend on the gauge). Nevertheless, unitarity is violated. According
to the authors, ``we learn that gauge invariance and gauge covariance of a
theory are not strong enough conditions to ensure unitarity." This result is
paradoxical since it comes into conflict with the proof of unitarity by
\cite{Kug79}. \cite{Del88} resolve this contradiction by pointing out that,
contrary to the Stueckelberg model \cite{Kun67}, the model in \cite{Del86} does
not tend to the massive
Yang--Mills theory when the Stueckelberg field is taken
to zero. They conclude that ``the original Stueckelberg model is just right to
ensure (one loop) unitarity, and any tampering leads to
non-unitarity.''

Finally, \cite{Del88} describe the completely different model of \cite{Bat74},
which avoids several of the problems of the previous ones. The lagrangian is
\be {\cal L} = -\frac14 F^2 + \frac{m^2}2 A^2 + \frac\xi2 (\partial \cdot A)^2
+ {\cal G}(\xi, A) \ee with ${\cal G}(0,A)=0$. Under a gauge transformation one
has in particular \ba && \xi \to \xi + \delta \xi \nn \\ && {\cal L} \to {\cal
L}
+ \frac{\delta\xi}2 (\partial \cdot A)^2 + {\cal G}(\xi+\delta\xi, A)  - {\cal
G}(\xi, A)\ea Batalin gives a procedure to calculate $\cal G$, but this model
is
not perturbatively renormalizable and, as far as we know, its unitarity has not
been  established either.

 In conclusion, \cite{Del88} notice that renormalizability and unitarity seem
 to be competing qualities of massive non Abelian theories. ``The original
 Stueckelberg formulation, with its inherent non-polynomiality, is unitary but
 not renormalizable. This is in itself quite interesting, implying that the
 naive massive
 Yang--Mills action is of the correct form to ensure unitarity,
 and as we have seen any tampering with this leads us astray."

 \cite{Bur86b} has argued that in the Stueckelberg model one can find
 gauges in which ultraviolet divergences are confined to vertices that always
 involve unphysical fields, so that these divergences may cancel in the
 physical sector. ``Finally, it must be admitted that the Higgs mechanism
 remains the most complete method for giving mass to the vector bosons."

\subsubsection{\label{subsec:vann}1995--2003: new viewpoints}

Van Nieuwenhuizen and collaborators  have reexamined several aspects of the
review of \cite{Del88} in a series of papers.

\cite{Nie95} emphasizes the importance of Veltman's and Slavnov's
work, comprising explicit computations of one-loop \cite{Vel68}
and two-loop \cite{Rei69} divergences, the relation between
massive and massless Yang-Mills theories \cite{Dam70,Sla71}, and
the generalized Ward identity  \cite{Vel70}.  The \cite{Cur76}
model is then  rederived in detail by requiring a (not nilpotent)
BRST invariance. Renormalizability holds, but not unitarity,
because one cannot enforce nilpotent BRST invariance.

\cite{Boe96} confirm \cite{Nie95}: the BRST transformation leaving invariant
the \cite{Cur76} model cannot be made nilpotent by the addition of a scalar
field Lagrange multiplier. Due to nilpotency, the Ward identity of non-Abelian
gauge theories acquires an additional term which calls for a more careful
enquiry into renormalizability. This is carried out using the full beauty of
the BRST formalism, with or without Lagrange multipliers. They find again that
the model is indeed renormalizable, with five multiplicative renormalization
$Z$ factors. This checks with an explicit one-loop computation. The authors
then ``determine the physical states, extending the work of \cite{Oji82}. Many
of these states have, for arbitrary values of the parameters of the theory, a
negative norm, and from this we conclude that the model is not unitary".  This
statement contradicts the curious claim of \cite{Per95} that the theory is
unitary and renormalizable, with three $Z$ factors. \cite{Boe96} conclude that
the \cite{Cur76} model might be useful as a regularization scheme for infrared
divergences, in particular in superspace.

 \cite{Hur97} analyzes both the \cite{Cur76} and the non-Abelian Stueckelberg
 models of \cite{Kun67,Sla72b}, using the causal Epstein-Glaser approach to
 quantum field theory \cite{Eps73,Eps76}. This allows to consider only the
 asymptotic (linear) BRST symmetry. Furthermore, the technical details
 concerning the well-known ultraviolet and infrared problems in field theory
 are separated and reduced to mathematically well-defined problems.
 Let us sketch a few of the salient points
 of the \cite{Hur97} analysis.

In the well-defined Fock space of the free asymptotic fields,  the
$S$--matrix
is constructed directly as a formal power series: \be S(g) = 1 +
\sum_{n=1}^\infty \frac1{n!} \int d^4x_1 \cdots\int d^4x_n \quad
T_n(x_1,\ldots,x_n) g(x_1) \cdots g(x_n) \label{vn1} \ee where $g(x)$ is a
tempered test function which switches on the interaction. The theory is defined
by the fundamental (anti)commutation relations of the free field operators,
their dynamical equations, and the specific coupling of the theory $T_1$.  The
$n$--point distributions $T_n$ ($n>1$) in \eq{vn1} are then constructed
inductively, making sure that they are compatible with causality and Poincar\'e
invariance.

The formalism can be applied to Yang-Mills theories. Non-Abelian
gauge invariance is introduced by a linear operator condition,
separately  at every order of perturbation theory. This is done
with the help of the generator $Q$ of the  BRST transformation of
the free asymptotic field operators, an operator which is both
linear and Abelian, by requiring that \be \left[ Q,
T_n(x_1,\ldots,x_n) \right]   \label{vn2}\ee be the derivative of
a local operator.  Physical unitarity, the decoupling of the
unphysical degrees of freedom, is a direct consequence of \eq{vn2}
and the nilpotency $Q^2=0$.

Normalizability of the theory means, in the Epstein-Glaser
approach, that the number of finite constants to be fixed by
physical conditions stays the same at any order of perturbation
theory. This is generally called power-counting renormalizability,
and it relies only on the scaling properties of the theory. If a
theory can be normalized in a gauge invariant way, then it is
renormalizable.

A normalizable theory can be established by a suitable choice of defining
equations. For example, the massive non-Abelian gauge potentials in a general
linear
$\xi$--gauge, transforming according to the adjoint representation of
$SU(N)$, satisfy \be  \left( \partial^2 +m^2 \right) A_\mu^i  - (1-\alpha^{-1})
\partial_\mu \partial^\nu A_\nu^i =0 \ee \be \left[ A_\mu^i(x) , A_\nu ^j (y)
\right] = i\, \delta^{ij} \left( g_{\mu\nu} + \frac{\partial_\mu
\partial_\nu}{m^2} \right)  \Delta_m(x-y)  - i\, \delta^{ij}
\frac{\partial_\mu \partial_\nu}{m^2}   \Delta_M(x-y) \ee where $\Delta_m(x)$
is  the massive
Pauli--Jordan commutation distribution \eq{27}  and \be M^2=
\alpha\; m^2 \ee The
Faddeev--Popov ghost fields are required to fulfill \be
\left\{ \omega^i (x) , \omega^{*j}(y)  \right\} = - i \delta_{ij} \Delta_M(x-
y) \ee \be (\partial^2 +M^2) \omega^i(x) =  (\partial^2 +M^2)  \omega^{*i}(x)
=0\ee Already \cite{Cur76} and \cite{Oji82} noticed that if one takes over the
formula for $Q$ from the massless case in the general
$\alpha$--gauge \be
Q_{CF}= \int \frac{\partial_\nu A^\nu}\alpha \partial_0 \omega \; {\rm d}^3x
\ee one gets an operator which is not nilpotent in the massive case. But, as
noted above,  nilpotency is a necessary condition for unitarity. So a different
$Q$ is needed.

The generalization in \cite{Kun67}, \cite{Fuk81,Fuk82,Fuk83} of
the Stueckelberg formalism is to add  scalar fields $B^i(x)$
satisfying \ba  && \left[ B^i(x) , B^j (y) \right] = i \delta^{ij}
\Delta_M(x-y)  \\ && (\partial^2 + M^2 ) B^i(x) = 0 \ea The BRST
generator is then \be Q_S= \int \eta (x) \partial_0 \omega (x) \;
{\rm d}^3x \ee where the local quantities \be \eta^i(x) =
\alpha^{-1} \partial^\mu A^i_\mu (x) + m B^i(x) \ee are the
algebraically solved Nakanishi--Lautrup ghosts or, equivalently,
the supplementary condition. The corresponding BRST transformation
looks familiar: \ba && \left[ Q_S, A^i_\mu \right]= i \partial_\mu
\omega^i \\ && \left[ Q_S,B^i
\right]= i \,m \omega^i \\ && \left[ Q_S,  \omega^i \right] =0\\
&&  \left[ Q_S,
\partial^\mu A^i_\mu \right] = -i \, M^2 \omega^i \\ &&  \left[ Q_S,
 \omega^{*i} \right] =-i \,\eta^i \\ &&  \left[ Q_S,  \eta^i \right] =0\ea

\cite{Hur97} proceeds to construct the most general
gauge--invariant coupling $T_1$ such that  $[ Q_S,  T_1] $ is a
total derivative as in \eq{vn2}, Lorentz-- and $SU(N)$--invariant,
with ghost number zero, and with maximal mass dimension equal to
four. There are a certain number of trilinear couplings in the
fields $A_\mu^i$, $B^i$, $\omega^i$, $\omega^{*i}$ and their
derivatives. This has thus defined a manifestly normalizable
theory which is gauge invariant to first order in perturbation
theory, and respects further symmetry conditions. Can one prove
the condition of gauge invariance, namely that $[Q_S,T_n]$  be a
total derivative, inductively to all orders in perturbation
theory? By explicit calculation, \cite{Hur97} shows that this
condition fails already at second order. The constraint of
normalizability is essential for this conclusion. Hence, the
Stueckelberg generalization for non-Abelian massive gauge theories
is not perturbatively renormalizable. More precisely: no
(non-linear) deformation of the (linear) asymptotic BRST
invariance (implemented on the asymptotic Fock space) can lead to
a renormalizable and unitary theory.


Non-Abelian Stueckelberg lagrangians are plagued by nonpolynomial
terms. Perhaps a correct understanding of these terms could lead
to a unitary and renormalizable non-Abelian Stueckelberg model.
The work of \cite{Dra97} rules out this possibility by showing
that the non-polynomial structure can be reduced algebraically to
a polynomial version of the Stueckelberg model. But then the
results of \cite{Hur97} are applicable, and thus this model is not
unitary and renormalizable.

\cite{Dra97} start with the \cite{Kun67} generalization of the Stueckelberg
model to a massive
Yang--Mills theory: the vector fields $A_\mu^i$ and scalar
fields $B^i$ belong to  the adjoint representation of a Lie  group $G$. The
kinetic and mass terms of the lagrangian are separately
gauge--invariant. To
compensate the unphysical degrees of freedom of $A_\mu$ and $B$, they introduce
Faddeev--Popov ghosts $\omega^i$ and $ \omega^{*i}$, also in the adjoint.  The
lagrangian, including a
gauge--fixing term, is invariant under a BRST operator
which is nilpotent if one adds   further
Nakanishi--Lautrup Lagrange multipliers $b^i$
to the set of fields. The
gauge--fixing term is such that the propagators of
$A_\mu$ and $B$ fall off like $k^{-2}$ for large momenta $k^\mu$.
Unfortunately, the exponential of $B$ appears in the lagrangian. They then
redefine the vector field $A_\mu$ into a new $\hat A_\mu$ such that the
$S$--matrix is unchanged \cite{Col68}. The BRST transformation $\bf s$ is then
given by
\ba && {\bf s} \hat A_\mu = 0\\ &&{\bf s}B= m  \omega\\ &&{\bf
s} \omega=0\\ &&{\bf s}\omega^*=b\\  &&{\bf s}b=0\ea
and the
BRST--invariant lagrangian is of the form \be {\cal L} ={\cal L}_{\rm
phys} (\hat A_\mu, \partial^\nu \hat A_\nu) + s( \omega X)\ee where one can
choose \be X= \frac1{2} \left[ b- \partial^\mu \hat A _\mu - (\partial^2 +m^2)
\frac{B}{m} \right] \ee so that the
Faddeev--Popov ghosts $\omega $ and
$\omega^*$ are
free and the
Nakanishi--Lautrup auxiliary field $b$ can be solved from $X=0$.
Redefining again $\bar A_\mu = \hat A_\mu + m^{-1} \partial_\mu B$, the
physical lagrangian containing  the gauge  and the Stueckelberg fields becomes
\be {\cal L}_{\rm phys} = -\frac{1}{4}\left( {\rm tr}\,\left[ F_{\mu\nu} (\bar A -m^{-1}
\partial B) \right] \right)^2 \label{dddD1}\ee
with the usual notation $ F_{\mu\nu}^i (Y) = \partial_\mu Y_\nu^i -
\partial_\nu Y_\mu^i +gf^i_{\; jk} Y_\mu^j Y_\nu^k $.

The result of all these redefinitions is to replace the original exponential in
$B$ by a polynomial in $\partial_\mu B^i$, of mass dimension eight. The
propagators of $\bar A_\mu$ and $B$ are
well--behaved at high energies, but the
derivative interaction of $B$ is still not
power--counting renormalizable.  Of
course, \eq{dddD1} reminds us very much of expressions in \cite{Ume61,Sal62}.

\subsection{\label{sec:pubel}Related applications}

We end this chapter mentioning a variety of applications of the Stueckelberg
formalism which fall outside our main line of exposition, but might be of
interest to the reader.

Early phenomenological applications of charged vector mesons in
the Proca and Stueckelberg theories are \cite{You63},
\cite{Bai64a,Bai64b,Bai65} and also \cite{Abe69,San68}.

\cite{Wat67} applied the Stueckelberg formalism to the
Rarita--Schwinger (spin~$3/2$) field, whereas
\cite{Del75b,Ark02,Lut03} did it for the graviton (spin~2) field.
These last two references illustrate nicely the rich possibilities
of $D$-brane backgrounds.

 The Stueckelberg theory was supersymmetrized ($N=1$) very early
  \cite{Del75a}. It was found that the condition for the
super-Stueckelberg mechanism to work was that $\bar D D J=0$, with
$J$ the external supercurrent coupled to the superphoton. In the
following, we use Wess-Bagger notation (i.e. two-component
fermions) but with our usual metric $+---$.
 The standard kinetic term (in terms of $W^\alpha = \bar D^2 D^\alpha V$, with
$V=V^\dagger$ a vector superfield)
\be {\cal L}_0 =\frac14 \int {\rm d}^2\theta W^2
+\frac14\int {\rm d}^2\bar \theta \bar W^2 \ee
which in the Wess-Zumino gauge reduces to
\be {\cal L}_0 =-\frac14 F_{\mu\nu}^2 -i\lambda \sigma^\mu \partial_\mu
\bar\lambda +\frac12 D^2\ee
is supplemented with
\be {\cal L}_m = -m^2 \int {\rm d}^2\bar\theta {\rm d}^2\theta [V
+\frac{i}{m}(\Phi -\Phi^\dagger)]^2 \ee
mimicking the (bosonic) Stueckelberg starting point. Note that we must
introduce
a chiral and an antichiral superfield. The mass term above  gives a mass to the
vector, yields the kinetic terms for the complex Stueckelberg scalar fields $a$
and $a^*$ and their spin 1/2  superpartners $\psi$ and $\bar\psi$, induces a
mixed mass term between the photino $\lambda$  and the Stueckelberg fermion
$\psi$, and provides the cross term $mA^\mu \partial_\mu (a+a^*)$. When the
auxiliary field $D$ is eliminated, a mass term for $(a-a^*)$ comes out as well.
We still must add a gauge-fixing term. In the massless case, it is of  the form
$\int {\rm d}^4\theta  (\bar D^2 V)(D^2 V) $, whereas now it is better to take
\be {\cal L}_{gf} = \xi \int {\rm d}^2\bar\theta {\rm d}^2  \theta (\bar D^2 V
+\frac1{m^2} \Phi)(D^2 + \frac1{m^2}\Phi^\dagger) \ee

The Stueckelberg mechanism has been
used extensively in the pedagogical presentations of \cite{Gat83,Sie99} where
the  Stueckelberg
or ''compensator'' fields are introduced as a simple example of Goldstone
fields, and numerous
applications to supersymmetry are discussed.
See also \cite{Gue91}.

The non-renormalizability of the non-Abelian Stueckelberg models
are  crucial to establish the non-renormalizability of the $N=2$
massive Yang--Mills theory \cite{Khe91}, and useful for its
analysis in harmonic superspace \cite{Vol94}.

Stueckelberg's trick has been used to reformulate chiral
\cite{Ban94b,Kul94a,Kul94b,Kul98,Kul01a,Kul01c,Kul02a}, and other
two-dimensional models \cite{Kul01b,Kul01d,Kul02b,Kul02c}, as well
as  the massive three-dimensional gauge theory \cite{Sch81,Dil95}.
\cite{Ban96}  established a three-dimensional duality similar to
the well--known duality between sine--Gordon and Thirring
\cite{Col75} exploiting the Stueckelberg formulation. Attempts to
extend the three-dimensional  topological Yang--Mills mass
\cite{Des82} to four dimensions used also the Stueckelberg trick
\cite{All91,Hwa97,Lah97}, as did a powerful no-go theorem that
pretty much  ended such attempts \cite{Hen97}, except perhaps for
the provocative \cite{Lah99} which claims, using the duality with
massive two-forms,  that the non-Abelian massive vector boson
theory is renormalizable although its  unitarity is not discussed.

 The two--dimensional Stueckelberg theory has been studied in   a
Robertson--Walker background, in a black hole metric, and in a
Rindler wedge \cite{Chi92,Chi93,Chi94}. \cite{Jan87} formulated
the  Stueckelberg theory in anti--de Sitter space.

 \cite{Deg94,Deg95a,Deg95b} derived  Stueckelberg's construction
starting from loop space, and extended it to a  classical
non-Abelian
 version for
 rank--two tensor fields. \cite{Deg95c} found, in particular,
that only null strings interact with massive vector fields, and no
strings interact with massive third--rank tensor fields at the
classical level.  A similar application of the Stueckelberg trick
to string theory is \cite{Ale03}, whereas a different one to
membrane theory is \cite{Pav98}.

The idea of inventing new fields in order to
uncover or make manifest hidden symmetries has been applied  in many contexts.
  The extension of the Stueckelberg formalism for a massive antisymmetric field
\cite{Kal74} has been the subject of intensive research
\cite{Saw95,Bar96,Ban96b,Deg99a,Sma01,Kuz02,Deg02}, approaching a
clear understanding \cite{Dia01}. The generalization to
 $p$--forms ($p=2$ for the
Kalb--Ramond field) has met with success \cite{Biz96,Biz98,Biz99}. \cite{Deg97}
 show that a $U(1)$ gauge theory defined in the
 configuration space for closed  $p$--branes yields the
 gauge theory of a massless  $(p + 1)$ antisymmetric tensor field and a Stueckelberg
  massive vector field.  The $p$--form extension of the
Stueckelberg formalism has been used to establish dualities
between  field theories \cite{Fre81,Saw96,Saw97,Sma00} and  to
study symmetry breaking in $D$--brane theories \cite{Ans00}.

Dualities between massive three--dimensional
 non-commutative field theories appear elegantly if
 care is taken to apply the Stueckelberg mechanism first
 to make explicit the gauge symmetries \cite{Gho00,Gho02}.

 The Stueckelberg theory has been analyzed in the
 Batalin--Fradkin--Vilkovisky formalism
\cite{Day88,Day93,Ban94a,Ban94b,Saw95,Saw97},  in axiomatic field
theory \cite{Mor86,Mor87} and in a new quantization procedure in
algebraic field theory \cite{Wie96}. The relationship between spin
zero and spin one has been emphasized in \cite{Kru01a,Kru01b}.

\cite{Ban97} consider the (classical) hamiltonian formulation of a
Higgs--free massive Yang--Mills theory with the Stueckelberg
trick, and \cite{Bar97} apply it to the standard model.
\cite{Deg99b} interpolates between various classical lagrangians,
with or without Higgs and Stueckelberg fields, and \cite{Deg03}
applies it to the Abelian projection of the simplest $SU(2)$.

A particularly nice application of the Stueckelberg mechanism has
appeared in the construction of string-derived particle field
theories \cite{Ald00}. In general, after compactifying a
ten-dimensional string, or choosing a suitable brane background,
there are too many $U(1)$s left. Non-Abelian gauge groups can be
broken by a variety of mechanisms, but the rank of the gauge group
does not change in general.  Polchinski's observation
 that all continuous symmetries in string theories
must be gauge symmetries [for a discussion, see \cite{Pol98}]
precludes the possibility of arranging moduli in such a way that
these spurious $U(1)$s become mere global symmetries. The
Stueckelberg mechanism is then applied to them, with a large
Stueckelberg mass, whereby the additional Abelian gauge vectors
become very massive (without breaking any gauge symmetry) and
disappear from the low-energy spectrum, not leaving even a global
symmetry behind.

\section{\label{sec:conc}Conclusions and outlook}

Stueckelberg's original idea was to introduce a physical scalar
field $B(x)$ into the Abelian massive vector field lagrangian to
make the theory as similar as possible to QED. It was shown by
several authors that this proposal facilitated the discussion of
the renormalizability of massive vector field theories and made
manifest some hidden symmetries. The neutral massive Abelian
vector field theory is gauge invariant and BRST invariant, because
the transformation of the  Stueckelberg $B$ field compensates the
transformation of the mass term.  This explains the
renormalizability of this theory. The field $B$ plays a role
 similar to that of the Goldstone boson in spontaneously broken theories.
Charged  or non-Abelian theories without the Higgs mechanism
 have  been shown to be not renormalizable, because the
 derivative couplings of the Stueckelberg field can only be eliminated at the
  expense  either of exponential couplings or of unitarity.
 Work on these issues has not been completely abandoned, however.

 Stueckelberg's mechanism, simple as it seems today and always elegant,
inspired numerous imitations, ranking from non-Abelian massive
vector theories without Higgs fields to supersymmetric,
topological and  string    theories.  In many cases, new
symmetries were discovered.

Additionally, in this paper we have constructed a standard
electroweak theory with a massive photon, preserving the
$SU(2)_L\times U(1)_Y$ BRST symmetry. The neutral scalar
Stueckelberg field $B$ appears together with a massive hypercharge
vector field, and the photon inherits a mass after the spontaneous
symmetry breaking.  This can be interpreted in two ways. The
photon mass can be considered as an infrared cut-off, a mere
calculational trick, allowing one to deal cleanly and separately
with the  infrared divergences.    This would require,  of course,
the zero mass limit to be smooth.  A less conservative point of
view
 calls for taking the photon mass seriously, albeit limited by empirical data
to  a very small value. In this case, new phenomena appear,
proportional to the photon's mass   (squared): neutrino photon
couplings, parity violation of the electron photon   couplings,
slightly different $Z$ mass, etc.  None of them seem comparable,
 in precision,  to the direct limits on the photon mass,
 but more research in this direction seems  warranted, in
 particular for neutrino cosmology. Finally, the physical definition
  of the electric current appears now in a new light.

\begin{acknowledgments} We are very grateful to Raymond Stora for lending us
the transparencies of his talk at the University of Leipzig in February 2000, where he
suggested to use
the Stueckelberg mechanism in the standard model, and for many illuminating
discussions, in particular on the distinction between asymptotic fields and
local fields perturbatively coupled to conserved currents. Any
errors, conceptual or otherwise, are ours and not his.  We thank
Tobias Hurth for making his bibliography available to us and for a careful
reading of the manuscript. We thank Tini Veltman for extensive constructive criticism of
the first draft of this paper.
\end{acknowledgments}

  \appendix
           \section{} 

We have collected here some long formulae on the Stueckelberg
modification of the standard model, section \ref{sec:sm}.

             \subsection{\label{a:nota}Notation}
             We use the metric $(+,-,-,-)$. Let us point out in passing that
\cite{Stu38I,Stu38II,Stu38III} used a real metric but with the opposite sign
(standard in modern, string--oriented, notation),
while Pauli and others used Euclidean metric with an imaginary fourth
component.

 We use throughout the notations
\be A_\pm=\frac1{\sqrt 2} \left(A_1 \pm i A_2\right) \ee
\be \vec A \cdot \vec B = \sum_{i=1}^3 A_i B_i = A_+B_-+A_-B_+ +A_3B_3 \ee
and
\be (\vec A \times \vec B)_i = \varepsilon_{ijk} A_j B_k \ee
with $\varepsilon_{123}=1$ and cyclic; in the $\{+,-,3\}$ basis the non-zero
elements of the $\varepsilon_{ijk}$ are
\be \varepsilon_{3+-}=-\varepsilon_{3-+}
=\varepsilon_{+3+}=-\varepsilon_{++3}=\varepsilon_{--3}=-\varepsilon_{-3-}=i
\ee
Sometimes we use the short-hand $\vec A^2=\vec A \cdot \vec A$.

 \subsection{\label{a:scal}Scalar lagrangian}
Explicitly,
\ba & \left| D_\mu \Phi\right|^2 =& \frac12 (\partial_\mu H)^2  +
\frac12 (\partial_\mu \phi_3)^2 +\partial _\mu \phi_+ \partial^\mu \phi_- \nn\\
&&+\frac12 \partial_\mu H (g W_3^\mu -g'V^\mu) \phi_3
-\frac12 \partial_\mu \phi_3 (g W_3^\mu -g'V^\mu) (H+f) \nn\\
&& +\frac{g}2 (\partial_\mu H -i\partial_\mu \phi_3) W_-^\mu \phi_+
 + \frac{g}2 (\partial_\mu H +i\partial_\mu \phi_3) W_+^\mu \phi_- \nn\\
 && -\frac{g}2 \partial_\mu  \phi_+ W_-^\mu (H+f-i\phi_3)
 -\frac{g}2 \partial_\mu  \phi_- W_+^\mu (H+f+i\phi_3) \nn\\
 &&+\frac14 \left[ (H+f)^2 +\phi_3^2 \right]
  \left[ g^2 W_+^\mu W_{-\mu} +\frac12 (gW_3^\mu -g'V^\mu)^2 \right] \nn \\
 && -i\frac{gg'}2 (H+f-i\phi_3) \phi_+ V_\mu W_3^\mu
 +i\frac{gg'}2 (H+f+i\phi_3) \phi_- V_\mu W_3^\mu \nn\\
 && +\frac12 \phi_+\phi_-
  \left[ g^2 W_+^\mu W_{-\mu} +\frac12 (gW_3^\mu +g'V^\mu)^2 \right]
 \ea

 \subsection{\label{a:quad}Quadratic lagrangian}
Using the   choices of
gauge--fixing functions \eq{kmkm}, and
dropping three total derivatives, the quadratic part of the lagrangian is
\ba & {\cal L}_{2} =& -\frac12 ( \partial_\mu \vec W_\nu )^2
+ \frac{f^2g^2}8  \vec  W_\mu^2
+\frac12\left(1-\frac1{\alpha}\right)  (\partial_\mu \vec W^\mu)^2
 - \frac{gg'f^2}4 W_3^\mu V_\mu \nn\\
&&-\frac12 (\partial_\mu V_\nu )^2 +
\frac12 \left( m^2 + \frac{f^2g^{\prime 2}}4 \right) V_\mu^2
+\frac12\left( 1-\frac1{\alpha'} \right) (\partial_\mu V^\mu)^2 \nn\\
&&  +\frac12 (\partial_\mu \vec \phi  )^2
- \frac{\alpha f^2g^2}8  \vec \phi^2
  - \frac{\alpha' f^2g^{\prime 2}}8  \phi_3^2  \nn\\
  &&+\frac12 (\partial_\mu B)^2
  - \frac{\alpha' m^2}2 B^2
  + \frac {\alpha' g' m f }2  \phi_3 B\nn\\
  && +\frac12 (\partial_\mu H)^2 -\lambda f^2 H^2 \nn\\
  && -\omega^* \left (\partial^2 + \alpha' m^2
  + \frac{\alpha' f^2g^{\prime 2}}4 \right) \omega
  + \frac{ f^2g g^{\prime }}4 \left( \alpha' \omega^* \omega_3
        + \alpha  \omega_3^* \omega \right) \nn\\
&& -\vec\omega^* \left (\partial^2  + \frac{\alpha f^2g^{ 2}}4 \right) \vec
\omega
\label{lacuadra}\ea

 It might be necessary to expand  $\epsilon$ in \eq{combini} to second order for some
computations, but here we restrict ourselves to first order.
The various rotations of $g$ and $g'$ to order $\epsilon$ are
\be  g {\rm c}_w
+g'{\rm s}_w \simeq \sqrt{g^2 +g^{\prime2}} +{\cal    O}(\epsilon^2)
\label{giol1}\ee
\be
g {\rm c}_w - g'{\rm s}_w \simeq \frac{g^2 -g^{\prime2}}{\sqrt{g^2
+g^{\prime2}}} \left(
1-2 \frac{g^2g^{\prime2}}{ g^4-g^{\prime4}} \epsilon   \right)
+{\cal    O}(\epsilon^2)  \ee
\be
 g{\rm s}_w - g'{\rm c}_w \simeq
 \epsilon \frac{gg'}{\sqrt{g^2 +g^{\prime2}}}   +{\cal    O}(\epsilon^2)  \ee
\be
g {\rm s}_w +g' {\rm c}_w \simeq
\frac{2gg'}{\sqrt{g^2 +g^{\prime2}}} \left(
1+\frac12 \frac{g^2-g^{\prime2}}{ g^2+g^{\prime2}} \epsilon   \right)
+{\cal    O}(\epsilon^2)  \label{giol4}\ee

 \subsection{\label{a:mass}Mass formulas}

The exact mass eigenvalues of the neutral vectors, scalars and
ghosts are \be M^2_{Z\atop A} = \frac{f^2}8 \left(g^2
+g^{\prime2}\right) \left(1 +\epsilon \pm \sqrt{1  -2\epsilon
\frac{g^2 -g^{\prime2} }{g^2 +g^{\prime2}} +\epsilon^2}
\right)\label{1mma} \ee

 \ba & M^2_{\chi_Z\atop
\chi_A} =M^2_{G\atop S} = &\frac{\alpha' f^2}8 \Bigl(
\frac\alpha{\alpha'} g^2 +g^{\prime2} +\epsilon  \left(  g^2
+g^{\prime2}\right) \nn \\ && \pm \sqrt{ \left (
\frac\alpha{\alpha'}  g^2 +g^{\prime2}\right)^2 -
2\epsilon\left(\frac\alpha{\alpha'} g^2 -g^{\prime2}\right) \left(
g^2 +g^{\prime2}\right) + \epsilon^2 \left( g^2
+g^{\prime2}\right)^2} \Bigr) \ea This formula is the same as
\eq{1mma} with the substitutions $g^2\to\alpha\,g^2$, $
g^{\prime2} \to\alpha'\,g^{\prime2} $, and $\epsilon\to
\alpha'\,\epsilon$. In general, to lowest order in $\epsilon$, the
masses of the neutral longitudinal scalars and the neutral ghosts
are \be M^2_{G} =  M^2_{\chi_Z} \simeq \frac{ f^2 }{4} \left(
\alpha g^2 +\alpha' g^{\prime2} \right) \left(1+ +\epsilon \frac {
g^{\prime2}\left( g^2 +g^{\prime2}\right) } {\left (
\frac\alpha{\alpha'}  g^2 +g^{\prime2}\right)^2 } \right) \ee \be
M^2_{ S} = M^2_{\chi_A} \simeq \epsilon \frac{\alpha f^2 g^2}{4}
\frac{ g^2 +g^{\prime2} } {  \frac\alpha{\alpha'}  g^2
+g^{\prime2} } \ee These expressions simplify to those in the main text
by setting $\alpha'=\alpha$.

\subsection{\label{a:inte}Interaction lagrangian}

We turn now to the interaction lagrangian ${\cal L}_{\rm int}$, which
added to the quadratic lagrangian \eq{lacuadra} makes up the full physical
gauge--fixed (but still matter--free) lagrangian useful for quantum
computations. We
quote it in terms of the original variables: these fields do not have a
well--defined propagator!

\ba & {\cal L}_{\rm int} =& -i g \left[ \partial_\mu W_\nu^+
 \left( W^{\mu -} W^{\nu 3}-W^{\mu 3}W^{\nu -} \right)
+{\rm cyclic} \right]
\nn\\  &&
+\frac12 \left( g W_\mu^3 -g' V_\mu \right)
  \left( \phi_3 \partial^\mu H -H \partial^\mu \phi_3\right)
\nn\\  &&
+\frac{g}2 W^-_\mu \left[ \left( \phi_+ \partial^\mu H -H \partial^\mu \phi_+
\right)
  +i \left( \phi_3 \partial^\mu \phi_+ - \phi_+ \partial^\mu \phi_3 \right)
\right]
\nn\\  &&
+\frac{g}2 W^-+\mu
\left[
\left( \phi_- \partial^\mu H -H \partial^\mu \phi_- \right)
  -i \left( \phi_3 \partial^\mu \phi_- -\phi_- \partial^\mu \phi_3\right)
\right]
\nn\\  &&
+ f H \left[
\frac{g^2}2 \left( W^+ W^- \right) +\frac14 \left(gW^3-g'V\right)^2
-\lambda \left( 2\phi_+\phi_- +H^2 +\phi_3^2 \right) \right]
\nn\\  &&
+i \frac{gg'f}2 \left(\phi_- -\phi_+\right) \left( V W^3 \right)
\nn\\ &&
+g^2 \left[
\left(W^3 W^+\right) \left( W^3 W^-\right)
-\left(W^3\right)^2 \left( W^+W^-\right)
-\frac12 \left(W^+W^-\right)^2
+\frac12 \left(W^+\right)^2 \left(W^- \right)^2
\right]
\nn\\ &&
+\frac{g^2} 2 \left( \phi_+\phi_- +\frac12 H^2 +\frac12 \phi_3^2 \right)
\left( W^+W^-\right)
\nn\\ &&
-\lambda \left( \phi_+\phi_- +\frac12 H^2 +\frac12 \phi_3^2 \right) ^2
\nn\\ &&
+\frac18 \left( H^2 +\phi_3^2 \right) \left(gW^3-g'V\right)^2
+\frac14  \phi_+\phi_-  \left(gW^3+g'V\right)^2
\nn\\ &&
-\frac{g'g}2 \left[ \left( \phi_3 +i H \right) \phi_+ +
\left( \phi_3 -i H \right) \phi_- \right] \left( V W^3 \right)
\ea

Of course, to do actual computations of $S$--matrix elements, this has to be
rewritten in terms of the
mass eigenstates. The general expressions are quite unwieldy, even in the
simplified case $\alpha'=\alpha$, whereby $\tilde\theta_w
=\theta_w$.

  \subsection{\label{a:ghos}Ghost lagrangian}

The interaction part of the ghost lagrangian is
\ba
  && -g \vec \omega ^*
\cdot
\partial_\mu (\vec W _\mu \times \vec \omega) -\frac{\alpha' f g^{\prime2}}4 H
\omega^* \omega  -\frac{\alpha f g^{2}}4  \vec \omega^*\cdot ( H \vec \omega +
\vec \phi \times \vec\omega ) \nn\\ && +
 \frac {f g g'}4 \left\{ \alpha'
\omega^*
[i(\phi_+ \omega_- - \phi_- \omega_+) + H \omega_3]  + \alpha [i(\omega_-^*
\phi_+
- \omega_+^* \phi_-) + \omega_3^* H] \omega \right\} \ea
whereas the full ghost lagrangian in the mass eigenstate basis, with the
tree--level simplification $\alpha'=\alpha$ is
\ba &{\cal L}_{gh}=& \omega_+^* (\partial^2 +\alpha
M_W^2
) \omega_- + \omega_-^* (\partial^2 +\alpha M_W^2 ) \omega_+ \nn\\ &&+\chi_Z^*
(\partial^2 +\alpha M_Z^2 ) \chi_Z +\chi_A^* (\partial^2 +\alpha M_A^2 ) \chi_A
\nn\\
&& + i \,\omega_+^* \partial_\mu \left[W_-^\mu ({\rm c}_w \; \chi_Z +
{\rm s}_w \; \chi_A) -({\rm c}_w \; Z^\mu + {\rm s}_w \; A^\mu)
\omega_-\right] \nn\\
&& -i\, \omega_-^* \partial_\mu \left[W_+^\mu ({\rm c}_w
\; \chi_Z + {\rm s}_w \; \chi_A) -({\rm c}_w \; Z^\mu + {\rm s}_w \;
A^\mu) \omega_+ \right] \nn\\ &&
+ i \left({\rm c}_w \; \chi_Z^* + {\rm s}_w \; \chi^*_A \right)\; \partial_\mu
(W_+^\mu \omega_- -W_-^\mu \omega_+) \nn\\ &&
+\frac{\alpha\, f\, g^2}4 \; \Bigl[ i (\cos\beta \; G + \sin
\beta \; S) \left(\omega_-^* \omega_+ -\omega_+^* \omega_- \right) +H
\left(\omega_+^* \omega_- +\omega_-^* \omega_+ \right) \nn\\
&&\qquad +
({\rm c}_w \; \chi_Z^* + {\rm s}_w \; \chi_A^* ) i \left(\omega_- \phi_+
- \omega_+ \phi_- \right)\Bigr] \nn\\
&& -i\frac{\alpha\, f\, g}{8} \;
\left(\phi_+ \omega_-^* -\phi_- \omega_+^* \right ) \left[ (g\,{\rm c}_w -
g'\, {\rm s}_w )\; \chi_Z +(g\,{\rm s}_w +g'\, {\rm c}_w )\; \;
\chi_A\right] \nn\\
&& +\frac{\alpha\, f\, g}4 \; H \; \Biggl[ (g\,{\rm c}_w
+g'\, {\rm s}_w ) {\rm c}_w\; \chi_Z^*\chi_Z +(g\,{\rm s}_w -g'\,
 {\rm c}_w ){\rm s}_w \; \chi_A^* \chi_A \nn\\
 &&\qquad (g\,{\rm c}_w
+g'\, {\rm s}_w ){\rm s}_w\; \chi_Z^*\chi_A +(g\,{\rm s}_w -g'\,
{\rm c}_w ){\rm c}_w \; \chi_Z^* \chi_A \Biggr] \ea

 \subsection{\label{a:coup}Neutral couplings}

   The neutral current lagrangian is
   \be {\cal L}_{nc} =\sum_\psi \bar \psi \left(
n_\psi^A \slash A +n_\psi ^Z \slash Z \right) \psi \ee where the
sum runs over all the fermionic fields with non-zero isospin,
$\psi\in\{\nu_L,e_L,e_R,d_L,d_R,u_L,u_R\}   $.
The couplings are  \ba & n_\nu^A & = -\frac12 g' {\rm c}_w +\frac12 g {\rm
s}_w \simeq {\frac\epsilon2} \frac{gg'}{\sqrt{g^2+g^{\prime2}}}
\label{nnnu}\\ & n_\nu^Z & = \frac12 g' {\rm s}_w +\frac12 g
{\rm c}_w \simeq \frac12 \sqrt{g^2+g^{\prime2}} \\ & n_{e_L}^A &
= -\frac12 g' {\rm c}_w -\frac12 g {\rm s}_w \simeq -
\frac{gg'}{\sqrt{g^2+g^{\prime2}} } \left(1+\frac\epsilon2
\frac{g^2-g^{\prime2}}{g^2+g^{\prime2}} \right) \\ & n_{e_R}^A &
= -g' {\rm c}_w \simeq - \frac{gg'}{\sqrt{g^2+g^{\prime2}}}
\left(1-\epsilon \frac{g^{\prime2}}{g^2+g^{\prime2}} \right) \\
& n_{e_L}^Z & = \frac12 g' {\rm s}_w -\frac12 g {\rm c}_w \simeq
-\frac12 \frac{g^2-g^{\prime2}}{\sqrt{g^2+g^{\prime2}}}
\left(1-2\epsilon \frac{g^2g^{\prime2}}{g^4-g^{\prime4}} \right)
\\ & n_{e_R}^Z & = g' {\rm s}_w \simeq
\frac{g^{\prime2}}{\sqrt{g^2+g^{\prime2}}} \left(1+\epsilon
\frac{g^{2}}{g^2+g^{\prime2}} \right) \\ & n_{u_L}^A & = \frac16
g' {\rm c}_w +\frac12 g {\rm s}_w \simeq \frac23
\frac{gg'}{\sqrt{g^2+g^{\prime2}} } \left(1+\frac\epsilon4
\frac{3g^2-g^{\prime2}}{g^2+g^{\prime2}} \right) \\ & n_{u_R}^A
& = \frac23 g' {\rm c}_w \simeq \frac23
\frac{gg'}{\sqrt{g^2+g^{\prime2}}} \left(1-\epsilon
\frac{g^{\prime2}}{g^2+g^{\prime2}} \right) \\ & n_{u_L}^Z & =
-\frac16 g' {\rm s}_w +\frac12 g {\rm c}_w \simeq \frac16
\frac{3g^2-g^{\prime2}}{\sqrt{g^2+g^{\prime2}}}
\left(1-4\epsilon \frac{g^2g^{\prime2}}{(g^2+g^{\prime2})(3g^2
-g^{\prime2})} \right) \\ & n_{u_R}^Z & = -\frac23 g' {\rm s}_w
\simeq - \frac23 \frac{g^{\prime2}}{\sqrt{g^2+g^{\prime2}}}
\left(1+\epsilon \frac{g^{2}}{g^2+g^{\prime2}} \right) \\ &
n_{d_L}^A & = \frac16 g' {\rm c}_w - \frac12 g {\rm s}_w \simeq
-\frac13 \frac{gg'}{\sqrt{g^2+g^{\prime2}}}
\left(1+\frac\epsilon4 \frac{3g^2+ g^{\prime2}}{g^2+g^{\prime2}}
\right) \\ & n_{d_R}^A & = -\frac13 g' {\rm c}_w \simeq -\frac13
\frac{gg'}{\sqrt{g^2+g^{\prime2}}} \left(1-\epsilon
\frac{g^{\prime2}}{g^2+g^{\prime2}} \right) \\ & n_{d_L}^Z & =
-\frac16 g' {\rm s}_w - \frac12 g {\rm c}_w \simeq - \frac16
\frac{3 g^2+g^{\prime2}}{\sqrt{g^2+g^{\prime2}}}
\left(1-2\epsilon \frac{g^2g^{\prime2}}{(g^2+g^{\prime2}) (3g^2
+g^{\prime2})} \right) \\ & n_{d_R}^Z & = \frac13 g' {\rm s}_w
\simeq \frac13 \frac{g^{\prime2}}{\sqrt{g^2+g^{\prime2}}}
\left(1+\epsilon \frac{g^{2}}{g^2+g^{\prime2}} \right)
\label{nnnv}\ea

These neutral currents can be rewritten in Dirac spinor notation as follows:
\be
{\cal L}_{nc} = \sum_\psi \left\{
\bar \psi \slash  A (v_\psi^A +a_\psi^A \gamma_5) \psi
+  \bar \psi \slash  Z (v_\psi^Z +a_\psi^Z \gamma_5) \psi
\right\}  \ee
where the sum runs over $\psi\in\{\nu,e,u,d\}$, and we recall that the
left--handed projector is $(1+\gamma_5)/2$. The various couplings are the
following:
\ba
&&v_\nu^A= a_\nu^A=-a_e^A=a_u^A=-a_d^A =
\frac{1}{4}\left( g'{\rm s} _w- g'{\rm c}_w  \right)
\simeq \frac\epsilon4 \frac{g g'}{\sqrt{g^2 +g^{\prime2}}} \\
&&v_\nu^Z= a_\nu^Z=-a_e^Z=a_u^Z=-a_d^Z=
\frac{1}{2}\left(g{\rm c} _w + g'{\rm s}_w \right)
\simeq \frac12 \sqrt{g^2 +g^{\prime2}} \\
&&v_e^A= -\frac14 g{\rm s}_w -\frac34 g' {\rm c}_w
\simeq  -\frac{gg'}{\sqrt{g^2 +g^{\prime2} } } \left(1 +
\frac\epsilon4 \frac{ g^2 -3g^{\prime2} } {g^2 +g^{\prime2} } \right) \\
&&v_u^A= \frac{g}{4}\sin\theta_w +\frac{5g'}{12}\cos\theta_w
\simeq \frac23 \frac{gg'}{\sqrt{g^2 +g^{\prime2} }} \left( 1+
\frac\epsilon8 \frac {3g^2-5g^{\prime2} }{g^2 +g^{\prime2} } \right) \\
&&v_d^A= -\frac{g}{4}\sin\theta_w -\frac{g'}{12}\cos\theta_w  \simeq
 - \frac13 \frac{gg'}{\sqrt{g^2 +g^{\prime2} }} \left( 1+
 \frac\epsilon4 \frac {3g^2-g^{\prime2} }{g^2 +g^{\prime2} } \right) \\
&&v_e^Z=  -\frac14 g{\rm c}_w +\frac34 g' {\rm s}_w
\simeq  \frac{1}{4 \sqrt{g^2 +g^{\prime2} } } \left(-g^2 +3g^{\prime2}
+  4\epsilon \frac{ g^2 g^{\prime2} } {g^2 +g^{\prime2} } \right)  \\
&&v_u^Z=\frac{g}{4}\cos\theta_w -\frac{5g'}{12}\sin\theta_w
\simeq   \frac{ 1 }{12 \sqrt{g^2 +g^{\prime2} }}
\left( 3g^2 -5g^{\prime2} -8\epsilon \frac {g^2g^{\prime2} }
{ g^2 +g^{\prime2} } \right) \\
&&v_d^Z= -\frac{g}{4}\cos\theta_w +\frac{g'}{12}\sin\theta_w
\simeq   \frac{ 1}{12 \sqrt{g^2 +g^{\prime2} }}
\left( -3g^2 +g^{\prime2} +4\epsilon \frac {g^2g^{\prime2} }
{  g^2 +g^{\prime2}  }\right)
\ea

\vfill\eject


\bibliography{stuck}

\end{document}